\documentclass[11pt, a4paper]{article}
\usepackage{jheppub}
\input{epsf.tex}
\usepackage{graphicx}
\usepackage{amsfonts}
\usepackage{amsmath}
\usepackage{amsbsy}
\usepackage{colortbl}
\usepackage{booktabs}
\usepackage{multirow}
\usepackage{anysize}
\usepackage{array}
\usepackage{latexsym, amscd, amsthm}
\usepackage{pdfsync}
\usepackage{epsf}
\usepackage[all]{xy}
\usepackage[usenames,dvipsnames]{xcolor}
\usepackage{float}
\restylefloat{table}
\usepackage{upgreek}
\usepackage{enumerate}

\DeclareSymbolFont{extraup}{U}{zavm}{m}{n}
\DeclareMathSymbol{\varheart}{\mathalpha}{extraup}{86}
\DeclareMathSymbol{\vardiamond}{\mathalpha}{extraup}{87}

\usepackage{fancyhdr}

\newtheorem{Theorem}{Theorem}[section]
\newtheorem{Lemma}[Theorem]{Lemma}
\newtheorem{Proposition}[Theorem]{Proposition}
\theoremstyle{definition}
\newtheorem{Definition}[Theorem]{Definition}
\theoremstyle{remark}

\newcommand{\Sym}{\mathrm{Sym}}

\newcommand{\bcE}{\boldsymbol{\check{E}}}

\newcommand{\bp}{\begin{Proposition}}
\newcommand{\ep}{\end{Proposition}}
\newcommand{\bl}{\begin{Lemma}}
\newcommand{\el}{\end{Lemma}}
\newcommand{\bt}{\begin{Theorem}}
\newcommand{\et}{\end{Theorem}}
\newcommand{\bd}{\begin{Definition}}
\newcommand{\ed}{\end{Definition}}
\newcommand{\End}{\mathrm{End}}

\newcommand{\g}{{\bf g}}

\newcommand{\ev}{\mathrm{ev}}
\newcommand{\eqdef}{\stackrel{{\rm def.}}{=}}

\newcommand{\cinf}{{{\cal \cC}^\infty(M,\R)}}

\DeclareFontFamily{U}{rsf}{}
\DeclareFontShape{U}{rsf}{m}{n}{<5> <6> rsfs5 <7> <8> <9> rsfs7 <10-> rsfs10}{}
\DeclareMathAlphabet\Scr{U}{rsf}{m}{n}

\newcommand{\KA}{K\"{a}hler-Atiyah~}

\def\cW{\mathcal{W}}
\def\hV{{\hat V}}

\def\N{\mathbb{N}}
\def\Q{\mathbb{Q}}
\def\Z{\mathbb{Z}}
\def\C{\mathbb{C}}
\def\R{\mathbb{R}}

\def\rk{{\rm rk}}

\def\dd{\mathrm{d}}

\def\vol{\mathrm{vol}}
\def\AdS{\mathrm{AdS}}

\def\tf{\mathrm{tf}}
\def\Tor{\mathrm{Tor}}

\def\cQ{{\cal Q}}

\newcommand{\be}{\begin{equation*}}
\newcommand{\ee}{\end{equation*}}
\newcommand{\ben}{\begin{equation}}
\newcommand{\een}{\end{equation}}
\newcommand{\beqa}{\begin{eqnarray*}}
\newcommand{\eeqa}{\end{eqnarray*}}
\newcommand{\beqan}{\begin{eqnarray}}
\newcommand{\eeqan}{\end{eqnarray}}

\newcommand{\nn}{\nonumber}

\newcommand{\id}{\mathrm{id}}

\newcommand{\tr}{\mathrm{tr}}

\def\Diff{\mathrm{Diff}}
\def\per{\mathrm{per}}
\def\cR{{\mathcal R}}
\def\cC{{\mathcal C}}

\def\cB{\Scr B}

\def\cK{\mathrm{\cal K}}
\def\odd{\mathrm{odd}}
\def\Spin{\mathrm{Spin}}

\def\SO{\mathrm{SO}}
\def \so{\mathfrak{so}}

\def\cD{\mathcal{D}}

\def\cA{\mathcal{A}}

\def\cP{\mathcal{P}}
\def\cN{\mathcal{N}}
\def\cG{\mathcal{G}}
\def\cT{\mathcal{T}}
\def\cF{\mathcal{F}}
\def\cC{\mathcal{C}}

\def\G_2{\mathrm{G_2}}
\def\g2{\mathfrak{g}_2}
\def\cO{\mathcal{O}}
\def\cL{\mathcal{L}}
\def\cS{\mathcal{S}}

\def\btu{\bigtriangleup}

\def\P{\mathbb{P}}

\def\Wh{\mathrm{Wh}}

\def\mb{\mathbf{b}}
\def\mV{\mathbf{V}}
\def\mf{\mathbf{f}}
\def\mF{\mathbf{F}}
\def\momega{{\boldsymbol{\omega}}}
\def\cQ{\check{Q}}
\def\f{\mathfrak{f}}

\title{Foliated eight-manifolds for M-theory compactification}

\author{Elena Mirela Babalic$^{1,2}$, Calin Iuliu Lazaroiu$^3$ }

\affiliation{
   $^1$ Department of Theoretical Physics, National
  Institute of Physics and Nuclear Engineering, Str. Reactorului
  no.30, P.O.BOX MG-6, Postcode 077125, Bucharest-Magurele, Romania  \\
 $^2$ Department of Physics, University
  of Craiova, 13 Al. I. Cuza Str., Craiova  200585, Romania\\
 $^3$ Center for Geometry and Physics, Institute for Basic Science (IBS), Pohang 790-784, Republic of Korea
}

\emailAdd{mbabalic@theory.nipne.ro, calin@ibs.re.kr} 

\abstract{We characterize compact eight-manifolds $M$ which arise as
  internal spaces in ${\cal N}=1$ flux compactifications of M-theory
  down to $\AdS_3$ using the theory of foliations, for the case when
  the internal part $\xi$ of the supersymmetry generator is everywhere
  non-chiral. We prove that specifying such a supersymmetric
  background is equivalent with giving a codimension one foliation
  $\cF$ of $M$ which carries a leafwise $G_2$ structure, such that the
  O'Neill-Gray tensors, non-adapted part of the normal connection and
  the torsion classes of the $G_2$ structure are given in terms of the
  supergravity four-form field strength by explicit formulas which we
  derive.  We discuss the topology of such foliations, showing that
  the $C^\ast$ algebra $C(M/\cF)$ is a noncommutative torus of
  dimension given by the irrationality rank of a certain cohomology
  class constructed from $\mathbf{G}$, which must satisfy the Latour
  obstruction. We also give a criterion in terms of this class for
  when such foliations are fibrations over the circle. When the
  criterion is not satisfied, each leaf of $\cF$ is dense in $M$.}

\begin{document}

\maketitle 

\pagebreak

\vskip .6in

\section*{Introduction}

$\cN=1$ compactifications of $M$-theory on eight-manifolds
\cite{Becker1, Becker2, Constantin, MartelliSparks, Tsimpis} hold particular
interest due to their potential relation to F-theory \cite{Bonetti}
and since they provide nontrivial testing grounds for many physical
and mathematical ideas. In this paper, we reconsider the class of
supersymmetric compactifications of eleven-dimensional supergravity
down to $\AdS_3$ spaces --- which was pioneered in
\cite{MartelliSparks} --- using the theory of foliations.  Our purpose
is to give a complete mathematical characterization of those oriented,
compact and connected eight-manifolds $M$ which satisfy the
corresponding supersymmetry conditions, in the case when the internal
part of the supersymmetry generator is everywhere non-chiral --- thus
providing a supersymmetric realization of some of the ideas proposed
in \cite{Grana}.

Using a combination of techniques from the theory of \KA algebras and
of $G$-structures, an everywhere non-chiral Majorana spinor $\xi$ on
$M$ can be parameterized by a one-form $V$ whose kernel distribution
$\cD$ carries a $G_2$ structure. We show that the condition that $\xi$
satisfies the supersymmetry equations is equivalent with the
requirement that $\cD$ is Frobenius integrable (namely, a certain
one-form proportional to $V$ must belong to a cohomology class
specified by the supergravity four-form field strength $\mathbf{G}$)
and that the O'Neill-Gray tensors of the codimension one foliation
$\cF$ which integrates $\cD$, the non-adapted part of the normal
connection of $\cF$ as well as the torsion classes of the $G_2$
structure of $\cD$, be given in terms of $\mathbf{G}$ through explicit
expressions which we derive. In particular, we find that the leafwise
$G_2$ structure is ``integrable'', in the sense
$\boldsymbol{\tau}_2=0$, i.e. that it belongs to the Fernandez-Gray
class $W_1\oplus W_3\oplus W_4$ (in the notation of \cite{FG}) 
--- a class of $G_2$ structures which was studied in detail in \cite{FriedrichIvanov1,
  FriedrichIvanov2}. More precisely, we find that this leafwise $G_2$
structure is conformally co-calibrated, thus being --- up to a
conformal transformation --- of the type studied in
\cite{GrigorianCoflow}.  Furthermore, the field strength $\mathbf{G}$
can be determined in terms of the geometry of the foliation $\cF$ and
of the torsion classes of its longitudinal $G_2$ structure, provided
that $\cF$ and this $G_2$ structure satisfy some purely geometric
conditions, the form of which we derive explicitly. These results
complete the analysis initiated by \cite{MartelliSparks}, giving a
full solution to the problem via three theorems which we prove
rigorously.

We point out that existence of a nowhere-chiral Majorana spinor on $M$
is obstructed by a certain class having its origin in Novikov theory.
We also discuss the topology of $\cF$, giving a criterion in terms of
$\mathbf{G}$ which allows one to decide when the leaves of $\cF$ are
compact or dense in $M$ and thus when it is possible to present $\cF$
as a fibration over the circle.  When $\cF$ has dense leaves, its leaf
space admits a non-commutative geometric description in terms of a
$C^\ast$ algebra $C(M/\cF)$ which is Morita equivalent with that of a
non-commutative torus whose dimension is determined by the four-form
$\mathbf{G}$.

The paper is organized as follows. Section 1 gives a brief review of
the class of compactifications we consider. Section 2 discusses a geometric
characterization of spin 1/2 Majorana spinors on $M$ which are
nowhere-chiral and everywhere of norm one. It also gives the
description of such spinors through the \KA algebra of $M$ and a certain
parameterization which is essential for the rest of the paper.
Section 3 gives our equivalent characterizations of the supersymmetry
conditions, thus providing a complete geometric description of such
supersymmetric backgrounds; it also describes the Latour obstruction
to the existence of solutions. Section 4 discusses the topology of the
foliation $\cF$, giving a flux criterion for compactness of the leaves
and the non-commutative geometric model of its leaf space. Section 5
provides a brief comparison with previous work, while Section 6
concludes. The appendices contain various technical details.

\paragraph{Notations and conventions. } 
Throughout this paper, $M$ denotes an oriented, connected and compact
smooth manifold (which will mostly be of dimension eight), whose
unital commutative $\R$-algebra of smooth real-valued functions we
denote by $\cinf$. All vector bundles we consider are smooth. We use
freely the results and notations of \cite{ga1,ga2,gf}, with the same
conventions as there. To simplify notation, we write the geometric
product $\diamond$ of loc. cit. simply as juxtaposition while
indicating the wedge product of differential forms through $\wedge$.
If $\cD\subset TM$ is a Frobenius distribution on $M$, we let
$\Omega(\cD)=\Gamma(M,\wedge \cD^\ast)$ denote the $\cinf$-module of
longitudinal differential forms along $\cD$.  When $\dim M=8$, then
for any 4-form $\omega\in \Omega^4(M)$, we let $\omega^\pm\eqdef
\frac{1}{2}(\omega\pm \ast \omega)$ denote the self-dual and
anti-selfdual parts of $\omega$ (namely, $\ast \omega^\pm=\pm
\omega^\pm$). This paper assumes some familiarity with the theory of
foliations, for which we refer the reader to
\cite{CC1,CC2,Tondeur,Moerdijk,Bejancu}.

\section{Basics}
\label{sec:basics}
We start with a brief review of the set-up, in order to fix notation. 
As in \cite{MartelliSparks, Tsimpis}, we consider 11-dimensional
supergravity \cite{sugra11} on an eleven-dimensional connected and paracompact 
spin manifold $\mathbf{M}$ with Lorentzian metric $\mathbf{g}$ (of `mostly plus'
signature).  Besides the metric, the classical action of the theory contains the
three-form potential with four-form field strength
$\mathbf{G}\in\Omega^4(\mathbf{M})$ and the gravitino
$\mathbf{\Psi}$, which is a Majorana spinor of spin $3/2$. The bosonic part of the action takes the form: 
\be
S_{\rm bos}[\mathbf{g}, \mathbf{C}]=
\frac{1}{2\mathbf{\kappa}_{11}^2}\int_{\mathbf M}R\boldsymbol{\nu}-
\frac{1}{4\mathbf{\kappa}_{11}^2}\int_{\mathbf M}\big(\mathbf{G}\wedge \star \mathbf{G}+\frac{1}{3}\mathbf{C}\wedge \mathbf{G}\wedge \mathbf{G}\big)~~,
\ee
where $\mathbf{\kappa}_{11}$ is the gravitational coupling constant in eleven dimensions, $\boldsymbol{\nu}$ and $R$ are the volume form and the scalar curvature of $\mathbf{g}$ and 
$\mathbf{G}=\dd \mathbf{C}$.
For supersymmetric bosonic classical backgrounds, both the gravitino and
its supersymmetry variation must vanish, which requires that there
exist at least one solution $\boldsymbol{\eta}$ to the equation:
\ben
\label{susy}
\delta_{\boldsymbol{\eta}} \mathbf{\Psi} \eqdef  \mathfrak{D} \boldsymbol{\eta} = 0~~,
\een
where $\mathfrak{D}$ denotes the supercovariant connection. The
eleven-dimensional supersymmetry generator $\boldsymbol{\eta}$ is a
Majorana spinor (real pinor) of spin $1/2$ on $\mathbf{M}$.

As in \cite{MartelliSparks, Tsimpis}, consider compactification down
to an $\AdS_3$ space of cosmological constant $\Lambda=-8\kappa^2$,
where $\kappa$ is a positive real parameter --- this includes the
Minkowski case as the limit $\kappa\rightarrow 0$.  Thus $\mathbf{
  M}=N\times M$, where $N$ is an oriented 3-manifold diffeomorphic
with $\R^3$ and carrying the $\AdS_3$ metric $g_3$ while $M$ is an oriented, 
compact and connected Riemannian eight-manifold whose metric we denote by $g$. The metric on
$\mathbf{M}$ is a warped product:
\beqan
\label{warpedprod}
\dd \mathbf{s}^2  & = & e^{2\Delta} \dd s^2~~~{\rm where}~~~\dd s^2=\dd s^2_3+ g_{mn} \dd x^m \dd x^n~~.
\eeqan
The warp factor $\Delta$ is a smooth real-valued function defined on $M$ while $\dd s_3^2$ is the
squared length element of the $\AdS_3$ metric $g_3$. For the field strength $\mathbf{G}$, we use the ansatz:
\ben
\label{Gansatz}
\mathbf{ G} = \nu_3\wedge \mathbf{f}+\mF~~,~~~~\mathrm{with}~~ 
\mF\eqdef e^{3\Delta}F~~,~~\mathbf{f}\eqdef e^{3\Delta} f~~,
\een
where $f\in \Omega^1(M)$, $F\in \Omega^4(M)$ and $\nu_3$ is the
volume form of $(N,g_3)$. For $\boldsymbol{\eta}$, we use the ansatz:
\be
\boldsymbol{\eta}=e^{\frac{\Delta}{2}}(\zeta\otimes \xi)~~,
\ee
where $\xi$ is a Majorana spinor of spin $1/2$ on the internal space
$(M,g)$ (a section of the rank 16 real vector bundle $S$ of indefinite chirality real 
pinors) and $\zeta$ is a Majorana spinor on $(N,g_3)$. 

Assuming that $\zeta$ is a Killing spinor on the $\AdS_3$ space $(N,g_3)$, the
supersymmetry condition \eqref{susy} is equivalent with the following
system for $\xi$:
\ben
\label{par_eq}
\boxed{\mathbb{D}\xi = 0~~,~~Q\xi = 0}~~,
\een 
where 
\be
\mathbb{D}_X=\nabla_X^S+\frac{1}{4}\gamma(X\lrcorner F)+\frac{1}{4}\gamma((X_\sharp\wedge f) \nu) +\kappa \gamma(X\lrcorner \nu)~~,~~X\in \Gamma(M,TM)
\ee
is a linear connection on $S$ (here $\nabla^S$ is the connection induced on $S$ by the Levi-Civita
connection of $(M,g)$, while $\nu$ is the volume form of $(M,g)$) and 
\be
Q=\frac{1}{2}\gamma(\dd \Delta)-\frac{1}{6}\gamma(\iota_f\nu)-\frac{1}{12}\gamma(F)-\kappa\gamma(\nu)
\ee 
is a globally-defined endomorphism of $S$. As in \cite{MartelliSparks,
  Tsimpis}, {\em we do not require that $\xi$ has definite chirality}.

The set of solutions of \eqref{par_eq} is a finite-dimensional
$\R$-linear subspace $\cK(\mathbb{D},{Q})$ of the infinite-dimensional
vector space $\Gamma(M,S)$ of smooth sections of $S$. Up to rescalings by
smooth nowhere-vanishing real-valued functions defined on $M$, the
vector bundle $S$ has two admissible pairings $\cB_\pm$ (see \cite{gf,
  AC1, AC2}), both of which are symmetric but which are distinguished
by their types $\epsilon_{\cB_\pm}=\pm 1$. Without loss of generality,
we choose to work with $\cB\eqdef \cB_+$.  We can in fact take $\cB$
to be a scalar product on $S$ and denote the corresponding norm by
$||~||$ (see \cite{ga1,ga2} for details). Requiring that the
background preserves exactly $\cN=1$ supersymmetry amounts to asking
that $\dim \cK(\mathbb{D},Q)=1$. It is not hard to check \cite{ga1} that
$\cB$ is $\mathbb{D}$-flat:
\ben
\label{flatness}
\dd \cB(\xi',\xi'')=\cB(\mathbb{D}\xi', \xi'')+\cB(\xi',\mathbb{D}\xi'')~~,
~~\forall \xi',\xi''\in \Gamma(M,S)~~.
\een
Hence any solution of \eqref{par_eq} which has unit $\cB$-norm at a
point will have unit $\cB$-norm at every point of $M$ and we can take
the internal part $\xi$ of the supersymmetry generator to be
everywhere of norm one. 

Besides the supersymmetry equations \eqref{par_eq}, one has the
Bianchi identity $\dd\mathbf{ G}=0$, which gives:
\ben
\label{Bianchi}
\dd\mF=\dd\mathbf{f}=0~~
\een
and one must impose the equations of motion\footnote{We use the
  conventions of \cite{ga1} for the Hodge operator, which are the
  standard conventions in the Mathematics literature; these
  conventions are recalled in Appendix \ref{app:GA}.}:
\ben
\label{eom11}
\mathbf{\dd} \star \mathbf{G}+\frac{1}{2}\mathbf{G}\wedge \mathbf{G}=0~~ 
\een
for the supergravity three-form potential, where $\star$ is the Hodge
operator of $(\mathbf{M},\mathbf{g})$. It is not hard to check that these amount to the following
conditions, where $\ast$ is the Hodge operator of $(M,g)$:
\beqan
\label{eom}
&&e^{-6\Delta}\dd(e^{6\Delta}\ast f)-\frac{1}{2}F\wedge F~~=0\nn\\
&&e^{-6\Delta}\dd(e^{6 \Delta}\ast F)-f\wedge F~~~~=0~~.
\eeqan
Together with the supersymmetry conditions, it follows from the arguments of
\cite{Eins} that \eqref{eom} imply the Einstein equations.  It was
noticed in \cite{MartelliSparks} that integrating the scalar part of
the Einstein equations:
\be
e^{-9 \Delta}\Box e^{9\Delta}+72\kappa^2=\frac{3}{2}||F||^2+3||f||^2
\ee
implies, when $\kappa=0$, that $F$ and $f$ must vanish identically on
$M$ (and thus $\mathbf{G}$ must vanish identically on $\mathbf{M}$)
while $\Delta$ must be constant on $M$. In that case $Q=0$ and
$\mathbb{D}=\nabla^S$ so \eqref{par_eq} reduce to the condition that
$\xi$ is covariantly constant on $M$, which implies that each of the
chiral components $\xi^\pm\eqdef \frac{1}{2}(1\pm \gamma(\nu))\xi$
must be covariantly constant (and thus of constant norm). When
$\xi$ is chiral (i.e. when $\xi^+=0$ or $\xi^-=0$), this means that
$(M,g)$ has holonomy contained in $\Spin(7)$ (equaling $\Spin(7)$ iff. 
$M$ is simply connected), while when $\xi$ is nowhere chiral
(i.e. when both $\xi^+$ and $\xi^-$ are nowhere vanishing), this
means that $(M,g)$ has holonomy contained in $G_2$ (equaling the latter iff. $M$ 
has finite fundamental group).

\paragraph{Remarks.}
\begin{enumerate}
\itemsep0em  
\item In early work on $\cN=1$ compactifications of M-theory on
  eight-manifolds \cite{Becker1, Becker2, Constantin}, it was assumed that the
  external space is Minkowski (thus $\kappa=0$) and that the internal
  part $\xi$ of the supersymmetry generator is chiral everywhere.  As
  recalled above following \cite{MartelliSparks}, classical
  consistency of the compactifications of \cite{Becker1} requires that
  $\mathbf{G}$ vanishes and that $\Delta$ is constant on $M$, while
  the supersymmetry conditions imply that $(M,g)$ has holonomy group
  contained in $\Spin(7)$. This conclusion is modified {\em at the quantum level} if
  one includes \cite{Becker1, Becker2, Constantin} the leading quantum correction in the
  right hand side of the equation of motion for $\mathbf{G}$ as well
  as the leading correction to the Einstein equation.  The first
  correction (which arises from the M5-brane anomaly \cite{Minasian})
  has the form $\frac{\beta}{c}{\hat X}_8$, where $c\eqdef \frac{1}{(2\pi)^4 3^2 2^{13}}$ 
is a dimensionless number while $\beta$ is a factor of order
  $\boldsymbol{\kappa}_{11}^{4/3}$ and ${\hat X}_8=\frac{1}{128
    (2\pi)^4}\left[\mathbf{R}^4-\frac{1}{4}(\tr
    \mathbf{R}^2)^2\right]$, where $\mathbf{R}$ is the curvature form
  of $\mathbf{g}$.  The second correction has the form
  $\beta\frac{1}{\sqrt{|\det \mathbf{g}|}}\frac{\delta}{\delta
    \mathbf{g}^{AB}}\left[\sqrt{|\det
      \mathbf{g}|}(J_0-\frac{1}{2}E_8)\right]$, where $J_0$ and $E_8$
  are polynomials in the curvature tensor of $\mathbf{g}$.  This
  allows one \cite{Becker2,Constantin} to turn on a {\em small} flux
  $\mathbf{G}$ of order $\boldsymbol{\kappa}_{11}^{2/3}$ (with higher corrections
  controlled by the ratio between the eleven-dimensional Planck length
  and the radius of the manifold $M$), thus {\em slightly perturbing} a
  fluxless classical solution of the type $\R^{1,2}\times M$, where
  $M$ is a $\Spin(7)$ holonomy manifold; the argument of
  \cite{MartelliSparks} no longer applies to the quantum-corrected
  Einstein equations.  Hence the framework of \cite{Becker1,
    Becker2,Constantin} only allows for small fluxes which are induced by
  quantum effects, fluxes which are suppressed by powers
  of $\boldsymbol{\kappa}_{11}$.
\item In the present paper, as in \cite{MartelliSparks, Tsimpis}, we
  do {\em not} require that $\kappa=0$ or that $\xi$ be chiral. This
  extension of the framework of \cite{Becker1, Becker2, Constantin} allows for
  non-vanishing fluxes which are already present in the classical
  limit and which {\em need not be small/suppressed by powers of
    $\boldsymbol{\kappa}_{11}$}.  Unlike the small fluxes considered in
  \cite{Becker1,Becker2, Constantin}, our fluxes do not have a quantum
  origin and hence are {\em not} constrained by the tadpole
  cancellation condition $\int_{M}{\mathbf{F}\wedge
    \mathbf{F}}=\frac{(2\pi^2)^{1/3}}{6}\boldsymbol{\kappa}_{11}^{4/3}\chi(M)$.  We will
  in fact be considering only the case when $\xi$ is everywhere
  non-chiral, so in this sense we will be `maximally far' from the
  classical limit of the framework discussed in \cite{Becker1, Becker2, Constantin}. As in
  \cite{MartelliSparks}, we do not need to (and will not) include
  quantum corrections in order to obtain flux solutions, since the
  class of backgrounds we consider is already consistent at the
  classical level in the presence of fluxes --- unlike
  compactifications with $\kappa=0$ and chiral $\xi$. Notice that one has 
  to consider compactifications down to spaces which are different from 3-dimensional 
  Minkowski space in order to have fluxes that are not suppressed in the manner 
  of those of \cite{Becker1, Becker2, Constantin}.
\item The seemingly innocuous relaxation of the framework of
  \cite{Becker1} obtained by allowing a non-chiral $\xi$ and a
  non-vanishing $\kappa$ increases dramatically the complexity of the
  problem. Unlike the common sector backgrounds of type IIA/B
  theories \footnote{Backgrounds where the Kalb-Ramond field strength
    $H$ is nonzero, while all RR field strengths vanish.}, presence of
  the terms induced by the four-form $F$ in the connection
  $\mathbb{D}$ prevents one from expressing the latter as the
  connection induced on $S$ by a torsion-full, metric-compatible
  deformation of the Levi-Civita connection of $(M,g)$ --- hence one
  cannot rely (as done, for example, in \cite{Puhle2}) on the
  well-understood theory of torsion-full metric connections (see
  \cite{Agricola} for an introduction). Also notice that equations
  \eqref{par_eq} are {\em not} of the form considered in
  \cite{FriedrichAgricola, Puhle1}. We will find, however, that $M$
  admits a foliation which carries a leafwise $G_2$ structure with
  $\boldsymbol{\tau}_2=0$ and hence the approach of
  \cite{FriedrichIvanov1, FriedrichIvanov2} can be applied along the
  leaves of this foliation, in order to produce a {\em leafwise}
  partial connection with totally antisymmetric torsion which governs
  the intrinsic geometry of the leaves.
\end{enumerate}

\section{Characterizing an everywhere non-chiral normalized Majorana spinor}
\label{sec:spinor}

\subsection{The inhomogeneous form defined by a Majorana spinor}

Fixing a Majorana spinor $\xi\in \Gamma(M,S)$ which is everywhere of
$\cB$-norm one, consider the inhomogeneus differential form:
\ben
\label{checkE}
\check{E}_{\xi,\xi}=\frac{1}{16} \sum_{k=0}^8 \bcE^{(k)}_{\xi,\xi}\in \Omega(M)~~,
\een
whose rescaled rank components have the following expansions in any
local orthonormal coframe $(e^a)_{a=1\ldots 8}$ of $M$ defined on some
open subset $U$:
\ben
\label{Exi}
\bcE^{(k)}_{\xi,\xi}=_U\frac{1}{k!}\cB(\xi,\gamma_{a_1...a_k}\xi)e^{a_1...a_k} \in \Omega^k(M)~~.
\een
The non-zero components turn out to have ranks $k=0,1,4,5$ and we have
$\cS(\check{E}_{\xi,\xi})=\bcE^{(0)}_{\xi,\xi}=||\xi||^2=1$, where
$\cS$ is the canonical trace of the \KA algebra (see Appendix
\ref{app:GA}). Hence:
\ben
\label{Eexp}
\boxed{\check{E}=\frac{1}{16}(1+V+Y+Z+b\nu)}~~,
\een
where we introduced the notations: 
\ben
\label{forms8}
V\eqdef \bcE^{(1)}  ~~,~~ Y\eqdef\bcE^{(4)} ~,~~ Z\eqdef \bcE^{(5)}~~,~~ b\nu\eqdef \bcE^{(8)}~~.
\een
Here, $b$ is a smooth real valued function defined on $M$ and $\nu$ is
the volume form of $(M,g)$, which satisfies $||\nu||=1$; notice the
relation $\cS(\nu\check{E}_{\xi,\xi})=b$. On a small enough open subset
$U\subset M$ supporting a local coframe $(e^a)$ of $M$, one has the
expansions:
\beqan
\label{forms8alt}
&&V=_U\cB(\xi,\gamma_a\xi)e^a ~~,~~ Y=_U\frac{1}{4!}\cB(\xi,\gamma_{a_1\ldots a_4}\xi) e^{a_1\ldots a_4}~,\nn\\
&& Z=_U\frac{1}{5!} \cB(\xi,\gamma_{a_1\ldots a_5}\xi) e^{a_1\ldots a_5}~~,
~~ b=_U\cB(\xi, \gamma(\nu)\xi)~~.
\eeqan
We have $b=||\xi^+||^2-||\xi^-||^2$ and
$||\xi^+||^2+||\xi^-||^2=||\xi||^2=1$, where $\xi^\pm\eqdef
\frac{1}{2}(1\pm \gamma^{(9)})\xi\in \Gamma(M,S^\pm)$ are the positive
and negative chirality components of $\xi$, which are global sections
of the positive and negative chirality sub-bundles $S^\pm$ of $S$ (we
have $S=S^+\oplus S^-$). Notice that the inequality $|b|\leq 1$ holds
on $M$, with equality only at those $p\in M$ where $\xi_p$ has
definite chirality.

\subsection{Restriction to Majorana spinors which are everywhere non-chiral}
In this paper, we study only the case when $\xi$ is everywhere
non-chiral on $M$, i.e. the case $|b|<1$ everywhere. This amounts to
requiring that each of the chiral components $\xi^\pm$ is
nowhere-vanishing --- an assumption which we shall make from now
on. It is well-known (see, for example, \cite{Isham1, Isham2}) that
the topological condition for existence on $M$ of a nowhere-vanishing
Majorana-Weyl spinor $\xi^\pm$ of chirality $\pm 1$
(i.e. corresponding to the representations ${\bf 8}_s$ and ${\bf 8}_c$
of $\Spin(8)$) is that the following relation holds between the Euler
number of $M$ and the first and second Pontryaghin classes $p_1$,
$p_2$ of its tangent bundle:
\be
\chi(M)=\pm \frac{1}{2}\int_{M}(p_2-\frac{1}{4}p_1^2)~~.
\ee
Since we require that both chiral components $\xi^\pm$ of 
$\xi$ must be everywhere non-vanishing, we must hence have: 
\ben
\label{SpinorExistenceCond}
\boxed{\chi(M)=\int_{M}(p_2-\frac{1}{4}p_1^2)=0}~~.
\een

\subsection{The Fierz identities}
The conditions:
\ben
\label{Esquare}
\check{E}^2=\check{E}~~,~~\cS(\check{E})=1~~,~~\tau(\check{E})=\check{E}~~,
~~ |\cS(\nu\check{E})|< 1
\een
amount \cite{ga1} to the requirement that 
an inhomogeneous form $\check{E}\in \Omega(M)$  
is given by \eqref{Exi} for some normalized Majorana spinor $\xi$ which is
everywhere non-chiral; that spinor is in fact determined up to a
global sign factor by an inhomogeneous form $\check{E}$ which
satisfies \eqref{Esquare}. Expanding the first condition in
\eqref{Esquare} into generalized products and separating ranks, one
can analyze the resulting system as in \cite{ga1}.  One finds
\cite{MartelliSparks,ga1} that \eqref{Esquare} is equivalent with the
following relations which hold on $M$:
\ben
\label{SolMS}
\boxed{
\begin{split}
& ||V||^2=1-b^2>0~~,\\
& \iota_V(\ast Z)=0~~,~~\iota_V Z=Y-b\ast Y~~~~\\
&(\iota_\alpha (\ast Z)) \wedge (\iota_\beta (\ast Z)) \wedge  (\ast Z) 
= - 6 \langle \alpha\wedge V, \beta\wedge V\rangle \iota_V \nu~~,~~\forall \alpha,\beta\in \Omega^1(M)~~.
\end{split}}
\een
In fact, these equations (which generate the algebra of Fierz
relations \cite{ga1} of $\xi$) also hold on $M$ in the general case
when the chiral locus is allowed to be non-empty \cite{ga1} and they
characterize $\cB$-norm one Majorana spinors up to sign also in that
case.  Notice that the first relation in the second row is equivalent
with $V\wedge Z=0$. 

\paragraph{Remark.}
Let (R) denote the second relation (namely $\iota_V Z=Y-b\ast Y$) on
the second row of \eqref{SolMS}.  Adding (R) to $b$ times its Hodge
dual and using the identity $\ast \iota_V Z=V\wedge \ast Z$ shows that
(R) implies the following condition which was given\footnote{The
  comparison with \cite{MartelliSparks} can be found in Appendix
  \ref{app:other}.} in
\cite{MartelliSparks}:
\ben
\label{YVZ}
(1-b^2)Y=\iota_V Z+bV\wedge (\ast Z)=\iota_VZ+b \ast(\iota_VZ)~~.
\een
Notice that this last condition is weaker than (R) unless $\xi$ is
required to be everywhere non-chiral.  Indeed, subtracting $b$ times
the Hodge dual of \eqref{YVZ} from \eqref{YVZ} gives
$(1-b^2)\iota_VZ=(1-b^2)(Y-b\ast Y)$, which implies relation (R) only
when $|b|$ is different from one on all of $M$.

\subsection{The Frobenius distribution and almost product structure defined by $V$}

Since $V$ is nowhere-vanishing, it determines a corank one Frobenius
distribution $\cD=\ker V\subset TM$ on $M$, whose rank one
orthocomplement (taken with respect to the metric $g$) we denote by
$\cD^\perp$. This provides an orthogonal direct sum decomposition:
\be
TM=\cD\oplus \cD^\perp
\ee
and thus defines an orthogonal almost product structure $\cP\in
\Gamma(M,\End(TM))$, namely the unique $g$-orthogonal and involutive
endomorphism of $TM$ whose eigenbundles for the eigenvalues $+1$ and
$-1$ are given by $\cD$ and $\cD^\perp$ respectively. Equivalently,
$V$ provides a reduction of structure group of $TM$ from $\SO(8)$ to
$\SO(7)$, where $\SO(7)$ acts on $T_p M$ at $p\in M$ as the isotropy
subgroup $\SO(\cD_p,g_p|_{\cD_p})$ of $V_p$ in $\SO(8)$.  For later
convenience, we introduce the normalized vector field:
\ben
\label{n}
n\eqdef \hV^\sharp=\frac{V^\sharp}{||V||}~~,~~||n||=1~~,
\een
which is everywhere orthogonal to $\cD$ and generates
$\cD^\perp$. Thus $\cD^\perp$ is trivial as a real line bundle and
$\cD$ is transversely oriented by $n$. Since $M$ itself is oriented,
this also provides an orientation of $\cD$ which agrees with that
defined by the longitudinal volume form
$\nu_\top=\iota_\hV\nu=n\lrcorner \nu\in
\Omega^7(\cD)=\Gamma(M,\wedge^7 \cD)$ in the sense that:
\be
\hV\wedge \nu_\top=\nu~~.
\ee
We let $\ast_\perp:\Omega(\cD)\rightarrow \Omega(\cD)$ be the Hodge
operator along $\cD$, taken with respect to this orientation of $\cD$:
\ben
\label{AstV}
\ast_\perp \omega=\ast(\hV\wedge \omega)=(-1)^{\rk \omega}\iota_\hV(\ast \omega)=
\tau(\omega)\nu_\top~~,~~\forall \omega\in \Omega(\cD)~~.
\een
We have: 
\be
\ast\omega=(-1)^{\rk \omega}{\hat V}\wedge \ast_\perp \omega~~.
\ee
Notice that $\cD$ is endowed with the metric $g|_\cD$ induced by $g$, which,
together with its orientation defined above, gives it an $\SO(7)$
structure as a vector bundle; this is the $\SO(7)$ structure mentioned
above.

\paragraph{Remark.} As mentioned above,  $\cD$ is also transversely orientable, an
orientation of its normal line bundle $\cD^\perp$ being given by
the image of ${\hat V}$ through the vector bundle epimorphism $\lambda_\cD:T^\ast
M\twoheadrightarrow (\cD^\perp)^\ast$ which is dual to the inclusion morphism
$\cD^\perp\hookrightarrow TM$. 

\paragraph{Proposition.} 
Relations \eqref{Esquare} are equivalent with the following conditions:
\ben
\label{fsol}
V^2=1-b^2~~,~~Y=(1+b\nu)\psi~~,~~Z=V\psi~~,
\een
where $\psi\in \Omega^4(\cD)$ is the canonically normalized
coassociative form of a $G_2$ structure on the distribution $\cD$
which is compatible with the metric $g|_\cD$ induced by $g$ and with the
orientation of $\cD$ discussed above.

\

\noindent We let $\varphi\eqdef \ast_\perp \psi\in \Omega^3(\cD)$ be
the associative form of the $G_2$ structure on $\cD$ mentioned in the
proposition. 

\paragraph{Remark.} We remind the reader that the canonical normalization condition for
the associative form $\varphi$ (and thus also for the coassociative
form $\psi=\ast_\perp \varphi$) of a $G_2$ structure on $\cD$ is:
\ben
\label{CanNorm}
||\psi||^2=||\varphi||^2=7~~.
\een
Notice the relations: 
\be
\varphi=\ast_\perp\psi=\ast ({\hat V}\wedge \psi)~~,~~
\ast\varphi=-{\hat V}\wedge \psi~~,~~\ast\psi=\hV\wedge \varphi~~,
\ee
where we used the fact that $\ast_\perp^2=\id_{\Omega(\cD)}$ while
$\ast^2=\pi$, where $\pi$ is the parity automorphism of the \KA
algebra of $(M,g)$ (see Appendix \ref{app:GA}).

\

\noindent{\bf Proof.} 
Since we already know that \eqref{Esquare} are equivalent with
\eqref{SolMS}, it is enough to show that \eqref{fsol} are equivalent
with the latter.

\

\noindent 1. (the direct implication)~Let us assume that \eqref{SolMS}
hold. The first relation in \eqref{fsol} coincides with the first
equation in \eqref{SolMS}.  Since $V$ is nowhere vanishing, it is
invertible as an element of the \KA algebra of $(M,g)$, with inverse:
\be
V^{-1}=\frac{1}{||V||^2}V=\frac{1}{1-b^2}V~~,
\ee
where we used the fact that $V^2=||V||^2$. Also notice that $1\pm
b\nu$ is invertible, with inverse:
\be
(1\pm b\nu)^{-1}=\frac{1}{1-b^2}(1\mp b\nu)~~.
\ee
We define the 3-form: 
\ben
\label{phidef}
\varphi\eqdef \frac{1}{||V||}\ast Z=\frac{1}{\sqrt{1-b^2}} Z\nu \in \Omega^3(\cD)~~.
\een
The first equation on the second row of \eqref{SolMS} gives
$\iota_V\varphi=0$, which means that $\varphi$ can be viewed as an
element of $\Omega^3(\cD)$.  Hence the condition on the third row of
\eqref{SolMS} is equivalent with:
\ben
\label{G2rel}
(\iota_\alpha \varphi) \wedge (\iota_\beta \varphi) \wedge  \varphi = 
-6 \langle \alpha, \beta\rangle \nu_\top~~,~~\forall \alpha,\beta\in \Omega^1(\cD)~~,
\een
which shows \cite{Kthesis} that $\varphi$ (when viewed as an element
of $\Omega^3(\cD)$) is the canonically normalized associative form of
a $G_2$ structure on the distribution $\cD$, which is compatible with
the metric induced by $g$ on $\cD$ and with the orientation of $\cD$
discussed above. The corresponding coassociative 4-form along $\cD$
is:
\ben
\label{psidef}
\psi\eqdef \ast_\perp\varphi=\ast(\hV\wedge \varphi)=-\iota_\hV(\ast \varphi)
=\frac{1}{||V||^2}\iota_VZ=\frac{1}{1-b^2} VZ\in \Omega^4(\cD)~~,
\een
where we used the fact that $VZ=\iota_VZ$ (since $V\wedge
Z=0$). Multiplying both sides of \eqref{psidef} with $V^{-1}$ gives
the third relation in \eqref{fsol}, which in turn implies
$\iota_VZ=(1-b^2)\psi$, where we used $||V||^2=1-b^2$ and the fact
that $\iota_V\psi=0$ (since $\psi\in \Omega^4(\cD)$). Hence
relation \eqref{YVZ} (which is equivalent with the second equation on
the second row of \eqref{SolMS}) becomes the second equation of
\eqref{fsol}.

\

\noindent 2. (the inverse implication)~Let us assume that \eqref{fsol}
holds, with $\varphi$ and $\psi$ defined by a $G_2$ structure on
$\cD$. Then $\iota_V\varphi=\iota_V\psi=0$ since $\varphi,\psi\in
\Omega(\cD)$ while \eqref{G2rel} holds (see \cite{Kthesis}). Since
$\iota_V\psi=0$, we have $V\psi=V\wedge \psi$ and the third relation
of \eqref{fsol} gives:
\ben
\label{astZ}
\ast Z=\ast(V\wedge \psi)=||V||\ast_\perp\psi=||V||\varphi~~.
\een
In particular, the first relation on the second row of \eqref{SolMS}
is satisfied (because $\iota_V\varphi=0$). Since $\iota_V\psi=0$, the
third relation in \eqref{fsol} implies $\iota_VZ=||V||^2\psi$, so the
second relation of \eqref{fsol} is equivalent with \eqref{YVZ}, which
in turn is equivalent with the second relation on the second row of
\eqref{SolMS}.  Since $V^2=||V||^2$, the first relation in
\eqref{fsol} is equivalent with $||V||^2=1-b^2$, which recovers the
first relation on the first row of \eqref{SolMS}.  The observations
above also immediately imply that \eqref{G2rel} is equivalent with the
relation on the third row of \eqref{SolMS}. $\blacksquare$

\paragraph{Corollary.} We have: 
\be
\label{Ynorms}
||Z||^2=7||V||^2~~,~~||Y||^2=7(1+b^2)~~.
\ee

\noindent{\bf Proof.} 
Since $\iota_V\psi=0$, the last relation of \eqref{fsol} gives
$||Z||^2=||V||^2||\psi||^2=7||V||^2$.  Since $\psi$ and $\nu$ commute
in the \KA algebra of $(M,g)$ and since $\nu^2=1$, the second relation
in \eqref{fsol} implies
$Y^2=(1+b\nu)^2\psi=(1+b^2+2b\nu)\psi^2$. Since $\psi$ is the
coassociative form of $G_2$ structure on $\cD$, identity
\eqref{psisquared} of Appendix \ref{app:G2} gives $\nu\psi^2=6\nu
\psi+7\nu$ and hence $\cS(\nu\psi^2)=0$. Using \eqref{NormCS} and
\eqref{CanNorm}, we find
$||Y||^2=\frac{1}{16}\cS(Y^2)=\frac{1}{16}(1+b^2)\cS(\psi^2)=(1+b^2)||\psi||^2=7(1+b^2)$. $\blacksquare$

\subsection{The two step reduction of structure group} 
Since $\cD$ is a sub-bundle of $TM$, the proposition shows that we
have a $G_2$ structure on $M$ which at every $p\in M$ is given by the
isotropy subgroup $G_{2,p}$ of the pair $(V_p,\varphi_p)$ in
$\SO(8)_p\eqdef\SO(T_pM,g_p)$. Hence we have a two step reduction
along the inclusions: 
\be 
G_{2,p}\hookrightarrow \SO(7)_p\hookrightarrow \SO(8)_p~~, 
\ee 
where $\SO(7)_p\eqdef \SO(\cD_p,g_p|_{\cD_p})$ is the stabilizer of $V_p$ in
$\SO(8)_p$. Since the first reduction (along $\SO(7)_p\hookrightarrow
\SO(8)_p$) corresponds to the almost product structure $\cP$, we can
equivalently describe the second step (along the inclusion
$G_{2,p}\hookrightarrow \SO(7)_p$) as a reduction of the structure
group of the distribution $\cD$ from $\SO(7)$ to $G_2$.

\subsection{Spinorial construction of the $G_2$ structure of $\cD$}
\label{subsec:spinorial}

The orthogonal decomposition $T^\ast M=\cD^\ast\oplus
(\cD^\perp)^\ast$ induces an obvious monomorphism of \KA bundles
$\wedge \cD^\ast\hookrightarrow \wedge T^\ast M$.  Composing this
with the structural morphism $\gamma:\wedge T^\ast
M\rightarrow \End(S)$ of $S$ gives a morphism of bundles of algebras
$\gamma_\cD:\wedge \cD^\ast \rightarrow \End(S)$ which makes $S$ into
a bundle of modules over the \KA algebra of $(\cD,g|_\cD)$ and thus
into a bundle of real pinors over the distribution $\cD$. We let:
\ben
\label{JD}
J\eqdef \gamma(\nu_\top)~~,~~D\eqdef \gamma({\hat V})~~. 
\een
Since $\nu_\top=\iota_{\hat V}\nu={\hat V}\nu$ while $\nu$ is twisted
central, we have $\nu_\top^2=-1$ and $\nu_\top$ anticommutes with
${\hat V}$ in the \KA algebra of $(M,g)$, namely ${\hat
  V}\nu_\top=-\nu_\top {\hat V}=\nu$. Furthermore, we have ${\hat
  V}^2=1$.  These observations imply that $J$ is a complex structure
on $S$ while $D$ is a real structure for $J$:
\be
J^2=-\id_S~~,~~D^2=\id_S~~,~~DJ=-JD~~\mathrm{with}~~\gamma(\nu)=DJ~~.
\ee
Using $J$ to view $S$ as a complex vector bundle with $\rk_\C S=8$, we
define the complex conjugate of a section $\xi\in \Gamma(M,S)$
through:
\be
\bar{\xi}\eqdef D(\xi)=\gamma({\hat V})\xi~~.
\ee
Since $\iota_{\hat V}\omega=0$ for any $\omega\in \Omega(\cD)$, we
have ${\hat V}\omega=\pi(\omega){\hat V}$ while $\nu_\top$ is central
in the \KA algebra of $(\cD,g|_\cD)$. This gives:
\ben
\label{JDgamma}
J\circ \gamma(\omega)=\gamma(\omega)\circ J~~,
~~D\circ \gamma(\omega)=\gamma(\pi(\omega))\circ D~~,~~\forall \omega\in \Omega(\cD)~~.
\een
It follows that $J$ and $D$ are the canonical complex and real
structures on the real pinor bundle $S$ over the seven-dimensional distribution 
$\cD$ in the sense discussed in \cite{gf}. In particular, the Majorana spinors (real spinors) 
over $\cD$, respectively the imaginary spinors over $\cD$ are those $\xi\in \Gamma(M,S)$ which
satisfy $\bar{\xi}=\pm \xi$; they are the sections of rank eight real
vector sub-bundles $S_\pm\subset S$ defined as the bundle of $\pm 1$
eigen-subspaces of $D=\gamma({\hat V})\in \Gamma(M,\End(S))$:
\be
\Gamma(M,S_\pm)=\{\xi\in \Gamma(M,S)|\gamma({\hat V})\xi=\pm \xi\}~~.
\ee
Relations \eqref{JDgamma} show that $\gamma(\omega)$ belongs to
$\Gamma(M,\End_\C(M))$ for all $\omega\in \Omega(\cD)$ and that it is
a real or purely imaginary endomorphism with respect to the real
structure $D$ according to whether the rank of $\omega$ is even or
odd:
\beqa
&&\gamma(\omega)(S_\pm)\subset (S_\pm)~~\mathrm{for}~~\omega\in \Omega^\ev(\cD)~~,\nn\\
&&\gamma(\omega)(S_\pm)\subset (S_\mp)~~\mathrm{for}~~\omega\in \Omega^\odd(\cD)~~.\nn
\eeqa
When viewed as a bundle of real pinors over $\cD$ using the module
structure given by $\gamma_\cD$, $S$ has four admissible pairings,
each of which is determined up to rescaling by a nowhere-vanishing
real-valued function defined on $M$ \cite{AC1,AC2}.  The bilinear
pairing $\cB$ of $S$ discussed in Section 1 (which arises when $S$ is
viewed as bundle of real pinors over $M$) has symmetry $\sigma_\cB=+1$
and type $\epsilon_\cB=+1$ and hence coincides with the second of
these four admissible pairings --- the one which is denoted by $\cB_2$
in \cite{gf}. Recall from \cite[Section 3.3.2]{gf} that $\cB_2$ has
the same restriction to $S_+$ as the basic admissible pairing $\cB_0$,
while its restriction to $S_-$ differs from that of $\cB_0$ by a minus
sign. The complexification of the restriction $\cB_0|_{S^+\otimes
  S^+}=\cB|_{S^+\otimes S^+}$ gives a $\C$-bilinear pairing $\beta$ on
the complexified bundle $S\simeq S_+\otimes \cO_\C$ (where $\cO_\C$
denotes the trivial complex line bundle on $M$) and we have \cite{gf}:
\be
\cB(\xi,\xi')=\beta(\xi_R,\xi'_R)+\beta(\xi_I,\xi'_I)~~,~~\forall \xi,\xi'\in \Gamma(M,S)~~,
\ee
where we used the decomposition into real and imaginary parts of $\xi\in \Gamma(M,S)$: 
\be
\xi=\xi_R+J\xi_I~~,~~\xi_R,\xi_I\in \Gamma(M,S_+)~~.
\ee
It is not hard to show that $\psi$ is given in terms of the spinor $\xi$ through the relation: 
\be
\psi=\frac{1}{1+b}Y^++\frac{1}{1-b}Y^-~~
\ee
where $Y^\pm=\bcE^{(4)}_{\xi^\pm,\xi^\pm}$ are the selfdual and
anti-selfdual parts of $Y$.  In terms of the unit norm spinors
$\eta^\pm\eqdef \sqrt{\frac{2}{1\pm b}}\xi^\pm\in \Gamma(M,S^\pm)$, we
have $Y^\pm=\frac{1}{2}(1\pm b)\bcE^{(4)}_{\eta^\pm,\eta^\pm}$
and since $\bcE^{(4)}_{\eta^\pm,\eta^\mp}=0$, we find:
\ben
\label{psieta}
\!\!\!\!\psi\!=\!\bcE^{(4)}_{\eta_0,\eta_0}\!=\!\frac{1}{4!}\cB(\eta_0,\gamma_{a_1\ldots a_4}\eta_0)e^{a_1\ldots a_4}\!=\!
\frac{1}{4!}\beta(\eta_0,\gamma_{a_1\ldots a_4}\eta_0)e^{a_1\ldots a_4}~~,
~\mathrm{with}~\eta_0
\!\eqdef \!\frac{1}{\sqrt{2}}(\eta^++\eta^-)\in \Gamma(M,S)~~
\een
where we used the fact that $\gamma_{a_1\ldots a_4}(S_+)\subset
S_+$. It is not hard to check the relation:
\be
\xi^\mp=\frac{1}{1\pm b}\gamma(V)\xi^\pm~~,
\ee
which  implies $\eta^\mp= D(\eta^\pm)$ and hence:
\ben
\label{eta}
\eta_0=\frac{1}{\sqrt{2}}(\eta^++\overline{\eta^+})=
\frac{1}{\sqrt{2}}(\eta^-+\overline{\eta^-})~~\in \Gamma(M,S_+)
\een
is a Majorana spinor (in the 7-dimensional sense) over $\cD$ which is
everywhere of norm one. It is well-known \cite{FKMS, Joyce} that such
a spinor determines a $G_2$ structure on $\cD$ which is compatible
with the metric and orientation of $\cD$ and whose canonically
normalized coassociative four-form is given by \eqref{psieta}. This
shows how the $G_2$ structure on $\cD$ can be understood directly in
terms of spinors. In this approach, the cubic relation on the third
row of \eqref{SolMS} can be seen as a mathematical consequence of the
fact that $\psi$ determines a $G_2$ structure on the distribution
$\cD$, which is compatible with its metric and orientation induced
from $M$.

\paragraph{Proposition.} The restriction of $\frac{1}{2}(\id_S+D)$ gives a bundle isomorphism 
from $S^+$ to $S_+$, whose inverse is given by the restriction of $\id_S+DJ$.

\

\noindent  We remind the reader that $S^\pm$ denote the positive and
negative chirality sub-bundles of $S$ when the latter is viewed as a
bundle of real pinors over $M$.

\

\noindent{\bf Proof.} 
Let $\xi\in \ker(1+D)\cap \Gamma(M,S^+)$. Then $D\xi=-\xi$ and
$\gamma(\nu)\xi=+\xi$. Thus $JD\xi=-\xi$, which implies $J\xi=\xi$ and
hence $-\xi=J^2\xi=\xi$ i.e. $\xi=0$. It follows that $\ker(1+D)\cap
\Gamma(M,S^+)=\{0\}$. Since $D(\id_S+D)=\id_S+D$, we have
$(1+D)(S)\subset S_+$ and rank comparison shows that the restriction
of $\id_S+D$ gives an isomorphism from $S^+$ to $S_+$.  Since $D|_{S_+}=\id_{S_+}$ while $DJ=-JD$, we
have
$\frac{1}{2}(\id_S+D)(\id_S+DJ)|_{S_+}=\frac{1}{2}(\id_S+D-JD+J)|_{S_+}=\id_{S_+}$.
$\blacksquare$

\

\noindent The proposition implies that $\eta^\pm$ are uniquely determined by
$\eta_0$ through relation \eqref{eta}:
\be
\eta^+=(1+DJ)\eta_0~~\mathrm{and}~~\eta^-=D(\eta^+)=(1-DJ)\eta_0~~.
\ee
As a consequence, $\xi$ is determined by $\eta_0$:
\be
\xi^\pm=\sqrt{\frac{1\pm b}{2}}(1\pm DJ)\eta_0\Longrightarrow \xi=\big[\sqrt{\frac{1+ b}{2}}(1+ DJ)+\sqrt{\frac{1- b}{2}}(1-DJ)\big]\eta_0~~.
\ee
Notice that $D$ and $J$ are known if the bundle $S_+$ of Majorana
spinors over $\cD$ is given, since $S$ is the complexification of
$S_+$.  Let us assume that $b$ is known. Since a $G_2$ structure on
$\cD$ determines \cite{FKMS} the orientation, metric and spin
structure of $\cD$ (thus also the vector bundle $S_+$ over $M$ and its
structure as a bundle of modules over the \KA algebra of $\cD$) as
well as (up to a global sign ambiguity\footnote{A $G_2$ structure determines
  a subgroup $G_{2,p}\subset \SO(\cD_p,g_p|_{\cD_p})$ for every $p\in
  M$. This has a unique lift ${\hat G}_{2,p}\subset
  \Spin(\cD_p,g_p|\cD_p)$ to a $G_2$ subgroup of the universal cover
  $\Spin(\cD_p,g_p|_{\cD_p})$. Since ${\hat G}_{2,p}$ acts
  transitively on the unit sphere in the real spinor representation
  $\Delta_{7,p}\simeq S_{+,p}\simeq \R^7$ of
  $\Spin(\cD_p,g_p|_{\cD_p})$, it follows that $G_{2,p}$ is the
  stabilizer of {\em two} unit norm spinors $\eta_p\in S_{+,p}$ and
  $-\eta_p\in S_{+,p}$ which differ by sign. By continuity, this
  implies that the $G_2$ structure of $\cD$ defines a spinor $\eta\in
  \Gamma(M,S_+)$ which is determined up to a global sign factor
  (recall that $M$ is connected). Notice that this sign ambiguity
  cannot be removed. We thank A.~Moroianu for correspondence on this aspect. })
the normalized Majorana spinor $\eta_0\in \Gamma(M,S_+)$ over
$\cD$, it follows that such a structure also determines the vector
bundle $S$ (as the complexification of $S_+$) and (up to a sign) the normalized
Majorana spinor $\xi\in \Gamma(M,S)$ over $M$. The module structure of
$S$ over the \KA algebra of $M$ is then determined by the module
structure of $S_+$ over the \KA algebra of $\cD$ and by the fact that
$\gamma({\hat V})$ equals the real structure of the complexification
$S$ of $S_+$. Notice that orientation of $M$ is determined by ${\hat V}$ and by 
the orientation of $\cD$ and that the metric of $M$ is determined by the
metric of $\cD$ and by the condition that ${\hat V}$ has norm one and
that it is orthogonal everywhere to $\cD$.

\subsection{A non-redundant parameterization of $\xi$} 

The original quantities $b,V,Y$ and $Z$ of
\eqref{Eexp} provide a redundant parameterization of the spinor
$\xi$; explicitly, the second and third relation in \eqref{fsol} can
be inverted as follows:
\ben
\label{YZinverse}
\psi=\frac{1}{1-b^2} VZ=\frac{1}{1-b^2}(1-b\nu)Y~~
\een
and hence $b,V,Y$ and $Z$ satisfy the cubic relation:
\be
(1-b^2)Y=(1+b\nu)VZ~~.
\ee  
A better parameterization (in terms of $b,V$ and $\psi$) is obtained by
substituting \eqref{fsol} into \eqref{Eexp}:
\ben
\label{Enr}
\boxed{\check{E}=\frac{1}{16}(1+V+b\nu)(1+\psi)=P\Pi}~~,
\een
where: 
\ben
\label{PiP}
P\eqdef \frac{1}{2}(1+V+b\nu)~~\mathrm{and}~~\Pi\eqdef \frac{1}{8}(1+\psi)
\een
are commuting idempotents in the \KA algebra.  Idempotency of $P$ is
equivalent with the relation $V^2=1-b^2$, while that of $\Pi$ is
equivalent with identity \eqref{psisquared} of Appendix \ref{app:G2},
which is satisfied by the coassociative form of any $G_2$ structure on
a vector bundle of rank $7$.  The condition that $P$ and $\Pi$ commute
in the \KA algebra is equivalent with the identity $\iota_V\psi=0$.
Knowing that $\psi$ is the canonically normalized coassociative 4-form
of a metric-compatible $G_2$ structure on $\cD$, equation
\eqref{G2rel} is satisfied by $\varphi=\ast_\perp \psi$ and
\eqref{Enr} solve the constraints \eqref{Esquare}, assuming that the
first condition in \eqref{fsol} holds. Substituting this condition
into \eqref{Enr} gives a non-redundant parameterization of $\xi$ in
terms of the quantities $(b,\hV, \psi)$:
\ben
\label{Eparam}
\check{E}=\frac{1}{16}(1+\sqrt{1-b^2}\hV+b\nu)(1+\psi)~~,
\een
which solves \eqref{Esquare} provided that $\psi$ is the canonically
normalized coassociative form of a $G_2$ structure on $\cD$ and that
$||\hV||=1$.

\subsection{Parameterizing the pair $(g,\xi)$}  

Let $\wedge_{\rm pos}^3 \cD^\ast $ be the principal
$\mathrm{SL(7,\R)}/G_2$-bundle of {\em positive} $\cD$-longitudinal
3-forms, whose fiber at a point is diffeomorphic with
$\mathbb{R}\mathbb{P}^7\times \R^{28}$ (see \cite{Bryant}). A $G_2$ structure on $\cD$ which is compatible
with the orientation of $\cD$ is specified by and specifies uniquely a
section $\varphi\in \Omega^3_{\rm pos}(\cD)=\Gamma(M,\wedge^3_{\rm
  pos}\cD^\ast)$. Every $\varphi\in \Omega_{\rm pos}^3(\cD)$ induces a
metric $g_\varphi$ on $\cD$ which is uniquely determined by the
condition \cite{Kthesis}:
\be
||X\wedge Y||^2=||X\lrcorner (Y\lrcorner \varphi)||^2~~,~~\forall X,Y\in \Gamma(M,\cD)~~.
\ee
To say that the restriction $g|_\cD$ is compatible with the $G_2$
structure induced by $\varphi$ on $\cD$ means that $g|_\cD$ coincides
with the metric $g_\varphi$.  If one is further given a vector field
$n$ which is everywhere transverse to $\cD$ (in the sense that
$\langle n\rangle\oplus \cD=TM$, where $\langle n \rangle$ is the unit
rank sub-bundle of $TM$ generated by $n$), then the metric $g$ is uniquely
determined by the triple $(n, \cD, \varphi)$ through the following
conditions:
\be
||n||=1~~,~~g|_\cD=g_\varphi~~\mathrm{and}~~g(n,X)=0~~\forall X\in \Gamma(M,\cD)~~.
\ee
The longitudinal Hodge operator $\ast_\varphi:\Omega(\cD)\rightarrow
\Omega(\cD)$ of $g_\varphi$ is completely determined by $\varphi$
(this is the restriction to $\cD$ of the operator \eqref{AstV}) and
hence $\psi=\ast_\varphi \varphi$ is also determined. Furthermore, the
volume form $\nu$ of $M$ is determined by $g$ and hence the
inhomogeneous form \eqref{Enr} is determined by the further choice of
a function $b\in \cC^\infty(M,(-1,1))$. As a consequence, the spinor
$\xi$ is determined up to sign. Since $\varphi$ determines \cite{FKMS} the
orientation and spin structure of $\cD$ (which, together with ${\hat
  V}$, determine\footnote{As shown above, we have $\gamma({\hat V})=D$
  where $D$ is the canonical real structure \cite{gf} of the bundle of
  complex spinors over $\cD$, which is the complexification of the bundle of 
Majorana spinors over $\cD$.} --- up to a global sign ambiguity --- the real 
spinor \eqref{eta}) and since $n$ and $\cD$ 
determine ${\hat V}$, we find:

\paragraph{Proposition.} The data $(b, n,\cD,\varphi)$ determine the 
metric $g$ on $M$, the spin structure and orientation of $M$ as well
as the spinor $\xi$, where the latter is determined up to a sign
ambiguity.

\

\noindent This proposition reduces the problem of finding pairs
$(g,\xi)$ such that $\xi$ is a nowhere-chiral Majorana spinor on
$(M,g)$ to the problem of finding quadruples $(n,\cD,\varphi,b)$ where
$n$ is a nowhere-vanishing vector field on $M$, $\cD$ is a corank one
Frobenius distribution on $M$ which is everywhere transverse to $n$
(and which is endowed with the orientation induced from that of $M$
using $n$), $\varphi$ is the associative form of a $G_2$ structure on
$\cD$ and $b$ is a smooth function defined on $M$ and satisfying $|b|<1$.

\paragraph{Remark.} Notice that the pair $(n,\cD)$ determines $\hV$ 
uniquely through the requirements $\cD=\ker \hV$ and $n\lrcorner {\hat
  V}=1$. However, the pair $({\hat V},\psi)$ does not
determine the metric $g$ since the set of solutions $n\in
\Gamma(M,TM)$ to the condition $n\lrcorner {\hat V}=1$ is an
infinite-dimensional affine space modeled on $\Gamma(M,\cD)$ (where
$\cD=\ker{\hat V}$).

\subsection{Two problems related to the supersymmetry conditions}
\label{subsec:p}

We shall consider two different (but related) problems regarding
equations \eqref{par_eq}:

\paragraph{Problem 1.} 
Given $f\in \Omega^1(M)$ and $F\in \Omega^4(M)$, find a set of
equations on the warp factor $\Delta$ and on the quantities $b,{\hat V},\psi$
appearing in the parameterization \eqref{Eparam} which is {\em
  equivalent} with the supersymmetry equations \eqref{par_eq}.

\paragraph{Problem 2.} 
Find the necessary and sufficient {\em compatibility conditions} on the quantities $\Delta$
and $b,{\hat V},\psi$ such that there exist at least one pair
$(f,F)\in \Omega^1(M)\times \Omega^4(M)$ for which
$\dim\cK(\mathbb{D},Q)>0$, i.e. such that \eqref{par_eq} admits at
least one non-trivial solution $\xi\neq 0$.

\

A solution of Problem 1 was already given in \cite{ga1}, but its
geometric meaning was not addressed in loc. cit.  In this paper, we
show that the equations on $\Delta, b, {\hat V}$ and $\psi$ which
solve Problem 1 can be expressed in geometric manner as equations
which determine the geometry of a codimension one foliation $\cF$ of
$M$ (which carries a longitudinal $G_2$ structure) in terms of $f$ and
$F$.  We also show that the compatibility conditions on $\Delta, b,
{\hat V}$ and $\psi$ which solve Problem 2 can be expressed as {\em
  admissibility conditions} on this foliation and that those pairs
$(f,F)$ for which $\dim\cK(\mathbb{D},Q)>0$ can be parameterized by
{\em admissible} foliations endowed with longitudinal $G_2$ structure.

\

\section{Encoding the supersymmetry conditions through foliated geometry}
\label{sec:susy}

In this section, we show that the supersymmetry conditions require
that $\cD$ is Frobenius integrable and hence that it determines a
codimension one foliation $\cF$ of $M$. As a consequence, the $G_2$
structure of $\cD$ becomes a leafwise $G_2$ structure on this
foliation. Furthermore, we show that the supersymmetry conditions are
{\em equivalent} with equations which determine (in terms of $F,f$ and
$\Delta$) the function $b$, the O'Neill-Gray tensors
\cite{ONeill,Gray, Tondeur} of $\cF$ (equivalently, they determine
the Naveira 3-tensor \cite{Naveira} of the almost product structure
$\cP$) as well as the torsion classes of the leafwise $G_2$ structure
and the normal covariant derivative of $\psi$.  The results of this
section provide an ``if and only if'' characterization of such
supersymmetric backgrounds (in the case when the spinor $\xi$ is
everywhere non-chiral), taking into account the {\em full} information
contained in the supersymmetry conditions. We also discuss some
topological obstructions for existence of a solution to the
supersymmetry conditions, which turn out to be encoded by the Latour
class \cite{Latour} known from Novikov theory \cite{Farber}.

\subsection{Expressing the supersymmetry conditions using the \KA algebra}

It was shown in \cite{ga1} that the supersymmetry conditions
\eqref{par_eq} are {\em equivalent} with the following equations for the
inhomogeneous form $\check{E}\eqdef \check{E}_{\xi,\xi}$ of
\eqref{Eexp}, where commutators $[~,~]_-$ are taken in the \KA algebra
of $(M,g)$:
\beqan
\label{NablaConstr}
&&\nabla_m\check{E}=-[\check A_m,\check{E}]_{-}~~,\\
\label{QConstr}
&&\check{Q} \check E=0~~.
\eeqan
The inhomogeneous differential forms $\check{A}_m$, $\check{Q}$ appearing in these relations are given by the following
expressions \cite{ga1} in a local orthonormal frame $e_m$ (defined over an open subset $U\subset M$) with dual
coframe $e^m$:
\beqa
&& \check{A}_m=_U\frac{1}{4}e_m\lrcorner F+\frac{1}{4}(e^{m}\wedge f) \nu +\kappa e^{m}\nu~~~,\\~~~
&&\check{Q}=_U\frac{1}{2}\dd \Delta-\frac{1}{6}f\nu-\frac{1}{12}F-\kappa\nu~~.
\eeqa
We shall refer to \eqref{NablaConstr} as the {\em covariant derivative
  constraints} and to \eqref{QConstr} as the {\em
  $\cQ$-constraints}. The fact that these relations are equivalent
with \eqref{par_eq} follows from the general theory of \cite{ga1,ga2,
  gf}, which clarifies the mathematical structure of the method of
bilinears \cite{Tod} and allows one to automatically translate
supersymmetry conditions (and generally any differential or algebraic
equation on spinors) into relations such as \eqref{NablaConstr} and
\eqref{QConstr}, without having to appeal to manipulations of gamma
matrices. Defining:
\ben
\label{Sdef}
S_m^{(k)}\eqdef [\Check{A}_m,\check{E}]_{-}^{(k)}~~,
\een
one finds upon separating ranks that the covariant derivative
constraints \eqref{NablaConstr} are equivalent with the system:
\beqan
\label{DS}
&&\partial_m b~~=-\ast S_m^{(8)}~~,~~ \nabla_m V=- S_m^{(1)}~~,\nn\\
&&\nabla_m Y=- S_m^{(4)}~~~~~,~~\nabla_m Z=- S_m^{(5)}~~.
\eeqan
The expanded form of these conditions can be found in \cite{ga1}.  Equations \eqref{DS} imply the {\em exterior
  differential relations}:
\ben
\label{EDS}
\dd \check E=-e^m\wedge[\check A_m,\check{E}]_{-}
\een
and the {\em exterior codifferential relations}:
\ben
\label{CDS}
\updelta \check{E}=\iota_{e^m} \Big([\check A_m,\check{E}]_{-}\Big)~~.
\een
We refer the reader to \cite{ga1} for the expanded form
of these.

\paragraph{Remark.} 
Notice that \eqref{EDS} and \eqref{CDS} are not equivalent (even
when taken together) with the initial differential system
\eqref{DS}. This is because specifying the differential and
codifferential of a form does not in general suffice to fix the
covariant derivative of that form; in particular, relations \eqref{DS}
determine the full covariant derivative of the one-form $V$, which is
not determined merely by the differential and codifferential of $V$.
We shall see explicitly how this occurs in subsection
\ref{sec:fundeq}.  Appendix \ref{app:other} contains a comparison of \eqref{EDS}
and \eqref{CDS} with certain exterior differential formulas which have
appeared previously in the literature.

\

\subsection{Integrability of $\cD$. The foliations $\cF$ and $\cF^\perp$}

As already noticed in \cite{MartelliSparks}, it turns out that the
covariant derivative constraints \eqref{NablaConstr}, taken together with the
$\cQ$-constraints \eqref{QConstr} imply the conditions (see the first
and second equations in \eqref{Diff}):
\ben
\label{dtV}
\boxed{
\begin{split}
&\dd \momega=0~~,~~\mathrm{where}~~\momega\eqdef 4\kappa e^{3\Delta} V~~,\\
&\momega=\mathbf{f}-\dd\mb~~,~~~\mathrm{where}~~\mb\eqdef e^{3\Delta} b~~.
\end{split}}
\een
In particular, the one-form $\mf$ must be closed, so the supersymmetry
conditions imply the second part of the Bianchi identities
\eqref{Bianchi}. Relations \eqref{dtV} imply that the closed form
$\momega$ belongs to the cohomology class of $\mf$. The first of these relations 
shows that the distribution $\cD=\ker V=\ker\momega$ is Frobenius integrable
and hence that it defines a codimension one foliation $\cF$ of $M$
such that $\cD=T\cF$. The complementary distribution is of course also
integrable (since it has rank one), determining a foliation
$\cF^\perp$ such that $\cD^\perp=T(\cF^\perp)$. The leaves of
$\cF^\perp$ are the integral curves of the vector field $n$, which are
orthogonal to the leaves of $\cF$ (i.e., they intersect the latter at
right angles). The 3-form $\varphi$ defines a leafwise $G_2$ structure
on $\cF$. The restriction $S|_L$ of the vector bundle $S$ to any given
leaf $L$ of $\cF$ becomes the bundle of real pinors of $L$, while the
restriction of $S_+$ becomes the bundle of Majorana spinors of the
leaf (cf. Subsection \ref{subsec:spinorial}).  The topology of such
foliations is discussed in Section \ref{sec:top}. Since the
considerations of the present section are local, we can ignore for the
moment the global behavior of the leaves\footnote{Recall that the
  leaves of any foliation are injectively immersed submanifolds of $M$
  (hence they cannot have self-intersections) and that every immersion
  is {\em locally} an embedding.}.
\begin{figure}[h!]
\begin{center}
\includegraphics[scale=0.5]{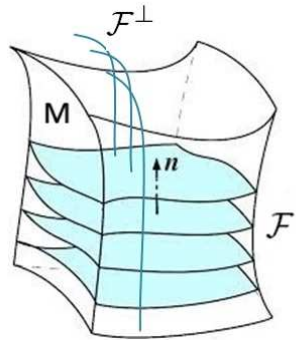}
\caption{Local picture of the plaques of the foliations $\cF$ and $\cF^\perp$ inside some open subset of $M$.}
\end{center}
\end{figure}

\subsection{Topological obstructions to existence of a nowhere-vanishing closed one-form in the cohomology class of $\mf$}

Notice that the cohomology class $\f\in H^1(M,\R)$ of $\mf$ cannot be
zero since otherwise the second condition in \eqref{dtV} would require
that $\momega=\dd \alpha $ for some smooth map $\alpha:M\rightarrow
\R$. Since $\alpha$ must attain its extrema on the compact manifold
$M$, this would imply that $\momega$ vanishes at those extrema,
contradicting our requirement that $V$ (and thus also $\momega$)
be nowhere-vanishing. Thus $\f$ must be a non-trivial cohomology class
and hence the first Betti number of $M$ cannot be zero:
\ben
\label{Betti}
\boxed{b_1(M)>0}~~.
\een
This condition is far from sufficient. 
To state further conditions on $\f$, let us recall some facts regarding the 
period morphism of an element of $H^1(M,\R)$. 

\paragraph{The period morphism and period group of $\f$.}
Recall that integration of any representative of $\f$ over closed
paths provides a group morphism (called {\em the period morphism})
from the unbased first homotopy group to the additive group of the reals:
\be
\per_\f:\pi_1(M)\rightarrow \R~~.
\ee
This factors through the map $\pi_1(M)\stackrel{[~~]}{\rightarrow} H_1^\tf(M,\Z)$
which associates to each homotopy class $\alpha\in \pi_1(M)$ of a path
its image $[\alpha]$ in the torsion-free part $H_1^\tf(M,\Z)$ of $H_1(M,\Z)$:
\be
\per_\f(\alpha)=\per'_\f([\alpha])~~.
\ee
The induced map $\per'_\f:H_1^\tf(M,\Z)\rightarrow \R$ is called the
{\em reduced period morphism}. The image $\Pi_\f$ of $\per_\f$ is a
(necessarily free) Abelian subgroup of $\R$ called the {\em group of
  periods} of $\f$ while the kernel $A_\f$ of $\per_\f$ is a normal
subgroup of $\pi_1(M)$ called the {\em $\f$-irrelevant subgroup}. It
is obvious that $A_\f$ contains the commutant subgroup
$[\pi_1(M),\pi_1(M)]$.  The corresponding subgroup $A'_\f=[A_\f]=\ker
(\per'\f)$ of $H_1^\tf(M,\Z)\subset H_1(M,\Z)$ is called the
$\f$-irrelevant subgroup of the group of one-cycles\footnote{Recall
  that we have a canonical direct sum decomposition
  $H_1(M,\Z)=H_1^{\rm torsion}(M,\Z)\oplus H_1^\tf(M,\Z)$ since $\Z$
  is a principal ideal domain (PID). This decomposition follows from
  the structure theorem for finitely generated modules over a PID
  ($\pi_1(M,Z)$ and hence its Abelianization
  $H_1(M,\Z)=\pi_1(M,\Z)/[\pi_1(M,Z),\pi_1(M,\Z)]$ are
  finitely-generated since $M$ is a compact manifold and thus has the
  homotopy type of a finite CW complex). Hence we have a natural
  embedding of $H_1^\tf(M,\Z)$ into $H_1(M,\Z)$.}. 

The rank $\rho(\f)\eqdef \rk \Pi_\f$ of the period group is called the
{\em irrationality rank} of $\f$. We have $\rho(\f)=\dim_\Q\Pi_\f$ if
$\Pi_\f$ is viewed as a finite-dimensional subspace of $\R$, when the
latter is viewed as an infinite-dimensional vector space over the
field $\Q$ of rational numbers. Notice that: 
\ben
\boxed{\rho(\f)\leq b_1(M)}~~. 
\een
Let us assume that $\f\neq 0$ (which, as explained above, is always
the case in our application).  Then $\Pi_\f$ is a discrete subgroup of
$(\R,+)$ iff. $\rho(\f)=1$, in which case $\Pi_\f$ is infinite-cyclic
(hence isomorphic with $\Z$), i.e. we have $\Pi_\f=\Z a_\f$ where
$a_\f$ is the {\em fundamental period} of $\f$, defined through:
\be
a_\f\eqdef\inf(\Pi_\f\cap \N^\ast)>0~~.
\ee
Here $\N^\ast$ denotes the set $\N\setminus \{0\}$ of positive integers.  
This happens iff. there exists a positive real
number $\lambda$ (for example, $\lambda=\frac{1}{a_\f}$) such that
$\lambda\f\in H^1(M,\Z)$ (equivalently, such that $\lambda\f\in
H^1(M,\Q)$), in which case we say that $\f$ is {\em projectively
  rational}. When $\rho(\f)>1$, the period group is a dense subgroup
of $(\R,+)$ and hence $\inf(\Pi_\f\cap \N^\ast)=0$.  In this case we
say that $\f$ is {\em projectively irrational}.

Letting $b_1\eqdef b_1(M)$ and picking a basis $c_1,\ldots,c_{b_1}$ of the
free Abelian group $H_1^\tf(M,\Z)$, we have (in both cases mentioned above):
\be
\Pi_\f=\Z\per_\f(c_1)+ \ldots + \Z \per_\f(c_{b_1})\subset \R~~.
\ee 
In the projectively rational case this sum equals $\Pi_\f=\Z a_\f$,
since in that case we have $\per_\f(c_i)=\nu_i a_\f$ for some
(setwise) coprime integers $\nu_1,\ldots,\nu_{b_1}$. For the general
case, let $b_1,\ldots , b_\rho\in \R$ (where $\rho\eqdef \rho(\f)$) be
a basis of the vector space $\Q\per_\f(c_1)+ \ldots + \Q
\per_\f(c_{b_1})\subset \R$ generated by $\per_\f(c_i)$ over
$\Q$. Then $\per_\f(c_i)=\sum_{k=1}^\rho q_{ik}b_k$ for some
uniquely-determined rationals $q_{ik}$. Clearing denominators, we have
$q_{ik}=q_k m_{ik}$ for some uniquely determined positive rationals
$q_k$ and integers $m_{ik}$ such that $m_{1k},\ldots,m_{b_1k}$ are
setwise coprime. Then $\Pi_\f=\sum_{k=1}^\rho {\Z a_k}$, where the
real numbers $a_k\eqdef q_k b_k\in \R$ are rationally independent and
thus they also form a basis of $\Pi_\f$ over $\Q$. It follows that the
last sum is direct, i.e.:
\ben
\label{PresPi}
\Pi_\f=\Z a_1\oplus \ldots \oplus \Z a_\rho~~.
\een
This shows how one can find a basis of $\Pi_\f$ when the latter is
viewed as a free Abelian group. 

\paragraph{The necessary and sufficient conditions.} 
Even on manifolds $M$ which satisfy \eqref{SpinorExistenceCond} and
\eqref{Betti}, finding a nowhere-vanishing closed one-form lying in a
cohomology class $\f$ imposes further restrictions on that class and
on the topology of $M$. The necessary and sufficient conditions are
known \cite{Farrell, Siebenmann, Latour} for manifolds $M$ of
dimension greater than $5$ (which is our case). Since they are rather
technical, we state them without giving any details, referring the
reader to loc. cit. as well as to \cite{Schutz1,Schutz2}. Let
$\hat{M}_\f$ be the integration cover of $\per_\f$, i.e. the Abelian
regular covering space of $M$ corresponding to the normal subgroup
$A_\f$ of $\pi_1(M)$.  When $\dim M\geq 6$, a class $\f \in
H^1(M,\R)\setminus \{0\}$ contains a nowhere-vanishing closed one-form
iff. $M$ is $(\pm \f)$-contractible and the {\em Latour obstruction}
$\tau_L(M,\f)\in \Wh(\pi_1(M),\f)$ vanishes. Here $\Wh(\pi_1(M),\f)$
is the Whitehead group of the Novikov-Sikorav ring
$\widehat{\Z\pi_1(M)}$ in the sense of \cite{Latour}, the
Novikov-Sikorav ring \cite{Sikorav} (see also \cite[Subsection
  3.1.5]{Farber}) being a completion of the group ring $\Z\pi_1(M)$
with respect to a certain norm induced by the period morphism
$\per_\f$. When $\f$ is a projectively rational class, these
conditions are equivalent \cite{Ranicki} with those found in
\cite{Farrell, Siebenmann}, namely that the integration cover
$\hat{M}_\f$ (which in that case is infinite cyclic) must be
finitely-dominated and that the {\em Farrell-Siebenmann obstruction}
$\tau_F(M, \f)\in \Wh(\pi_1(M))$ must vanish, where $\Wh(\pi_1(M))$ is
the Whitehead group of $\pi_1(M)$.

\subsection{Solving the $\cQ$-constraints}

Recall form \cite{ga1} that any inhomogeneous form decomposes uniquely
as $\omega=\omega_\perp+\hV\wedge \omega_\top$, where $\omega_\perp$
and $\omega_\top$ are orthogonal to $\hV$ and thus belong to 
$\Omega(\cD)$. Since $\cF$ carries a leafwise
$G_2$ structure, we can parameterize $\omega_\perp$ and $\omega_\top$
for any pure rank form as recalled in Appendix \ref{app:G2}.  In
particular, we have $F=F_\perp+\hV\wedge F_\top$ and
$f=f_\perp+\hV\wedge f_\top$, where $f_\top\in \Omega^0(M)$,
$f_\perp\in \Omega^1(\cD)$, $F_\top\in \Omega^3(\cD)$ and $F_\perp\in
\Omega^4(\cD)$. Relations \eqref{omega_param}, \eqref{eta_param} of
Appendix \ref{app:G2} give the parameterizations:
\beqan
\label{Fdecomp}
\boxed{
\begin{split}
&F_\perp=F_\perp^{(7)}+F_\perp^{(S)}~~\mathrm{where}~
~F_\perp^{(7)}=\alpha_1\wedge\varphi\in \Omega^4_7(\cD)~~,
~~F_\perp^{(S)}=-\hat{h}_{kl}e^k\wedge\iota_{e^l}\psi\in \Omega^4_S(\cD)~~~\\
&F_\top=F_\top^{(7)}+F_\top^{(S)}~~\mathrm{where}~
~F_\top^{(7)}=-\iota_{\alpha_2}\psi\in \Omega^3_7(\cD)~~,
~~F_\top^{(S)}=\chi_{kl}e^k\wedge\iota_{e^l}\varphi\in \Omega^3_S(\cD)~
\end{split}
}~~.~~~~~~~
\eeqan
Here $\alpha_1,\alpha_2\in\Omega^1(\cD)$ and ${\hat h}, \chi$ are
leafwise covariant symmetric tensors, i.e. sections of the bundle
$\Sym^2(\cD^\ast)$. Also recall from Appendix \ref{app:G2} that
$F_\top^{(S)}=F_\top^{(1)}+F_\top^{(27)}$ with $F_\top^{(1)}\in
\Omega^3_1(\cD)~,~F_\top^{(27)}\in \Omega^3_{27}(\cD)$, with a similar
decomposition for $F_\perp^{(S)}$.  The last relations correspond to
the decompositions of $\chi$ and ${\hat h}$ into their homothety and
traceless parts $\chi^{(0)}$ and ${\hat h}^{(0)}$. 
Since $\psi=\ast_\perp\varphi=\ast_\varphi\varphi$ is
determined by $\varphi$, relations \eqref{Fdecomp} determine $F$ in
terms of ${\hat V}$, $\psi$ and of the quantities
$\alpha_1,\alpha_2,{\hat h}$ and $\chi$. The following result shows
that the $\check{Q}$ constraints are equivalent with equations which
determine $\alpha_1,\alpha_2$ and $\tr_g({\hat h})$, $\tr_g(\chi)$ in
terms of $\Delta$, $b$ and $f$.

\paragraph{Theorem 1.} 
Let $||V||=\sqrt{1-b^2}$. Then the $\cQ$-constraints \eqref{QConstr}
are {\em equivalent} with the following relations, which determine (in
terms of $\Delta, b, {\hat V}, \psi$ and $f$) the components of
$F_\top^{(1)}$, $F_\perp^{(1)}$ and $F^{(7)}_\top$, $F_\perp^{(7)}$:
\beqan
\label{SolF}
\boxed{
\begin{split}
&\alpha_1=\frac{1}{2||V||}(f-3b\dd\Delta)_\perp  ~,\\
&\alpha_2= -\frac{1}{2||V||}(bf-3\dd\Delta)_\perp ~,\\
&\tr_g(\hat h)=-\frac{3}{4}\tr_g(h)=\frac{1}{2||V||}(bf-3\dd\Delta)_\top  ~,\\
&\tr_g(\hat \chi)=-\frac{3}{4}\tr_g(\chi)= 3\kappa- \frac{1}{2||V||}(f-3b\dd\Delta)_\top~.
\end{split}
}
\eeqan

\paragraph{Remark.} 
Notice that the $\cQ$-constraints \eqref{QConstr} do {\em not}
determine the components $F_\top^{(27)}$ and $F_\perp^{(27)}$.

\paragraph{Definition.} 
We say that a pair $(f,F)\in \Omega^1(M)\times \Omega^4(M)$ is {\em
  consistent} with a quadruple $(\Delta, b,{\hat V},\psi)$ if
conditions \eqref{SolF} hold, i.e. if the $\check{Q}$-constraints are
satisfied.

\

\noindent{\bf Proof.} 
Writing $\check{E}=\frac{1}{16}(\alpha+\beta)$ and
$\check{Q}=\frac{1}{12}(T-x)$, where:
\beqa
&&\alpha\eqdef V+Z=V(1+\psi)~\in \Omega^\odd(M)~~,
~~\beta\eqdef 1+Y+b\nu=(1+b\nu)(1+\psi)~\in \Omega^\ev(M)~,\\
&&  x\eqdef F+12 \kappa \nu~\in \Omega^\ev(M)~~~~~~~~~~~~~~~,~~~ T\eqdef 2(3 \dd\Delta -\ast f)~\in \Omega^\odd(M)
\eeqa
and using the fact that the geometric product is even with respect to
the $\Z_2$-grading of $\Omega(M)$ given by the decomposition
$\Omega(M)=\Omega^\ev(M)\oplus \Omega^\odd(M)$, the $\cQ$-constraints
\eqref{QConstr} can be brought to the form:
\beqan
\label{q}
&& \big[xV-T(1+b\nu)\big]\Pi=0~~,\nn\\
&& \big[x(1+b\nu)-TV\big]\Pi=0~~,
\eeqan
where (as mentioned before) the inhomogeneous form $\Pi\eqdef \frac{1}{8}(1+\psi)$ is an idempotent in the \KA algebra:
\be
\Pi^2=\Pi~\Longleftrightarrow~(1+\psi)^2=8(1+\psi)~.
\ee
Since $\nu$ is twisted central in the \KA algebra \cite{ga1} while $\iota_V\psi=0$, we have: 
\be
[\nu,\psi]_-=[V,\psi]_-=0 ~~\Longrightarrow~~[\nu,\Pi]_-=[V,\Pi]_-=0~~.
\ee
On the other hand, $V$ is invertible in the \KA algebra, while $\nu$ is involutive ($\nu^2=1$). Using these observations, we compute:
\beqa
V(1+b\nu)^{-1}=\frac{V(1-b\nu)}{||V||^2}=\frac{(1+b\nu)V}{||V||^2}~~,~~\big[ (1+b\nu)V \big]^{-1}=\frac{V(1-b\nu)}{||V||^4}
\eeqa
and find that the two equations of \eqref{q} (and thus the
$\cQ$-constraints \eqref{QConstr}) are both equivalent with the single condition:
\ben
\label{q2}
\Big[x-\frac{T(1+b\nu)V}{||V||^2} \Big]\Pi=0~~.
\een
Let
\be
y\eqdef\frac{T(1+b\nu)V}{||V||^2}=\frac{TV(1-b\nu)}{||V||^2}~~.
\ee
Separating \eqref{q2} into components parallel and orthogonal to $V$, it becomes:
\ben
\label{qsys}
(x_\parallel-y_\parallel)\Pi=(x_\perp-y_\perp)\Pi=0~~,
\een
where:
\be
y_\parallel=-\frac{1}{||V||}\big[ \hV\wedge T+b (\iota_\hV T)\nu \big]~~, ~~y_\perp =\frac{1}{||V||}\big[\iota_{_\hV} T+b (\hV\wedge T)\nu \big]~~.
\ee
Using the properties of Hodge duality, orthogonality and parallelism
given in Appendix \ref{app:GA}, the system \eqref{qsys} is found to be
equivalent with:
\ben
\label{q3}
\Big[x_\top+\frac{1}{||V||}(T+b T\nu)_\perp\Big]\Pi=\Big[x_\perp-\frac{1}{||V||}(T+b T\nu)_\top \Big]\Pi=0~~.
\een
Since $x_\perp=F_\perp$ while $x_\top=\iota_{_{\hV}}x=F_\top+12\kappa\ast\hV$, we find that \eqref{q3} amounts to:
\beqa
\label{q3bis}
&& F_\top\Pi=-\frac{1}{||V||}\Big[(T+b T\nu )_\perp+12\kappa\ast V \Big]\Pi~~,\nn\\
&& F_\perp\Pi= \frac{1}{||V||}\iota_{_{\hV}}(T+b\ast T)\Pi~~.
\eeqa
One computes:
\be
T+b\ast T=2\big[3\dd\Delta-bf+\nu(f-3b\dd\Delta) \big]~~,
\ee
so that the system finally becomes:
\beqan
\label{q4}
&& F_\top\Pi=-\frac{1}{||V||}\Big[ 2(3\dd\Delta-bf)_\perp+2\big[ 6\kappa||V||
-(f-3b\dd\Delta)_\top \big] \nu_{_{\top}} \Big]\Pi~,\nn\\
&& F_\perp\Pi=\frac{1}{||V||}\Big[ 2(3\dd\Delta-bf)_\top+
2(f-3b\dd\Delta)_\perp \nu_{_{\top}} \Big]\Pi~~.
\eeqan
Using the decomposition \eqref{Fdecomp} of $F$ and the right action of
$\psi$ (in the \KA algebra) on 3- and 4-forms given in
\eqref{Psi4}-\eqref{Psi3} of Appendix \ref{app:G2}, equations
\eqref{q4} reduce to:
\beqan
\label{q4bis}
&& F_\top(1+\psi)=-4\alpha_2+4F_\top^{(7)}+3\tr_g(\chi)\varphi
-4\alpha_2\wedge\psi+3\tr_g(\chi)\nu_\top~,\nn\\
&& F_\perp(1+\psi)=-4\tr_g(\hat h)+4\iota_{\alpha_1}\varphi+4F_\perp^{(7)}
-4\tr_g(\hat h)\psi+4\ast_\perp\alpha_1~~.
\eeqan
Identifying the terms of equal ranks, we find that \eqref{q4bis} (and
hence the $\cQ$-constraints \eqref{QConstr}) are equivalent with
relations \eqref{SolF} of the Theorem. $\blacksquare$

\subsection{Extrinsic geometry of $\cF$}
\label{subsec:extrinsic}

As explained in Appendix \ref{app:FE}, the extrinsic geometry of $\cF$
is described by the {\em fundamental equations}:
\beqan
\label{FE}
\boxed{
\begin{split}
&\nabla_n n=H ~~~(\perp n) ~,\\
&\nabla_{X_\perp} n=-A X_\perp ~~~(\perp n)~,\\
&\nabla_n (X_\perp)=-g(H,X_\perp) n+D_n(X_\perp)~,\\
&\nabla_{X_\perp} (Y_\perp)=\nabla^\perp_{X_\perp} (Y_\perp)+g(AX_\perp,Y_\perp)n~,
\end{split}~~
}
\eeqan
where $H\in \Gamma(M, \cD^\perp)$ encodes the second fundamental form
of $\cF^\perp$, $A\in \Gamma(M,\End(\cD))$ is the Weingarten operator
of the leaves of $\cF$ and $D_n:\Gamma(M,\cD)\rightarrow
\Gamma(M,\cD)$ is the derivative along the vector field $n$ taken with
respect to the normal connection of the leaves of $\cF^\perp$. The
first and third relations are the Gauss and Weingarten equations for
$\cF^\perp$ while the second and fourth relations are the Weingarten
and Gauss equations for $\cF$.  Notice that $D_n$ tells us how to
transport tensors (co)tangent to the leaves of $\cF$ in the direction
orthogonal to its leaves, while preserving the metric induced on
$\cD=T\cF=N(\cF^\perp)$. The O'Neill-Gray tensors \cite{ONeill,Gray,
  Tondeur} of the foliation $\cF$ can be expressed in terms of $H$ and
$A$ through the relations:
\ben
\label{OG}
\begin{split}
&\cT_XY = \cT_{X_\perp}Y\eqdef (\nabla_{X_\perp}(Y_\perp))_\parallel +(\nabla_{X_\perp}(Y_\parallel))_\perp=[ X_\perp(g(n,Y))+B(X_\perp,Y_\perp)]n+g(n,Y)H\\
&\cA_XY = \cA_{X_\parallel}Y\eqdef (\nabla_{X_\parallel}(Y_\parallel))_\perp+(\nabla_{X_\parallel}Y_\perp)_\parallel =g(n,X)[-g(H,Y)n+g(n,Y)H]~~,
\end{split}
\een
where: 
\ben
\label{B}
B(X_\perp,Y_\perp)\eqdef g(AX_\perp,Y_\perp)=B(Y_\perp,X_\perp)~~
\een
is the scalar second fundamental form of $\cF$ (see Appendix
\ref{app:FE}). The Naveira tensor \cite{Naveira} of the orthogonal
almost product structure $\cP$ defined by the pair of distributions
$(\cD,\cD^\perp)$ can also be expressed in
terms of $H$ and $A$ through the formulas given in Appendix
\ref{app:FE}. Notice that $H$ and $A$ contain the same information as
the O'Neill-Gray tensors/Naveira tensor and hence these quantities
fully characterize the extrinsic geometry of $\cF$. Let us examine 
some consequences of the fundamental equations \eqref{FE}. 

\paragraph{The covariant derivatives of $\hV$ and $V$.}
The covariant derivative of the one-form $\hV\in \Omega(\cD^\perp)$
(which is {\em transverse} to $\cF$) can be computed using the
relation $\hV=n_\sharp$, which implies $(\nabla_X\hV)=(\nabla_X
n)_\sharp$ for any vector field $X$ on $M$. Using the fundamental
equations, we find:
\ben
\label{nablaV}
\boxed{\nabla_n \hV ~~= H_\sharp~~,~~\nabla_{X_\perp}\hV=-(AX_\perp)_\sharp}~~.
\een
These relations imply:
\ben
\label{extV}
\dd \hV =\hV\wedge H_\sharp~~,~~\updelta \hV =-\tr A~~.
\een
Since $\iota_\hV(H_\sharp)=g(n,H)=0$, the first equation above gives
$H_\sharp=(\dd \hV)_\top=\iota_\hV\dd \hV$. A simple computation now
gives the following relations which express the covariant derivative
of $V$:
\ben
\label{DerV}
\begin{split}
&(\nabla_n V)_\top=\partial_n ||V||~~,~~(\nabla_jV)_\top=\partial_j ||V||\\
&(\nabla_n V)_\perp=||V||H_\sharp~~,~~(\nabla_jV)_\perp=-||V||(Ae_j)_\sharp~~.
\end{split}
\een
Equations \eqref{DerV} give:
\ben
\label{ExtV}
\dd V={\hat V}\wedge (||V||H_\sharp-\dd_\perp ||V||)=V\wedge (H_\sharp-\dd_\perp\ln ||V||)~~,
~~\updelta V=-\partial_n ||V||+||V||\tr A~~.
\een
\paragraph{Remark.} 
Notice that the differential and codifferential \eqref{extV} of $\hV$
determine $H$ and $\tr A$ but they fail to determine the traceless
part of $A$ and hence they do not fully determine the covariant
derivative \eqref{nablaV} of ${\hat V}$. If $H$ and $A$ are known,
then the space of solutions of \eqref{nablaV} is an affine space
modeled on the kernel $\cK_B$ of the Bochner Laplacian
$\nabla^\ast\nabla$ on $\Omega^1(M)$, thus $\cK_B$ is the space of
parallel one-forms on $M$. On the other hand, the space of solutions
of \eqref{extV} is an affine space modeled on the kernel $\cK_H$ of
the Hodge Laplacian $\dd \updelta+\updelta \dd$ on $\Omega^1(M)$, thus
$\cK_H$ is the space of harmonic one-forms. Recall that the Bochner
and Hodge Laplacians are related through the Weitzenbock identity:
\be
\nabla^\ast \nabla=\dd\updelta+\updelta\dd+\cW~~,
\ee 
where the Weitzenbock operator $\cW$ depends on the Riemann curvature
tensor of $g$. We have $\cK_B\subseteq \cK_H$, but, in general, the
inclusion is strict. Hence, given $H$ and $A$, the space of solutions
of \eqref{nablaV} is generally\footnote{If the Ricci tensor of $M$ is positive
  semidefinite then all harmonic one-forms are covariantly constant by
  Bochner's theorem. This, however, need not be the case for our
  eight-manifolds $M$. Remember that we are dealing with a flux
  compactification (hence the 11-manifold $\mathbf{M}$ is
  not Ricci flat, in fact its Ricci tensor is in general indefinite by
  Einstein's equations) and that we are considering warped product
  backgrounds (hence the components of the Ricci tensor of $\mathbf{M}$
  along $TM$ differ from those of the Ricci tensor of $TM$ by terms
  involving the Hessian of the warp factor $\Delta$ --- and that Hessian is in
  general an indefinite bilinear form).} smaller than the space of solutions to
\eqref{extV}. Similar remarks of course also apply to $V$.

\

\noindent The covariant derivative, exterior derivative and codifferential of arbitrary forms decompose into components 
parallel and perpendicular to ${\hat V}$ according to the formulas given in Appendix \ref{app:FE}.

\paragraph{The normal covariant derivatives of $\varphi$ and $\psi$.} It is shown in Appendix \ref{app:FE} that the following relations hold: 
\ben
\label{NormalPhiPsi}
\boxed{D_n\varphi=3\iota_\vartheta\psi~~~,~~~D_n\psi=-3\vartheta\wedge\varphi}~~.
\een
where $\vartheta\in \Omega^1(\cD)$ can be determined using the first relations in each of the two columns of \eqref{Krel}:
\ben
\label{thetaD}
\vartheta=-\frac{1}{12}\ast_\perp[\varphi\wedge \ast_\perp (D_n\psi )]=-\frac{1}{12}\ast_\perp(\varphi\wedge D_n\varphi)~~.
\een

\subsection{Encoding the covariant derivative constraints through foliated geometry}

In this Subsection, we prove the following result, which provides a solution to Problem 1 of Subsection \ref{subsec:p}:

\paragraph{Theorem 2.} 
Let $||V||=\sqrt{1-b^2}$ and suppose that $(F,f)$ is consistent with
the quadruple $(\Delta,b,{\hat V},\psi)$, i.e.  that the
$\cQ$-constraints are satisfied.  Then the covariant derivative
constraints \eqref{NablaConstr} are {\em equivalent} with the
following conditions:
\begin{enumerate}
\itemsep0em 
\item The function $b\in \cC^\infty(M,(-1,1))$ satisfies:
\ben
\label{cohom2}
\boxed{e^{-3\Delta} \dd(e^{3\Delta}b)=f-4\kappa \sqrt{1-b^2}\hV}
\een
\item The fundamental tensors $H$ and $A$ of $\cF$ and $\cF^\perp$ are given by the following expressions in terms of $b,\psi$ and $f,F$: 
\beqan
\label{FunForms}
\boxed{
\begin{split}
&H_\sharp=\frac{2}{||V||}\alpha_2=-\frac{1}{||V||^2}(bf-3 \dd\Delta )_\perp~,\\
&AX_\perp=\frac{1}{||V||}\Big[(b\chi^{(0)}_{ij} -h^{(0)}_{ij})X_\perp^j e^i+\frac{1}{7}\big(14\kappa b-8\tr_g(\hat h)-6b~\tr_g(\hat\chi)\big)X_\perp \Big]=\\
&~~~~=\frac{1}{||V||}\Big[(b\chi^{(0)}_{ij} -h^{(0)}_{ij})X_\perp^je^i+\frac{1}{7}\Big(-4\kappa b+9||V||(\dd\Delta)_\top-\frac{1}{||V||}(bf-3\dd\Delta)_\top\Big) X_\perp \Big]~~,
\end{split}}~~~~~~~~~
\eeqan
i.e. the covariant derivative of ${\hat V}$ is given by \eqref{nablaV}, where $H$ and $A$ are given by \eqref{FunForms}.
\item The one-form $\vartheta\in \Omega(\cD)$ of \eqref{ThetaParam} is given by the following relation in terms of $\Delta, b$ and $f$:
\ben
\label{vartheta}
\boxed{\vartheta=\frac{b\alpha_2-\alpha_1}{3||V||}=\frac{1}{6||V||^2}\big[-(1+b^2)f+6b\dd\Delta\big]_\perp}
\een
\item The torsion classes of the leafwise $G_2$ structure (in the
  conventions of \cite{Bryant, Kflows}) are given by the following
  expressions in terms of $\Delta, b$ and $f,F$: 
\beqan
\label{TorsionClasses}
\boxed{
\begin{split}
&\boldsymbol{\tau}_0=\frac{4}{7||V||}(b~\tr_g(\hat h)-\tr_g(\hat
  \chi)+7\kappa)=\frac{4}{7||V||}\Big[ 4\kappa
    +\frac{(1+b^2)f_\top-6b(\dd\Delta)_\top}{2||V||}
    \Big]~,\\ 
&\boldsymbol{\tau}_1=-\frac{3}{2}(\dd\Delta)_\perp~,\\ 
&\boldsymbol{\tau}_2=0~,\\ 
&\boldsymbol{\tau}_3=\frac{1}{||V||}(\chi^{(0)}_{ij}-b
  h^{(0)}_{ij})e^i\wedge\iota_{e_j}\varphi=\frac{1}{||V||}(F_\top^{(27)}-b\ast_\perp
  F_\perp^{(27)})~.
\end{split}}
\eeqan
In particular, the leafwise $G_2$ structure is integrable (we have
$\boldsymbol{\tau}_2=0$), i.e. it belongs to the class $W_1\oplus
W_3\oplus W_4$ of the Fernandez-Gray classification \cite{FG}.
\end{enumerate}

\paragraph{Remarks.} 
\begin{enumerate}
\itemsep0em 
\item Notice that Condition 2 of the theorem constrains the {\em
  covariant derivative} of ${\hat V}$ and not simply its exterior
  differential and codifferential (which are given by
  \eqref{extV}). As remarked in Subsection \ref{subsec:extrinsic},
  conditions \eqref{extV} (with $H$ and $A$ given in \eqref{FunForms})
  are {\em weaker} than Condition 2 itself and hence they do {\em
    not} suffice to insure that the background is supersymmetric if
  $F$ and $f$ are fixed.
\item If $e_a$ is a local orthonormal frame of $M$ such that
  $e_1=n\eqdef \hV^\sharp$ and with dual coframe $e^a$ (thus
  $e^a=\hV)$), then the second relation in \eqref{FunForms} gives:
\ben
B(e_i,e_j)\eqdef g(e_i,Ae_j)=A_{ij}=\frac{1}{||V||}\big( -h^{(0)}_{ij}+b\chi^{(0)}_{ij}\big)
+\frac{1}{7}\tr(A)g_{ij}~,\nn
\een
where:
\be
\tr A=\frac{1}{||V||} \big(14\kappa b-8\tr_g(\hat h)-6b~\tr_g(\hat\chi)\big)
\ee
and, from \eqref{Symtens1}, one has:
\beqa
&& h^{(0)}_{ij}=-\frac{1}{4}\left[ \langle \iota_{e^i}\varphi , \iota_{e^j}(\ast_\perp F^{(27)})\rangle + (i\leftrightarrow j) \right]~~,\nn\\
&&\chi^{(0)}_{ij}=-\frac{1}{4}\left[ \langle \iota_{e^i}\varphi , \iota_{e^j}F^{(27)}\rangle + (i\leftrightarrow j) \right]~~.\nn
\eeqa
Notice that the Weingarten tensor $A$ is completely
determined in terms of $F,f$ and $b$. 

\item In general, neither $H$ nor $A$ (equivalently, $B$)
  vanish. Hence Reinhart's criterion (see \cite[Theorem 5.17,
    p. 46]{Tondeur}) tells us that, in general, neither $\cF$ nor
  $\cF^\perp$ are Riemannian foliations (i.e. $g$ is {\em not} a
  bundle-like metric for any of these foliations).

\item Since the leafwise $G_2$ structure has $\boldsymbol{\tau}_2=0$, the results
  of \cite{FriedrichIvanov1} insure existence of a unique metric but
  torsion-full leafwise partial connection
  $\nabla^c:\Gamma(M,\cD)\times \Gamma(M,\cD)\rightarrow
  \Gamma(M,\cD)$ which has `totally antisymmetric torsion tensor'
  (corresponding through the musical isomorphism to a 3-form $T\in
  \Omega^3(\cD)$) and which is adapted to the $G_2$ structure:
\be
\nabla^c_{X_\perp}\varphi=0~~,~~\forall X_\perp\in \Gamma(M,\cD)~~.
\ee
 Furthermore, the spinor $\eta_0$ of \eqref{eta} satisfies
 $\nabla^c\eta_0=0$ (see \cite{FKMS}) and the torsion form $T$ and
 curvature of $\nabla^c$ can be computed using the formulas given in
 \cite{FriedrichIvanov1,FriedrichIvanov2}. Since $\boldsymbol{\tau}_1$
 is exact, the leafwise $G_2$ structure is conformally co-calibrated (a.k.a.
 conformally co-closed). In fact, the conformal transformation
 \eqref{G2conf} with $\alpha=\frac{3}{2}\Delta$ gives:
\be
\begin{split}
& g'_{ij}=e^{3\Delta}g_{ij}~~~,~~\varphi'=e^{\frac{9\Delta}{2}}\varphi~~~~,~~~\psi'=e^{6\Delta}\psi~~,\\
&\boldsymbol{\tau}'_0=e^\frac{3\Delta}{2}\boldsymbol{\tau}_0~~~,~~\boldsymbol{\tau}'_1=\boldsymbol{\tau}'_2=0~~,~~\boldsymbol{\tau}'_3=e^\frac{3\Delta}{2} \boldsymbol{\tau}_3~~,
\end{split}
\ee
so the conformally transformed $G_2$ structure satisfies $\dd_\perp \psi'=0$
i.e. $\updelta'_\perp\varphi'=0$. Co-calibrated $G_2$ structures were
studied in \cite{GrigorianCoflow}.
\end{enumerate}

\paragraph{Proof of Theorem 2.}

The rest of this Subsection is devoted to proving Theorem 2. We warn
the reader that we give only the major steps of most computations and
that performing some of the simplifications afforded by the $G_2$
structure identities of Appendix \ref{app:G2} is very tedious. We used
the package {\tt Ricci} \cite{Ricci} for {\tt
  Mathematica}$^{\textregistered}$, which we acknowledge
here. Throughout the proof, we consider a local orthonormal frame of
$M$ such that $e_1=n=\hat V^\sharp$.

\paragraph{Step 1. The covariant derivative constraints in the non-redundant parameterization.}

Using the identities of Appendix \eqref{app:GA} and \eqref{app:G2},
one can compute the explicit forms of $S_m^{(1)}$ and $S_m^{(8)}$,
finding that the two equations of \eqref{DS} which determine
$\partial_mb$ and $\nabla_m V$ take the following form, in which $F$
was eliminated using the solution of the $\check{Q}$-constraints given
in Theorem 1:
\ben
\label{Nablab}
\begin{split}
&\partial_n b=-\ast S_1^{(8)}=~||V||\Big(2\kappa-2\tr_g(\hat\chi)\Big)~,\\
&\partial_j b=-\ast S_j^{(8)}=~2||V||e_j\lrcorner\alpha_1~,
\end{split}
\een
respectively:
\beqan
\label{NablaV}
\begin{split}
&\nabla_n V=-S_1^{(1)}= 2\alpha_2- \Big(2\kappa b-2b~\tr_g(\hat\chi)\Big)\hat V~,\\
&\nabla_j V=-S_j^{(1)} =\Big[h^{(0)}_{ij}-b\chi^{(0)}_{ij}
-\frac{1}{7}\left(14\kappa b-8\tr_g(\hat h)-6b~\tr_g(\hat\chi)\right)g_{ij}\Big]
e^i- 2b(e_j\lrcorner \alpha_1)\hat V~.
\end{split}~~~~~~
\eeqan
In the non-redundant parameterization \eqref{Enr}, one finds, after
some computations, that the two equations of \eqref{DS} which express
$\nabla_m Y$ and $\nabla_m Z$ are equivalent with the relations:
\ben
\label{nabla_psi}
\nabla_m\psi=-\frac{V}{||V||^2}[-S_m^{(1)}\psi+S_m^{(3)}+S_m^{(5)}]~~,
~~\nabla_m\psi=-\frac{V}{||V||^2}(1-b\nu)[ S_m^{(4)}-S_m^{(8)}\psi ]~~,
\een
which appear to impose the algebraic integrability condition:
\be
-S_m^{(1)}\psi+S_m^{(3)}+S_m^{(5)}=(1-b\nu)[ S_m^{(4)}-S_m^{(8)}\psi ]~~.
\ee
A rather lengthy direct computation shows that this integrability
condition is in fact automatically satisfied and thus provides no new
conditions on the fluxes. Then \eqref{nabla_psi} can be written in
the equivalent form \eqref{nablapsi} given in Appendix \ref{app:other}
upon separating the parts orthogonal and
parallel to $V$. Using the solution \eqref{Fdecomp}, \eqref{SolF} of the
$Q$-constraints and the $G_2$ structure identities given in Appendix
\ref{app:other}, one finds after a lengthy computation that \eqref{nablapsi}
simplifies to:
\beqan
\label{NablaPsi}
\begin{split}
&(\nabla_{n}\psi)_\top=~-\frac{2}{||V||}\iota_{\alpha_2}\psi=\frac{1}{||V||^2}\iota_{(bf_\perp-3(\dd \Delta)_\perp)}\psi~,\\
&(\nabla_{n}\psi)_\perp=\frac{\alpha_1-b\alpha_2}{||V||}\wedge\varphi=\frac{(1+b^2)f_\perp-6b(\dd\Delta)_\perp}{2||V||^2}\wedge\varphi~,\\
&(\nabla_{j}\psi)_\top=\frac{1}{||V||}\Big[ - h^{(0)}_{ij}+b\chi^{(0)}_{ij}
+\frac{1}{7}\left(14\kappa b-8\tr_g(\hat h)-6b~\tr_g(\hat\chi)\right)g_{ij}\Big]\iota_{e_i}\psi~,\\
&(\nabla_{j}\psi)_\perp=\frac{3}{2}(\dd\Delta)_\perp\wedge\iota_{e^j}\psi 
-\frac{3}{2}e^j\wedge\iota_{(\dd\Delta)_\perp}\psi 
-\frac{1}{||V||}\Big[b h^{(0)}_{ij}-\chi^{(0)}_{ij}+\frac{1}{7}\left(b~\tr_g(\hat h)-\tr_g(\hat\chi)+7\kappa\right)g_{ij}\Big]              e^i\wedge\varphi ~.
\end{split}
\eeqan
In conclusion, the covariant derivative constraints \eqref{NablaConstr} are
equivalent, modulo the $\check{Q}$-constraints, with equations
 \eqref{Nablab}, \eqref{NablaV} and \eqref{NablaPsi}.  Direct
computation using the first and last relation in \eqref{SolF} shows
that the system \eqref{Nablab} is equivalent with relation
\eqref{cohom2}.

\paragraph{Remark.} 
Using these equations, one can also compute the covariant derivative
of $\varphi$ and the explicit form of the exterior differential and
codifferential constraints \eqref{EDS} and \eqref{CDS}, which are
given in Appendix \ref{app:other}.

\paragraph{Step 2. Extracting $H$ and $A$.}
\label{sec:fundeq}

\paragraph{Lemma.} Assume that $||V||^2=1-b^2$. Then: 

\

\noindent 1. The second equation on the first row of \eqref{DS} (the
covariant derivative constraint for $V$) is equivalent with the
following relations:
\ben
\label{DerVTopPerp}
H_\sharp=-\frac{1}{||V||}[S_1^{(1)}]_\perp~~,~~(Ae_j)_\sharp=\frac{1}{||V||}[S^{(1)}_j]_\perp~~,
~~\frac{b\partial_mb}{||V||}=[S^{(1)}_m]_\top~~.
\een
\noindent 2. Modulo the first equation in \eqref{DS} (i.e. the covariant
differential constraints for $b$), the last relation in \eqref{DerVTopPerp} is 
equivalent with the following algebraic condition for $S^{(1)}$ and
$S^{(8)}$:
\ben
\label{bVcond}
[S_m^{(1)}]_\top=-\frac{b}{||V||}\ast S_m^{(8)}~~.
\een
\noindent 3. Condition \eqref{bVcond} is automatically satisfied when
$S_m^{(1)}$ and $S_m^{(8)}$ are given by expressions \eqref{Nablab}
and \eqref{NablaV}.  Furthermore, the first two equations in
\eqref{DerVTopPerp} take the form \eqref{FunForms} when substituting
these expressions for $S_m^{(1)}$ and $S_m^{(8)}$.  Hence the first
row of \eqref{DS} is equivalent, modulo the $\check{Q}$-constraints,
with the first two equations in \eqref{FunForms} and the first equation in 
\eqref{DS}, which in turn is equivalent with \eqref{Nablab} and with \eqref{cohom2}.

\

\noindent{\bf Proof.} The first statement follows by separating the
`top` and `perp` parts of the covariant derivative constraint for $V$
given in \eqref{DS} and comparing with \eqref{DerV} while using the
relation $\partial_m||V||=-\frac{b\partial_mb}{||V||}$, which is
implied by the condition $||V||^2=1-b^2$. The second statement now
follows upon eliminating $\partial_m b$ from the third relation in
\eqref{DerVTopPerp} by using the differential constraint for $b$ given
in \eqref{DS}. The remaining statements of the lemma follow by direct
computation. $\blacksquare$

\paragraph{Step 3. Extracting the normal and longitudinal covariant derivatives of $\psi$.}

Applying \eqref{NablaLongit} for $\omega=\psi$, we find the following:
\begin{itemize}
\itemsep0em 
\item The first and third equation of \eqref{NablaLongit} for
  $\omega=\psi$ are equivalent respectively with the first and third
  equation of \eqref{NablaPsi} upon using expressions \eqref{FunForms}
  for $H$ and $A$.
\item The second equation of \eqref{NablaLongit} for $\omega=\psi$
  agrees with the second equation of \eqref{NablaPsi} provided that
  the normal covariant derivative of $\psi$ is given by:
\ben
\label{NorDerPsi}
D_n\psi=~\frac{\alpha_1-b\alpha_2}{||V||}\wedge\varphi~~.
\een
Relation \eqref{NorDerPsi} determines the one-form $\vartheta$ of
\eqref{ThetaParam}.  Comparing with the second equation of
\eqref{NormalPhiPsi} gives \eqref{vartheta}.
\item The last equation of \eqref{NablaLongit} for $\omega=\psi$
  agrees with the second equation of \eqref{NablaPsi} provided that
  the induced covariant derivative of $\psi$ along the leaves of the
  foliation is given by:
\!\!\!\!\!\!\!\!\!\!\!\!\!\!\!\!\!\!\!\!\!\!\!\!\ben
\label{IndDerPsi}
\!\!\!\!\!\!\!\!\!\!\!\!\nabla_{j}^\perp\psi=\frac{3}{2}(\dd\Delta)_\perp\wedge\iota_{e^j}\psi 
-\frac{3}{2}e^j\wedge\iota_{(\dd\Delta)_\perp}\psi 
-\frac{1}{||V||}\Big[b h^{(0)}_{ij}-\chi^{(0)}_{ij}+\frac{1}{7}\left(b~\tr_g(\hat h)-
\tr_g(\hat\chi)+7\kappa\right)g_{ij}\Big] e^i\wedge\varphi~.~~~~~~~~~~~~
\een~~~~~~~~~~~
\end{itemize}
Hence the entire system \eqref{NablaPsi} of covariant derivative
constraints for $\psi$ is {\em equivalent} with equations
\eqref{FunForms} taken together with \eqref{NorDerPsi} and
\eqref{IndDerPsi}. 

\paragraph{Step 4. Encoding the longitudinal covariant derivative of $\psi$ through the torsion forms of the leafwise $G_2$-structure.}
Relation \eqref{IndDerPsi} can be expressed in a simpler equivalent
form using the fact \cite{FG} that the covariant derivative of the
associative and/or coassociative forms of a manifold with $G_2$
structure (taken with respect to the Levi-Civita connection of the
corresponding metric) is completely specified by the torsion classes
of that $G_2$ structure.  The torsion forms
$\boldsymbol{\tau}_0\in\Omega^0_1(\cD)$, $\boldsymbol{\tau}_1\in
\Omega^1_7(\cD)$, $\boldsymbol{\tau}_2\in \Omega^2_{14}(\cD)$ and
$\boldsymbol{\tau}_3\in \Omega^3_{27}(\cD)$ of the leafwise $G_2$
structure (in the conventions of \cite{Bryant, Kflows}) are uniquely
determined by relations \eqref{TorsionEqs} and hence can be extracted
by computing the differentials of $\psi$ and $\varphi$ starting from
equation \eqref{IndDerPsi}, which gives:
\be
\dd_\perp\psi=e^j\wedge(\nabla_j\psi)_\perp=-6(\dd\Delta)_\perp\wedge\psi~~,
\ee
and:
\beqa
\dd_\perp\varphi&=&e^j\wedge(\nabla_j\varphi)_\perp=
-\ast_\perp[\iota_{e^j}(\nabla_j\psi)_\perp]=\ast_\perp\updelta_\perp\psi=
\nn\\
&=&-\frac{9}{2}(\dd\Delta)_\perp\wedge\varphi+\frac{4}{7||V||}\left(b~\tr(\hat h)-\tr_g(\hat\chi)+7\kappa\right)\psi+\frac{1}{||V||}\ast_\perp(F_\top^{(27)}-b\ast F_\perp^{(27)})~.
\eeqa
Comparing with \eqref{TorsionEqs} gives relations
\eqref{TorsionClasses}. The results of \cite{FG} assure us that
equation \eqref{IndDerPsi} is {\em equivalent} with conditions \eqref{TorsionEqs}, 
where the torsion classes are given by \eqref{TorsionClasses}. Theorem
2 now follows by combining the previous statements. $\blacksquare$

\subsection{The exterior derivatives of $\varphi$ and $\psi$ and the differential and codifferential of $V$}
Applying \eqref{ExtOmega} to the longitudinal forms $\varphi\in
\Omega^3(\cD)$ and $\psi\in \Omega^4(\cD)$ gives:
\beqan
&&(\dd \varphi)_\top=D_n\varphi-A_{jk}e^j\wedge \iota_{e^k}\varphi~~,~~~(\dd \varphi)_\perp=\dd_\perp\varphi=\boldsymbol{\tau}_0\psi+3\boldsymbol{\tau}_1\wedge \varphi+\ast_\perp\boldsymbol{\tau}_3\nn~~,\\
&& (\dd \psi)_\top=D_n\psi-A_{jk}e^j\wedge \iota_{e^k}\psi~~,~~~(\dd \psi)_\perp=\dd_\perp\psi=4\boldsymbol{\tau}_1\wedge \psi+\ast_\perp \boldsymbol{\tau}_2~~.
\eeqan
The second relations in each row above show that $(\dd \varphi)_\perp$
and $(\dd \psi)_\perp$ determine the torsion classes of the leafwise
$G_2$ structure.  Decomposing the first relation in each row according
to the irreducible components of the $G_2$ action on $\wedge^3
\cD^\ast$ and $\wedge^4 \cD^\ast $, we find:
\beqan
&& (\dd\varphi)_\top^{(1)}=-\frac{3}{7}(\tr A) \varphi~~,~~
(\dd\varphi)^{(7)}_\top=D_n\varphi~~,~~(\dd\varphi)^{(27)}_\top=-A_{jk}^{(0)}e^j\wedge \iota_{e^k}\varphi~~,\nn\\
&& (\dd\psi)_\top^{(1)}=-\frac{4}{7}(\tr A) \psi~~,~~(\dd\psi)^{(7)}_\top=D_n\psi~~,~~(\dd\psi)^{(27)}_\top=-A_{jk}^{(0)}e^j\wedge \iota_{e^k}\psi~~.
\eeqan
which shows that any of $(\dd\varphi)_\top$ or $(\dd\psi)_\top$
suffices to determine the Weingarten tensor $A$ as well as
$\vartheta$.

\paragraph{Remark.} 
One might imagine that the remark above obsoletes the need for the
analysis of the full covariant derivatives which we carried out in the
proof of Theorem 2 --- since the differential and codifferential of
$\varphi$ and $\psi$ determine $A, \vartheta$ and the torsion classes
of the leafwise $G_2$ structure, while the differential of ${\hat V}$
determines $H$ (see \eqref{extV}), one might be tempted to think that
one could use them from the outset and forget about the full covariant
derivatives of $\varphi$ and $\psi$.  This, however, would be
insufficient to {\em prove} a result such as Theorem 2, since it is
not clear a priori that there are no supplementary algebraic
constraints imposed by the full supersymmetry conditions
\eqref{NablaConstr} and \eqref{QConstr}. The point of Theorem 2 is
that it gives an {\em equivalent} geometric characterization of the
supersymmetry conditions, without loosing any information contained in
the latter.  Notice also that the supersymmetry conditions determine
the covariant derivative \eqref{DerV} of $V$ if $f$ and $F$ are given
(since they determine $H$ and $A$ as well as $\partial_m b$).  This is
stronger than simply determining the differential and codifferential
of $V$. In fact, the covariant derivatives:
\beqan
\nabla_n V&=&2\alpha_2+2b\big(\tr_g(\hat\chi)-\kappa\big)\hat V~, \\
\nabla_j V&=&-2b(e_j\lrcorner\alpha_1)\hat V+\Big[ h^{(0)}_{ij}-b\chi^{(0)}_{ij}-\frac{1}{7}\big( 14\kappa b-8\tr_g (\hat h)-6b~\tr_g(\hat\chi) \big)g_{ij} \Big]e^i ~\nn
\eeqan
depend explicitly on the fluxes $f$ and $F$ (through the tensors $\hat
h,\hat\chi$) while the differential and codifferential of $V$:
\beqan
\dd V&=&3 V\wedge(\dd\Delta)_\perp~,\nn\\
\updelta V& =&-8\kappa b+12||V||(\dd\Delta)_\top~~
\eeqan
depend only on $b$ (equivalently, on $||V||)$ and on $\Delta$ and $n$.
\subsection{Eliminating the fluxes} 

The following result gives the set of conditions equivalent with the
existence of at least one non-trivial solution $\xi$ of \eqref{par_eq}
which is everywhere non-chiral, while expressing $f$ and $F$ in terms
of $\Delta$ and of the quantities $b,{\hat V}$ and $\varphi$ of
\eqref{Eparam}. This solves Problem 2 of Subsection \ref{subsec:p}.

\paragraph{Theorem 3.} 
The following statements are equivalent: 
\begin{enumerate}[(A)]
\itemsep0em
\item There exist $f\in \Omega^1(M)$ and $F\in \Omega^4(M)$ such that
  \eqref{par_eq} admits at least one non-trivial solution $\xi$ which
  is everywhere non-chiral (and which we can take to be everywhere of
  norm one).
\item There exist $ \Delta\in \cinf$, $b\in \cC^\infty(M,(-1,1))$, 
  ${\hat V}\in \Omega^1(M)$ and $\varphi\in \Omega^3(M)$ such that:
\begin{enumerate}[1.]
\itemsep0em
\item $\Delta$, $b$, ${\hat V}$ and $\varphi$ satisfy the conditions:
\ben
\label{bVphi}
||{\hat V}||=1~~,~~\iota_{\hat V} \varphi=0~~.
\een
Furthermore, the Frobenius distribution $\cD\eqdef \ker {\hat V}$ is
integrable and we let $\cF$ be the
foliation which integrates it.
\item The quantities $H$, $\tr A $ and $\vartheta$ of the foliation $\cF$ are given by:
\beqan
\label{TgeomHB}
\boxed{
\begin{split}
&H_\sharp=\frac{2}{||V||}\alpha_2=-\frac{b}{||V||^2}(\dd b)_\perp+3(\dd\Delta)_\perp =\frac{\dd_\perp ||V||}{||V||} +3(\dd \Delta)_\perp ~, \\
& \tr A =12(\dd\Delta)_\top-\frac{b(\dd b)_\top}{||V||^2}-8\kappa \frac{b}{||V||}=12\partial_n \Delta +\frac{\partial_n ||V||-8\kappa b}{||V||}~,\\
& \vartheta=-\frac{1+b^2}{6||V||^2}(\dd b)_\perp+\frac{b}{2}(\dd\Delta)_\perp~~.
\end{split}}
\eeqan
\item $\varphi$ induces a leafwise $G_2$ structure on $\cF$ whose
  torsion classes satisfy:
\beqan
\label{TgeomTorsion}
\boxed{
\begin{split}
&\boldsymbol{\tau}_{0}=\frac{4}{7||V||}\Big[  2\kappa(3+ b^2)-\frac{3b}{2}||V||(\dd\Delta)_\top+\frac{1+b^2}{2||V||}(\dd b)_\top  \Big]~,\\
&\boldsymbol{\tau}_{1}= -\frac{3}{2}(\dd\Delta)_\perp~,\\
&\boldsymbol{\tau}_{2}=0~~.
\end{split}}
\eeqan
\end{enumerate}
\end{enumerate}
In this case, the forms $f$ and $F$ are uniquely determined by
$b,\Delta, V$ and $\varphi$. Namely, the one-form $f$ is given by:
\ben
\label{Tf}
\boxed{f=4\kappa V+e^{-3\Delta}\dd(e^{3\Delta} b)}~~,
\een
while $F$ is given as follows:
\begin{enumerate}[(a)] 
\itemsep0em
\item We have $F^{(1)}_\top=\frac{3}{7}\tr_g(\chi)\varphi=-\frac{4}{7}\tr_g(\hat\chi)\varphi$ and $
  F^{(1)}_\perp=-\frac{4}{7}\tr_g(\hat h)\psi$, with:
\ben
\label{TF1}
\boxed{
\tr_g(\hat h)=-\frac{3||V||}{2}(\dd\Delta)_\top+2\kappa b+
\frac{b}{2||V||}(\dd b)_\top~~~,~~~\tr_g(\hat\chi)=\kappa-\frac{1}{2||V||}(\dd b)_\top
}
\een
\item We have $F^{(7)}_\top=-\iota_{\alpha_2}\psi$ and
  $F^{(7)}_\perp=\alpha_1\wedge\varphi$, with:
\ben
\label{TF7}
\boxed{
\alpha_1=\frac{1}{2||V||}(\dd b)_\perp~~~,~~~\alpha_2=-\frac{b}{2||V||}(\dd b)_\perp+\frac{3||V||}{2}(\dd\Delta)_\perp=\frac{\dd_\perp ||V||}{||V||}+\frac{3||V||}{2}(\dd\Delta)_\perp
}
~~~
\een
\item We have:
\ben
\boxed{
\begin{split}
h^{(0)}_{ij}=-\frac{b}{4||V||}[\langle e_i\lrcorner \varphi, e_j\lrcorner\boldsymbol{\tau}_{3}\rangle + (i\leftrightarrow j)]-\frac{1}{||V||}A^{(0)}_{ij}=\frac{b}{||V||}t_{ij}-\frac{1}{||V||}A^{(0)}_{ij}~~,\nn\\
\chi^{(0)}_{ij}=-\frac{1}{4||V||}[\langle e_i\lrcorner \varphi, e_j\lrcorner\boldsymbol{\tau}_{3} \rangle+ (i\leftrightarrow j)] - \frac{b}{||V||}A^{(0)}_{ij}=\frac{1}{||V||}t_{ij}-\frac{b}{||V||}A^{(0)}_{ij}~~,
\end{split}}
\een
(where in the last equalities we used relation \eqref{t}), i.e.:
\beqan
\label{TF27}
&& F^{(27)}_\perp=-h^{(0)}_{ij} e^i\wedge\iota_{e^j}\psi =\frac{b}{||V||}\ast_\perp\boldsymbol{\tau}_3+\frac{1}{||V||}A^{(0)}_{ij} e^i\wedge\iota_{e^j}\psi~~,\nn\\
&& F^{(27)}_\top=\chi^{(0)}_{ij} e^i\wedge\iota_{e^j}\varphi=\frac{1}{||V||}\boldsymbol{\tau}_3-\frac{b}{||V||}A^{(0)}_{ij}e^i\wedge\iota_{e^j}\varphi~~,
\eeqan
where $A^{(0)}$ is the traceless part of the Weingarten tensor of $\cF$ while $\boldsymbol{\tau}_3$ is the rank 3 torsion class of the leafwise $G_2$
structure.
\end{enumerate}

\paragraph{Remarks.} 
\begin{enumerate}
\itemsep0em
\item We show in Appendix \ref{app:other} that \eqref{TF1} and \eqref{TF7}
are equivalent with \cite[eqs. (3.20), (3.21)]{MartelliSparks}. Notice
that loc. cit. does not give the component $F^{(27)}$, which we give
here explicitly (see relation \eqref{TF27}).
\item Using \eqref{ExtV} as well as the identities: 
\be
e^{-3\Delta}\dd(e^{3\Delta}V)=\dd V-3V\wedge \dd \Delta~~,~~
e^{-12\Delta}\updelta(e^{12\Delta}V)=\updelta V-12||V||\partial_n\Delta~~,
\ee
it is easy to check that the first and second conditions in
\eqref{TgeomHB} are equivalent with the following two equations for
$V$:
\beqan
\label{EV}
\boxed{
\begin{split}
&&\dd (e^{3\Delta} V)=0\\
&&e^{-12\Delta}\updelta(e^{12\Delta}V)+8\kappa b=0
\end{split}}~~,
\eeqan
where the second equation in \eqref{EV} can also be written as follows
upon using an orthonormal frame with $e_1=n$:
\be
e^{-12\Delta}\partial_n(e^{12\Delta}\sqrt{1-b^2})=8\kappa b~~.
\ee
Since the first relation in \eqref{EV} implies integrability of $\cD$,
it follows that this condition stated in point (B.1) of the theorem is
in fact implied by the conditions stated in point (B.2).  It turns out
that conditions \eqref{EV} coincide with \cite[eqs. (3.16)]{Tsimpis},
since it is possible to show \footnote{The full comparison with the
  approach of \cite{Tsimpis} can be found in \cite{g2s}.} that the quantity denoted by $L$ in loc. cit. is given
by $L=\frac{1}{1+b}V$. 

\item The theorem allows one to determine the metric $g$ as follows. 
First notice that \eqref{EV} can be written as:
\ben
\begin{split}
\label{beq}
&-\partial_n||V||+8\kappa b=12||V||\partial_n \Delta~~,\\
& -\frac{\dd_\perp ||V||}{||V||}=3(\dd \Delta)_\perp-n\lrcorner \dd {\hat V}~~.
\end{split}
\een
If $n$ and $\cD$ are given, then ${\hat V}$ is uniquely determined by the 
conditions: 
\ben
\label{hatV}
\ker {\hat V}=\cD~~,~~n\lrcorner {\hat V}=1
\een
and \eqref{beq} can be used to determine $\Delta$ if $b$ is given or
to determine $b$ if $\Delta$ is given. Now suppose that $\Delta$,
$\cF$, $n$ and a leafwise $G_2$ structure along $\cF$ are given, where
$n$ is a vector field on $M$ which is everywhere transverse to $\cF$
and where the torsion classes of the leafwise $G_2$ structure satisfy
\eqref{TgeomTorsion}. Then $\cD=T\cF$ and ${\hat V}$ is determined by
\eqref{hatV}. The system \eqref{beq} can be used to determine $b$ and
hence $||V||^2$ and $V$, which in turn fixes the restriction of the
metric $g$ to the foliation $\cF^\perp$ which integrates the vector
field $n$. The restriction of the metric on $\cF$ (and hence the
metric on $M$) is then determined by the associative form $\varphi$ of
the leafwise $G_2$ structure through relation \eqref{G2rel}. Using
these observations, one can formulate the mathematical problem of
studying our backgrounds in a metric-free manner, namely as a problem
of foliations which can be defined by a closed one form and which are
endowed with longitudinal $G_2$ structures satisfying the version of
the conditions listed in point 3 of the theorem which is obtained by
expressing $b$ as the solution of \eqref{beq}. This approach could be
used to construct examples of such foliations.
\end{enumerate}
\paragraph{Proof.} 
Using relations \eqref{Nablab}, we extract $\alpha_1$ and $\tr_g(\hat\chi)$:
\ben
\label{alpha1}
\alpha_1=\frac{1}{2||V||}(\dd b)_\perp~~~,
~~~\tr_g(\hat\chi)=\kappa-\frac{1}{2||V||}(\dd b)_\top~~.
\een
Substituting these relations into the first and fourth relations of
\eqref{SolF}, we find the following expressions for the
components of the 1-form flux:
\beqan
\label{geomf}
&& f_\perp =(\dd b)_\perp+3b(\dd \Delta)_\perp~,\nn\\
&&  f_\top=3b(\dd \Delta)_\top+(\dd b)_\top +4\kappa ||V||~.
\eeqan
The second and third relations in \eqref{SolF} become:
\beqan
\label{alpha2}
&&\alpha_2=-\frac{b}{2||V||}(\dd b)_\perp+\frac{3||V||}{2}(\dd\Delta)_\perp~,\nn\\
&&\tr_g(\hat h)=-\frac{3||V||}{2}(\dd\Delta)_\top+2\kappa b+\frac{b}{2||V||}(\dd b)_\top~.
\eeqan
Substituting \eqref{geomf} and \eqref{alpha2} into \eqref{FunForms} gives:
\beqan
\label{geomHB}
&&H_\sharp=\frac{2}{||V||}\alpha_2=-\frac{b}{||V||^2}(\dd b)_\perp+3(\dd\Delta)_\perp ~~,\\
&& B(e_i,e_j)\eqdef g(e_i,Ae_j)=A_{ij}=\frac{1}{||V||}(b\chi^{(0)}_{ij}-h^{(0)}_{ij})+\frac{1}{7}\Big[12(\dd\Delta)_\top-\frac{b(\dd b)_\top}{||V||^2}-\frac{8\kappa b}{||V||} \Big] g_{ij} \nn~~,
\eeqan
where the traceless symmetric tensors can be expressed from relations \eqref{TF27} and \eqref{Symtens1} as follows:
\beqan
h^{(0)}_{ij}&=&-\frac{1}{4}\left[\langle \iota_{e^i}\varphi, \iota_{e^j}(\ast_\perp F^{(27)}_\perp)\rangle + (i\leftrightarrow j) \right]
=-\frac{b}{4||V||}\left[\langle \iota_{e^i}\varphi, \iota_{e^j}\boldsymbol{\tau}_3\rangle + (i\leftrightarrow j) \right] -\frac{1}{||V||}A^{(0)}_{ij}~,\nn\\
\chi^{(0)}_{ij}&=&-\frac{1}{4}\left[ \langle \iota_{e^i}\varphi, \iota_{e^j}F^{(27)}_\top \rangle +(i\leftrightarrow j) \right]
=-\frac{1}{4||V||}\left [\langle \iota_{e^i}\varphi, \iota_{e^j}\boldsymbol{\tau}_3\rangle +(i\leftrightarrow j) \right] -\frac{b}{||V||}A^{(0)}_{ij}~.
\eeqan 
Using \eqref{alpha1}, \eqref{geomf} and  \eqref{alpha2}, the covariant
derivatives \eqref{NablaPsi} and \eqref{NablaPhi} of $\psi$ become:
\beqan
\label{geomNablaPsi}
&&(\nabla_{n}\psi)_\top=~-\frac{2}{||V||}\iota_{\alpha_2}\psi=\frac{b}{||V||^2}\iota_{(\dd b)_\perp}\psi-3\iota_{(\dd \Delta)_\perp}\psi~,\nn\\
&&(\nabla_{n}\psi)_\perp=\frac{\alpha_1-b\alpha_2}{||V||}\wedge\varphi=\Big[\frac{(1+b^2)}{2||V||^2}(\dd b)_\perp-\frac{3b}{2}(\dd\Delta)_\perp\Big]\wedge\varphi~,\\
&&(\nabla_{j}\psi)_\top=\frac{1}{||V||}\Big[ - h^{(0)}_{ij}+b\chi^{(0)}_{ij}
+\frac{1}{7}\Big(12||V||(\dd\Delta)_\top-8\kappa b-\frac{b}{||V||}(\dd b)_\top\Big)g_{ij}\Big]\iota_{e_i}\psi~,\nn\\
&&(\nabla_{j}\psi)_\perp=\frac{3}{2}(\dd\Delta)_\perp\wedge\iota_{e^j}\psi 
-\frac{3}{2}e^j\wedge\iota_{(\dd\Delta)_\perp}\psi 
-\frac{1}{||V||}(b h^{(0)}_{ij}-\chi^{(0)}_{ij})e^i\wedge\varphi \nn\\
&&~~~~~~~~~~~-\frac{1}{7||V||}\big[ 2\kappa(3+ b^2)-\frac{3b}{2}||V||(\dd\Delta)_\top+\frac{1+b^2}{2||V||}(\dd b)_\top\big]e^j\wedge\varphi ~.\nn
\eeqan
These expressions allow us to compute:
\beqa
&& \dd_\perp\psi=e^j\wedge(\nabla_j\psi)_\perp=6(\dd\Delta)_\perp\wedge\psi~,\nn\\
&&\dd_\perp\varphi=e^j\wedge\ast_\perp(\nabla_j\psi)_\perp=-\frac{9}{2}(\dd\Delta)_\perp\wedge\varphi+\frac{4}{7||V||}\Big[  2\kappa(3+ b^2)-\frac{3b}{2}||V||(\dd\Delta)_\top+\frac{1+b^2}{2||V||}(\dd b)_\top  \Big]\psi \nn\\
&&~~~~~~~~~~~~~~~~~~~~~~~~~~~~~~~+\frac{1}{||V||}\ast_\perp(F^{(27)}_\top-b\ast_\perp F^{(27)}_\perp)~.  \nn
\eeqa
Comparing with \eqref{TorsionEqs} gives \eqref{TgeomTorsion}. Finally,
notice that the second relation in \eqref{FunForms} can be written as:
\be
e^j\wedge(\nabla_j\varphi)_\top=e^j\wedge\iota_{Ae_j}\varphi=\frac{1}{||V||}\Big[-\ast_\perp F^{(27)}_\perp+b F^{(27)}_\top 
+\frac{3}{7}\Big(12||V||(\dd\Delta)_\top-8\kappa b -\frac{b}{||V||}(\dd b)_\top) \Big)\varphi \Big]~~.\nn
\ee
Combining this with the last relation in \eqref{TorsionClasses} gives
expressions \eqref{TF27}. $\blacksquare$

\section{Topology of $\cF$}
\label{sec:top}

Recall our assumptions that $M$ is compact and connected,  $V$
is nowhere-vanishing and that our foliation $\cF$ integrates the
distribution $\cD=\ker\omega$ defined by the closed nowhere-vanishing
one-form $\momega\eqdef 4\kappa e^{3\Delta} V=4\kappa \mV$, where:
\be
\mV\eqdef e^{3\Delta}V~~.
\ee
The topology of foliations defined by a closed nowhere-vanishing
one-form is well understood.  We recall some relevant results
\cite{Reeb1, Haefliger, Moussu, Tischler, Novikov, Imanishi, S}, stressing
aspects which are of interest for this paper. The entire discussion of
this section applies to any codimension one foliation $\cF$ which
is defined by a closed nowhere-vanishing one-form $\momega$ on a
compact, connected and boundary-less manifold $M$ of arbitrary positive
dimension. We let $\f\in H^1(M,\R)$ denote the cohomology class of
$\momega$.

\subsection{Basic properties}

We already noticed above that $\cF$ is transversely orientable.  By a
result of Reeb \cite{Reeb1}, the holonomy group of each leaf of $\cF$
is trivial. The following argument of loc. cit. shows that all leaves
of $\cF$ are diffeomorphic.  Since $\momega$ is nowhere-vanishing,
there exists a smooth vector field $v$ on $M$ (determined up to
addition of a vector field lying in the kernel of $\momega$) such that
$v\lrcorner \momega=1$ everywhere; in our application, we can take
$v=\frac{e^{-3\Delta}}{||V||}n$. In particular, $v$ is transverse to
the leaves of $\cF$. Since $M$ is compact, the vector field $v$ is
complete and its flow $\phi_t\in\Diff(M)$ is defined for all $t\in
\R$. The Lie derivative $\cL_v\momega=\dd(v\lrcorner
\momega)+v\lrcorner \dd\momega$ is identically zero, which means that
$\phi_t$ preserves $\momega$: \be \phi_t^\ast(\momega)=\momega~~.  \ee
Thus $\phi_t$ is an automorphism of $\cF$ (it diffeomorphically maps
leaves into leaves) for any $t\in \R$. Since $M$ is connected, this
immediately implies that all leaves are diffeomorphic with each
other. Notice that this conclusion does not depend on whether the
leaves are compact or not. It is not hard to check (see, for example,
Exercise 1.3.18 of \cite[page 41]{CC1} or \cite{GMP}) that the group
of periods $\Pi_\f$ coincides with the set of those $t\in \R$ for
which the flow $\phi_t$ stabilizes any (and hence all) leaves $L$ of
$\cF$:
\ben
\label{PerPhi}
\Pi_\f=\{t\in \R|\phi_t(L)=L\}~~.
\een
Hence an integral curve $\ell:\R\rightarrow M$ of $v$ which
is parameterized such that $\ell(0)$ belongs to $L$ will meet $L$
exactly at the points $\ell(t)$ for which $t\in \Pi_\f$. 

Another useful fact (which also holds \cite{Novikov} for any foliation
of $M$ having trivial holonomy) is that the map
$j_\ast:\pi_1(L)\rightarrow \pi_1(M)$ induced by the inclusion of
$j:L\rightarrow M$ is injective and that $j_\ast(\pi_1(L))$ coincides
with the kernel $A_\f$ of the period map $\per_\f$; hence $\pi_1(L)$
can be identified with $A_\f$. In fact, the universal covering space
${\tilde M}$ of $M$ is diffeomorphic \cite{Novikov} with the product
${\tilde L}\times \R$ where ${\tilde L}$ is the universal covering
space of $L$.  Further, the integration cover $\hat{M}_\f$ of
$\per_\f$ is diffeomorphic with the cylinder $L\times \R$, hence $M$
can be presented as a quotient of the latter by an action of $\Pi_\f$
which maps $L_t\eqdef L\times\{t\}$ into $L_{t+s}$ for each $s\in
\Pi_\f$.

\subsection{The projectively rational and projectively irrational cases}

\paragraph{The case when $\momega$ is projectively rational.} 
In this case, one has the following result, which is essentially due to Tischler \cite{Tischler}: 

\paragraph{Proposition.} Let $\momega$ be projectively rational. Then the 
leaves of $\cF$ are compact and coincide with the fibers of a
fibration $\mathfrak{h}:M\rightarrow S^1$.  Moreover, $M$ is diffeomorphic with the
mapping torus $\mathbb{T}_{\phi_{a_\f}}(M)\eqdef M\times[0,1]/\{(x,0)\sim
(\phi_{a_\f}(x),1)\}$, where $a_\f\eqdef \inf(\Pi_\f\cap \N^\ast)$ is the fundamental 
period of $\f$. 

\

\noindent The construction of $\mathfrak{h}$ is given in Appendix \ref{app:fibration}.

\paragraph{The case when $\momega$ is projectively irrational.}
In this case, each leaf of $\cF$ is non-compact and dense in $M$ and
hence $\cF$ cannot be a fibration. The quotient topology on $\R/\Pi_\f$
(which is the leaf space of $\cF$) is the coarse topology.
One way to approach this situation is to approximate $\cF$ by a
fibration as follows \cite{Tondeur}. Let $g$ be an arbitrary
Riemannian metric on $M$ and let $||~||$ denote the ${\rm L}^2$ norm
induced by $g$ on $\Omega^1(M)$.  Then given any $\epsilon>0$, one can
find a closed one-form $\omega_\epsilon$ on $M$ which is projectively
rational and which satisfies $||\momega-\omega_\epsilon||<\epsilon$,
which implies that the distribution $\cD_\epsilon\eqdef \ker \omega_\epsilon$
approximates $\cD$ when $\epsilon\rightarrow 0$. Then the foliation
$\cF_\epsilon$ (which is a fibration) defined by $\omega_\epsilon$
``approximates'' $\cF$. A similar result holds when approximating
$\momega$ in the $\cC^\infty$ topology \cite{CC1}.

\subsection{Noncommutative geometry of the leaf space} 

Since the quotient topology on $M/\cF$ is extremely poor in the
projectively irrational case, a better point of view is to consider
the $C^\ast$ algebra $C(M/\cF)$ of the foliation (the convolution
algebra of the holonomy groupoid of $\cF$), which encodes the
`noncommutative topology' \cite{ConnesFol, ConnesNG} of the leaf space. 
Since in our case the leaves of $\cF$ have no holonomy, the explicit
form of this $C^\ast$ algebra can be determined up to stable equivalence.

Consider a presentation of $\Pi_\f$ of the form \eqref{PresPi}, where $\rho\eqdef \rho(\f)$. 
Then the Abelian group $\Z^{\rho-1}$ has an action on the unit circle given by: 
\be
\Xi(m_2, \ldots, m_\rho )(e^{\frac{2\pi i }{a_1} x})\eqdef e^{\frac{2\pi i}{a_1} (x+m_2 a_2+ \ldots + m_\rho a_\rho)}~~(x\in [0,a_1))~~,
\ee
which induces an action through $\ast$-automorphisms on the $C^\ast$
algebra $C(S^1)$ of continuous complex-valued functions defined on $S^1$:
\ben
\label{Xi}
\Xi'(m_2, \ldots, m_\rho)(\sigma)=\sigma\circ [\chi(-m_2,\ldots, -m_\rho)]~~(\sigma \in C(S^1))~~.
\een
The transformation group $C^\ast$ algebras $C_0(\R)\rtimes
\Pi_\f\simeq C_0(\R)\rtimes_\Xi \Z^\rho$ and $C(S^1)\rtimes_{\Xi'}
\Z^{\rho-1}$ are stably isomorphic \cite{Stadler}, where $C_0(\R)$
denotes the algebra of continuous complex-valued functions on $\R$
which vanish at infinity.

\paragraph{Proposition \cite{Natsume, Stadler}}  $C(M/\cF)$ is separable and 
strongly Morita equivalent (hence also \cite{BGR} stably
isomorphic) with the crossed product algebra $C_0(\R)\rtimes
\Pi_\f\simeq C(S^1)\rtimes_{\Xi'} \Z^{\rho-1}$, which is isomorphic with 
$C(S^1)$ when $\rho=1$ and with a $\rho$-dimensional noncommutative
torus when $\rho>1$.

\

\noindent Thus $C(M/\cF)$ is isomorphic with the $C^\ast$ algebra
$C(T^\rho/\cF_T)\simeq C(T^\rho)\rtimes \R^{\rho-1}\simeq C(S^1)\rtimes \Z^{\rho-1}$ of the
codimension one linear foliation $\cF_T$ which is defined on the
$\rho$-dimensional torus $T^\rho$ by the one-form $a_1 \dd x_1+\ldots
+a_\rho\dd x_\rho=0$. In this sense, $\cF_T$ models the geometry of
the leaf space of $\cF$ (it is a `classifying foliation' for the
latter in the sense of \cite{Stadler}). As a consequence of this
description, foliations defined by a closed one form are among the
cases for which the Baum-Connes conjecture is known to be true (see
\cite{NatsumeBC} for the smooth case and \cite{Stadler} for the $C^0$
case).
\begin{figure}[h!]
\begin{center}
\includegraphics[scale=0.7]{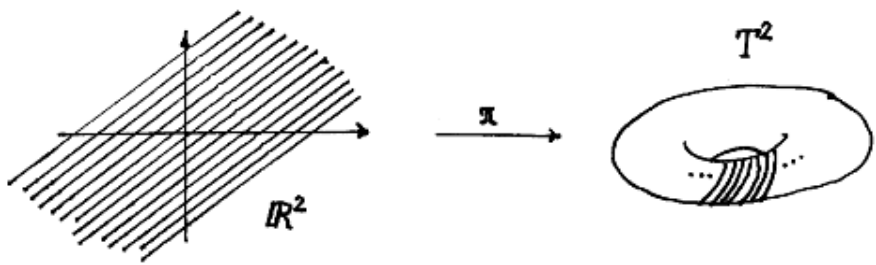}
\caption{The linear foliations of $T^2$ model the noncommutative geometry of the leaf space of $\cF$ in the case $\rho(\f)\leq 2$.}
\end{center}
\end{figure}

\subsection{A ``flux'' criterion for the topology of $\cF$}

The criterion given above for deciding when the foliation is a fibration is
expressed directly in terms of a component of the 4-form flux of
eleven-dimensional supergravity, which takes the form (see
\eqref{Gansatz}):
\ben
\label{tG}
\mathbf{G} =\nu_3\wedge \mf+\mF~~.
\een
The Bianchi identity for $\mathbf{G}$ amounts to $\dd \mathbf{f}=0$ (which we already know to be a
consequence of the supersymmetry conditions) plus the supplementary
condition $\dd \mF=0$. Thus:

\paragraph{Proposition.} If there exists a positive scaling factor $\lambda\in \R^\ast_+$ such that $\lambda \f\in H^1(M,\Z)$, 
then all leaves of $\cF$ are compact and $\cF$ is a fibration over
$S^1$, while $M$ is diffeomorphic with a mapping torus. If such a
scaling factor does not exist, then the foliation $\cF$ is minimal, i.e. each leaf of $\cF$ is dense in
$M$ and $\cF$ cannot be a fibration (since $M$ is compact).

\paragraph{Remarks.}
\begin{enumerate}
\itemsep 0.0em
\item When passing to M theory, quantum consistency requires  \cite{Witten} the flux of $\mathbf{G}$ to satisfy the condition:
\ben
\label{Gq}
\int_{D}\frac{\mathbf{G}}{2\pi} -\frac{1}{4} \int_{D} p_1(\mathbf{M})\in \Z ~~,~~\forall D\in H_4(\mathbf{M},\Z)~~,
\een
where $H_\ast(\mathbf{M},\Z)$ denotes singular homology while
$\int_{D}p_1(\mathbf{M})\in 2\Z$ since $\mathbf{M}$ is spin. One might
naively imagine that this condition could constrain $\f$ to be
projectively rational, thus ruling out foliations $\cF$ whose leaves
are dense in $M$. This is in fact not the case, for the following
reason. Since the ordinary de Rham cohomology groups of the contractible manifold 
$N\simeq \R^3$ are given by:
\be
H^0(N,\R)\simeq \R~~,~~H^1(N,\R)=H^2(N,\R)=H^3(N,\R)=0~~,
\ee
the Kunneth theorem for de Rham cohomology gives
$H^4(\mathbf{M},\R)\simeq H^0(N,\R)\otimes_\R H^4(M,\R) \simeq
H^4(M,\R)$ and hence $[\frac{\mathbf{G}}{2\pi}]\equiv
[\frac{\mathbf{F}}{2\pi}]$ in de Rham cohomology since $[\nu_3\wedge
  \mathbf{f}]=0$ (because the de Rham cohomology class of $[\nu_3]\in
H^3(N,\R)$ vanishes). On the other hand, we have
$2p_1(\mathbf{M})=2\Pi_2^\ast(p_1(M))$ since $T{\mathbf M}\simeq
\Pi_1^\ast(TN)\oplus \Pi_2^\ast(TM)$ (where $\Pi_i$ are the canonical
projections of the product $N\times M$) and $TN$ is trivial. By the
Kunneth theorem for singular homology, we also have:
\be
H_4(\mathbf{M},\Z)\simeq H_0(N,\Z)\otimes_\Z H_4(M,\Z)\oplus H_3(N,\Z)\otimes_\Z H_1(M,\Z) \simeq H_4(M,\Z) 
\ee
since $H_0(N,\Z)\simeq \Z$ while $H_1(N,\Z)=H_2(N,\Z)=H_3(N,\Z)=0$ and $\Tor_1^\Z(\Z, H_3(M,\Z))=0$. 
Hence \eqref{Gq} is equivalent with:
\be
\int_{D}\left[\frac{\mathbf{F}}{2\pi}-\frac{1}{4}p_1(M)\right]\in \Z ,~~\forall D\in H_4(M,\Z)~~,
\ee
a relation which does not involve $\f$. Thus \eqref{Gq} does {\em not}
constrain the cohomology class $\f$ and hence one can
expect that foliations whose leaves are dense in $M$ provide
consistent backgrounds in M theory. One would of course need to
perform a careful analysis of all known quantum corrections
\cite{Meff} around such backgrounds in order to verify this
expectation, but such analysis is outside the scope of the present
paper.

\item Recall that $\cF$ is a fibration over the circle iff. $\f$ is
  projectively rational, in which case $M$ is the mapping torus of the
  diffeomorphism $\phi_a$ of some arbitrarily chosen leaf $L$. In this
  case, one can view our M-theory background as a two step
  compactification, where in the first step one compactifies on $L$
  down to four dimensions while in the second step one further
  compactifies the resulting four-dimensional theory by ``fibering''
  it over the circle. When using this perspective, the diffeomorphism
  $\phi_a$ induces a symmetry of the corresponding supergravity action
  in four dimensions. Then the second step can be described as
  compactifying this four-dimensional theory over the circle using the
  ``duality twist'' provided by that symmetry.  This point of view was
  used, for example, in reference \cite{Vandoren} to describe
  compactifications of $M$-theory on certain seven-manifolds which are
  fibrations over the circle (with six-manifold fibers), where such
  compactifications were called ``Scherk-Schwarz compactifications
  with a duality twist''. In our situation, the case when $\f$ is
  projectively rational could be described in the same manner, except
  that one has to start with the four-dimensional supergravity theory
  obtained by reducing M-theory over the seven-manifold $L$ (which
  carries a non-parallel $G_2$ structure), while taking the effect of
  $F$ and $f$ into account (in particular, the relevant actions in
  both three and four dimensions will be gauged supergravity
  actions). When $b_1(M)\geq 2$, such generalized
  Scherk-Schwarz reductions can describe only a negligible subset of
  all $\cN=1$ compactifications of $M$-theory down to $\AdS_3$, since in
  that case the projectively rational classes correspond to a
  countable subset of the projectivization $\P H^1(M,\R)\simeq
  \P^{b_1(M)-1}$ of $H^1(M,\R)$, which is an uncountable set.
  The generic compactification in our class corresponds to a {\em
    minimal} foliation $\cF$ --- thus being very
  different in nature from compactifications of generalized
  Scherk-Schwarz type.
\end{enumerate}

\section{Comparison with previous work}
\label{sec:comparison}

\paragraph{Relation with \cite{MartelliSparks}.} 
The class of compactifications analyzed in this paper was pioneered in
\cite{MartelliSparks}, where the solution of the Fierz identities as
well as certain combinations of the {\em exterior} differential and
codifferential constraints analyzed in our paper were first
given. Appendix C of loc. cit also lists, in a
different form, a set of `useful relations' which turn out, after some
work \cite{ga1}, to be equivalent  with what we call the
$\check{Q}$-constraints. Here are some points of difference with
\cite{MartelliSparks} regarding the techniques that allowed us to
extract the full solution. They concern the local analysis of such
geometries (since the topology of the foliation $\cF$ was not
previously discussed in the literature).
\begin{itemize}
\itemsep0em 
\item We made systematic use of the non-redundant parameterization
  \eqref{Enr}, which eliminates those rank components of $\check{E}$
  that are determined by the Fierz relations.
\item We solved the $\check{Q}$-constraints \eqref{QConstr}
  explicitly.  We found that they determine certain components of $F$
  as algebraic combinations of $\Delta, b, V$ and $f$, while imposing
  no further conditions.
\item We fully encoded the supersymmetry conditions \eqref{par_eq}
  through the extrinsic geometry of the foliation $\cF$, through the
  non-adapted part of the normal connection $D_n$ and through
  the torsion classes of its longitudinal $G_2$ structure, all of
  which we extracted explicitly. 
\item We used directly the covariant derivative constraint
  \eqref{NablaConstr} rather than the exterior differential and
  codifferential constraints \eqref{EDS} and \eqref{CDS} which it
  implies. This allowed us to give the full set of conditions
  characterizing supersymmetric solutions and to prove rigorously that
  they do so. In particular, we determined the covariant derivative of
  $V$, which is completely specified by the supersymmetry conditions
  when $\mathbf{G}$ is given.
\item Details of the comparison of our results with certain relations
  given in \cite{MartelliSparks} can be found in Appendix
  \ref{app:other}.
\end{itemize}

\paragraph{Relation with \cite{Tsimpis}.}
The class of backgrounds discussed in this paper can also be
approached using the method proposed in \cite{Tsimpis}, which relies
on the fact that existence of an everywhere non-vanishing Majorana
spinor on $M$ implies that both $M$ and its orientation opposite
${\bar M}$ admit $\Spin(7)$ structures --- an approach which uses
explicitly only part of the full symmetry of the problem.  The
relation between the description presented here and that of
\cite{Tsimpis} is given in \cite{g2s}, in
the general context when $\xi$ is not required to be everywhere
non-chiral. As it turns out, that case can be described using the
theory of singular foliations.

\section{Conclusions and further directions}

We analyzed $\cN=1$ compactifications of eleven-dimensional
supergravity down to $\AdS_3$ using the theory of foliations.  We
found that, in the nowhere chiral case, the compactification manifold
can be described through a codimension one foliation carrying a
leafwise $G_2$ structure and that the supersymmetry conditions are
{\em equivalent} with explicit equations determining the extrinsic
geometry of this foliation and the torsion classes of the $G_2$
structure.  In particular, we found that the leafwise $G_2$ structure
is integrable (in fact, conformally co-calibrated), belonging to the
Fernandez-Gray class $W_1 \oplus W_3\oplus W_4$. We also discussed the
topology of such foliations, including the non-commutative topology of
their leaf space, giving a criterion which distinguishes the cases
when the leaves are compact and dense, respectively.  We also showed
that existence of solutions requires vanishing of the Latour
obstruction for the cohomology class of a certain component of the
supergravity 4-form field strength. The case when $\xi$ is not
everywhere non-chiral is discussed in \cite{g2s} using the theory of
singular foliations.

Foliations also feature in the proposals of \cite{Grana} where,
however, the conditions imposed by supersymmetry were not
considered. It would be interesting to explore further the connection
of the backgrounds discussed in this paper with the abstract classes
of foliations which were discussed in loc. cit. starting from the framework of
extended generalized geometry. The {\em supersymmetric}
foliated backgrounds discussed in our paper could serve to realize
explicitly part of the picture proposed in that reference. It is also
likely that proposals such as \cite{Bonetti} may be understood better
by enlarging the framework \cite{Becker1} of $\Spin(7)$ {\em holonomy}
which was used in loc. cit. to backgrounds of the type considered
in this paper.

We mention that the problem of finding explicit solutions to our
equations is of the type considered in \cite{Rovenski}, so 
it could be approached through the theory of geometric flows. We also expect that a
modification of the methods of \cite{FriedrichAgricola, Puhle1} (which
would adapt them to the spinor equations \eqref{par_eq}) may
allow one to draw conclusions about the existence of solutions and the
dimensionality of their moduli space.

As pointed out in \cite{MartelliSparks} and recalled in Section
\ref{sec:basics}, the class of backgrounds we considered are consistent
at the classical level. It would be interesting to study quantum
corrections, using the known subleading terms of the effective action
of M-theory \cite{Meff}. While the appearance of non-commutative
geometry in Section \ref{sec:top} is of purely mathematical nature
(being a general phenomenon in the theory of foliations), we suspect
that it in fact has a physical interpretation through the general idea
of reducing quantum theories along foliations. Non-commutativity (and
even non-associativity) in closed string theories was previously
observed in studies of topological \cite{TopTDuality} and classical
\cite{TDuality} T-duality and it would therefore not be surprising
should its IIA incarnation turn out to have an M-theoretic origin.  It
would indeed appear that this is a much more general phenomenon having
to do with certain limits of field theories.

\acknowledgments{The work of C.I.L. is supported by the research
  grant IBS-R003-G1, while E.M.B. acknowledges the support from the 
  strategic grant POSDRU/159/1.5/S/133255, Project ID 133255 (2014),
  co-financed by the European Social Fund within the Sectorial
  Operational Program Human Resources Development 2007--2013.  
This work was also financed by the CNCS-UEFISCDI grants PN-II-ID-PCE 121/2011 
and 50/2011, and also by PN 09  37 01 02.  The authors thank M.~Gra\~na, S.~Grigorian,
  D.~Martelli, A.~Moroianu, J.~Sparks and D.~Tsimpis for
  correspondence.}

\appendix

\section{Some \KA algebra relations}
\label{app:GA}

We summarize a few useful relations from \cite{ga1}.  Given two pure
rank forms $\omega\in\Omega^{\tilde\omega}(M)$ and $\eta\in\Omega^{\tilde\eta}(M)$ on an oriented $d$-dimensional Riemannian manifold $M$, 
one defines their geometric product through:
\be
\omega\eta \eqdef \sum_{m=0}^{d} (-1)^{\left[\frac{m+1}{2}\right]+m{\tilde \omega}}\omega\btu_m\eta~~,
\ee
where $\btu_m$ denotes the generalized product of order $m$, with ~$\rk(\omega\btu_m\eta)={\tilde \omega}+{\tilde \eta}-2m$~.

The {\em reversion} $\tau$ of the \KA algebra of $(M,g)$ is the
anti-automorphism defined through
$\tau(\omega)=(-1)^{\frac{\tilde\omega(\tilde\omega-1)}{2}}\omega$,
while the {\em signature} $\pi$ is the automorphism defined through
$\pi=\oplus_{k}{(-1)^k\id_{\Omega^k(M)}}$.

\paragraph{The Hodge operator and codifferential for pseudo-Riemannian manifolds.} 
For a (not necessarily compact) pseudo-Riemannian manifold $(M,g)$ of
dimension $d$ and whose metric has exactly $q$ negative eigenvalues,
the Hodge operator is defined through \cite{ga1}:
\be
\ast\omega\eqdef \tau(\omega)\nu~~,
\ee
and we have
\be
\ast^2=(-1)^q\pi^{d-1}~~,~~\nu^2=(-1)^{q+\left[\frac{d}{2}\right]} ~~
\ee
as well as: 
\be
\omega\wedge \ast \eta=\langle \omega,\eta\rangle\nu~~,~~\forall \omega,\eta\in \Omega(M)~~\mathrm{with}~~\rk \omega=\rk \eta~~.
\ee
The codifferential is defined through:
\be
\updelta \omega=(-1)^{d({\tilde \omega}+1)+q+1}\ast \dd \ast\omega=-\iota_{e^a}\nabla_{e^a} \omega~~,~~\forall ~\omega\in \Omega(M)~~\mathrm{with}~~\rk\omega={\tilde \omega}~~
\ee
and is the formal adjoint of $\dd$ with respect to the non-degenerate
bilinear pairing $\int_M\langle~,~\rangle\nu$ on the subspace
$\Omega_c(M)\subset \Omega(M)$ of compactly-supported differential
forms:
\be
\int_M\langle \dd\omega,\eta\rangle\nu=\int_M\langle \omega,\updelta\eta\rangle \nu~~\forall \omega,\eta\in \Omega_c(M)~~.
\ee
Under a conformal transformation $g\rightarrow g'\eqdef e^{2\alpha}g$, the Hodge operator changes as $\ast\rightarrow \ast'$, where: 
\be
\ast'(\omega)=e^{(d-2~\!\rk \omega)\alpha}\ast \omega~~,~~\mathrm{for}~\omega~\mathrm{of~pure~rank.}
\ee
\paragraph{Identities for Riemannian manifolds.} For the rest of this Appendix, we assume that $M$ is Riemannian. If
$\nu$ is the volume form of $(M,g)$, while $\ast$ is the Hodge
operator, then the following identities hold:
\beqan
\ast\omega&=&\iota_\omega\nu=\tau(\omega)\nu~,\nn\\
\nu\omega&=&\pi^{d-1}(\omega)\nu~,\nn\\
\ast^2=\pi^{d-1}~&,&~~~\nu^2=(-1)^{[\frac{d}{2}]}~,\\
\omega\btu_{\tilde\omega}\eta &=& \iota_{\tau(\omega)}\eta =\eta\btu_{\tilde\omega} \omega~,\nn
\eeqan
\beqan
\omega\wedge\ast\eta=(-1)^{\tilde\omega(\tilde\eta-1)}\ast(\iota_{\tau(\omega)}\eta)~,
~~{\rm when}~~\tilde\omega\leq\tilde\eta~,\nn\\
\iota_{\omega}(\ast\eta)=(-1)^{\tilde\omega\tilde\eta}\ast(\tau(\omega)\wedge\eta)~,
~~{\rm when}~~\tilde\omega+\tilde\eta\leq d~,
\eeqan
\beqan
(-1)^{[\frac{m+1}{2}]}\pi^m(\omega)\btu_m[\ast\tau(\eta)]
 &=& (-1)^{[\frac{m'+1}{2}]}[\ast\tau(\omega)]\btu_{m'}\pi^{d-1}(\eta) \nn\\
 &=& (-1)^{[\frac{m''+1}{2}]}\ast\tau\left[\pi^{m''}(\omega)\btu_{m''}\eta\right]~,\\
 \mathrm{for}~~~\tilde\omega-m &=& \tilde\eta-m'=m''~,~~\mathrm{where}~~m,m',m''>0~~.\nn
\eeqan
Any pure rank form decomposes uniquely into parallel and orthogonal components w.r.t any 1-form $u$
of unit norm $||u||=1$~~,
\be
\omega=\omega_\perp+\omega_\parallel=\omega_\perp+u\wedge\omega_\top~~,
\ee
where $\omega_\top\eqdef \iota_u \omega$ and $\omega_\perp\eqdef
\omega-u\wedge \iota_u \omega$ are the top and orthogonal parts of
$\omega$ discussed in \cite{ga1}.  If $\ast_\perp \eta\eqdef \ast(u\wedge
\eta)$ denotes the Hodge operator along the Frobenius distribution
$\cD$ transverse to $u$ (the Hodge operator defined on $\cD$ by the induced metric, when $\cD$ is endowed 
with the orientation given by the volume form $\nu_\top=\iota_u\nu$ along $\cD$), then one has the identities, which can be used to 
decompose various formulas into components along $\cD$ and $\cD^\perp$:
\beqa
(u \wedge\omega)_\perp=0~~~&,&~~~(u \wedge\omega)_\top=\omega_\perp~,\nn\\
(\iota_u\omega)_\perp=\omega_\top~~~&,&~~~(\iota_u\omega)_\top=0~, \nn\\
\ast \omega=\ast_\perp(\omega_\top) &+&u\wedge\ast_\perp\pi(\omega_\perp)~,\\
(\ast \omega)_\perp=\ast_\perp(\omega_\top)~~~&,&~~~ (\ast \omega)_\top=\ast_\perp \pi(\omega_\perp)~, \\
\tau(\omega)_\perp=\tau(\omega_\perp)~~~&,&~~~\tau(\omega)_\top=\pi(\tau(\omega_\top))~~\\
\pi(\omega)_\perp=\pi(\omega_\perp)~~~&,&~~~\pi(\omega)_\top=-\pi(\omega_\top)~~\\
(\omega\eta)_\perp=\omega_\perp\eta_\perp+\pi(\omega_\top)\eta_\top~~&,&~~(\omega\eta)_\top=\omega_\top\eta_\perp+\pi(\omega_\perp)\eta_\top~~\\
(\nu\omega)_\perp=-\nu_\top\omega_\top~~&,&~~(\nu\omega)_\top=\nu_\top\omega_\perp\\
(\omega\nu)_\perp=\pi(\omega_\top)\nu_\top~~&,&~~(\omega\nu)_\top=\pi(\omega_\perp)\nu_\top~~.
\eeqa
For $\alpha\in \Omega^1(M)$ such that $\alpha\perp u$, we have: 
\beqa
(\alpha\wedge \omega)_\perp=\alpha\wedge(\omega_\perp)~~&,&~~(\alpha\wedge \omega)_\top=-\alpha\wedge (\omega_\top)~~,\nn\\
(\iota_\alpha\omega)_\perp=\iota_\alpha(\omega_\perp)~~&,&~~(\iota_\alpha\omega)_\top=-\iota_\alpha(\omega_\top)~~.
\eeqa
When $\dim M=8$, the canonical trace of the \KA algebra is given by \cite{ga1}:
\be
\cS(\omega)=16 \omega^{(0)}~~,~~
\ee
where $\omega^{(0)}$ denotes the rank zero component of the
inhomogeneous form $\omega\in \Omega(M)$. Furthermore:
\be
\ast^2=\pi~~,~~\nu^2=+1~~
\ee
and $\nu$ is twisted central (i.e. $\omega\nu=\nu\pi(\omega)$) in the \KA algebra of $M$. We also
mention the following relation which holds when $(M,g)$ is an
eight-dimensional Riemannian manifold.
\ben
\label{NormCS}
\cS(\omega^2)=16 (-1)^{\left[\frac{k}{2}\right]}||\omega||^2~~,~~\forall \omega \in \Omega^k(M)~~.
\een

\section{Useful identities for manifolds with $G_2$ structure}
\label{app:G2}
On the 7-dimensional leaves of the foliation $\cF$ we have a $G_2$
structure. Our computations rely on various identities for such
structures (see, for example, \cite{Kthesis,Kflows}) and on the
decomposition of the exterior bundle of the leaves into vector
sub-bundles carrying fiberwise irreps. of $G_2$ \cite{FG, BryantEH, Bryant}. Notice that all
statements of this Appendix generalize trivially to $G_2$ structures
defined on (necessarily orientable) vector bundles of rank seven, such as 
the Frobenius distribution $\cD$ of this paper.

Let $L$ be a 7-manifold endowed with a a $G_2$ structure described by
the canonically normalized associative 3-form $\varphi$. Since $G_2$
is a subgroup of $\SO(7)$, $\varphi$ determines both a metric $g\eqdef
g_\varphi$ and an orientation of $L$, the metric being fixed
uniquely by the following condition, where $\nu_L$ is the corresponding
volume form of $L$ (see \cite{Kthesis}):
\be
(v\lrcorner\varphi)\wedge(w\lrcorner\varphi)\wedge\varphi=-6 g(v,w)\nu_L~~,~~v,w\in\Gamma(L,TL)~~.
\ee
The $G_2$ structure is also determined by the coassociative 4-form
$\psi=\ast_L\varphi$, where $\ast_L\eqdef \ast_\varphi$ is the Hodge
operator defined by $g_\varphi$ and by the orientation induced by
$\varphi$. The canonical normalization conditions are:
\be
||\varphi||^2=||\psi||^2=7~~.
\ee
The bundles of 1- and 6-forms (which are Hodge dual to each other)
carry irreducible fiberwise representations, while the bundles of 2,
3, 4 and 5-forms decompose canonically into vector bundles carrying
fiberwise irreps.  This leads to the following decompositions of the
$\cinf$-modules of pure rank forms \cite{FG, BryantEH, Bryant, Kflows}:
\beqa
\Omega^2(L)&=&\Omega^2_7(L)\oplus\Omega^2_{14}(L) ~~,~~\Omega^3(L)=\Omega^3_1(L)\oplus\Omega^3_7(L)\oplus\Omega^3_{27}(L) ~~,\\
\Omega^5(L)&=&\Omega^5_7(L)\oplus\Omega^5_{14}(L) ~~,~~\Omega^4(L)=\Omega^4_1(L)\oplus\Omega^4_7(L)\oplus\Omega^4_{27}(L) ~~,
\eeqa
where the subscript indicates the dimension of the corresponding
irrep.  of $G_2$. For convenience, we set $\Omega^k_S(L)\eqdef
\Omega^k_1(L)\oplus\Omega^k_{27}(L)$ for $k=3,4$.  The Hodge operator
$\ast_L$ maps $\Omega^2(L)$ into $\Omega^5(L)$ and $\Omega^3(L)$ into
$\Omega^4(L)$ preserving these decompositions.  The modules
$\Omega^3_7(L)$ and $\Omega^4_7(L)$ are both isomorphic with
$\Omega^1(L)$, while $\Omega^3_S(L)$ is isomorphic with the $\cinf$-module of
symmetric 2-tensors, with $\Omega^k_{27}(L)$ corresponding to
traceless symmetric tensors. More precisely, any four-form $\omega\in\Omega^4(L)$
decomposes as:
\be
\omega=\omega^{(7)}+\omega^{(S)}~~,~~{\rm with}~~\omega^{(7)}\in\Omega^4_7(L)~,
~\omega^{(S)}\in\Omega^4_{S}(L)~~,
\ee 
while the components can be parameterized as follows \cite{BryantEH, Bryant, Kflows}:
\ben
\label{omega_param}
\omega^{(7)}=\alpha\wedge\varphi~, ~~\alpha\in \Omega^1(L)~~,
~~\omega^{(S)} =-\hat{h}_{ij}e^i\wedge\iota_{e^j}\psi~,
\een
where $\hat{h}_{ij}$ is a symmetric tensor. The decomposition ${\hat
  h}_{ij}=\frac{1}{7}\tr_g({\hat h})g_{ij}+{\hat h}_{ij}^{(0)}$ with
$\tr_g({\hat h}^{(0)})=0$ gives $\omega^{(S)}=\omega^{(1)}+\omega^{(27)}$, where
$\omega_1=-\frac{4}{7}\tr_g({\hat h})\psi\in\Omega^4_1(L)$ and
$\omega^{(27)}=-\hat{h}^{(0)}_{ij}e^i\wedge\iota_{e^j}\psi\in
\Omega^4_{27}(L)$. 

\noindent Similarly, any 3-form $\eta\in\Omega^3(L)$ decomposes as:
\be
\eta~=\eta^{(7)}+\eta^{(S)}~~,~~{\rm with}~~\eta^{(7)}\in\Omega^3_7(L)~,~\eta^{(S)}\in\Omega^3_{S}(L)~,
\ee
where the components can be parameterized through \cite{BryantEH,Bryant, Kflows}:
\beqan
\label{eta_param}
\eta^{(7)}=-\iota_\alpha\psi~, ~~\alpha\in \Omega^1(L)~~,~~\eta^{(S)} =h_{ij}e^i\wedge\iota_{e^j}\varphi~,
\eeqan
with $h_{ij}$ a symmetric tensor. The decomposition
$h_{ij}=\frac{1}{7}\tr_g(h)g_{ij}+h_{ij}^{(0)}$ with
$\tr_g(h^{(0)})=0$ gives $\eta^{(S)}=\eta^{(1)}+\eta^{(27)}$ with
$\eta_1=\frac{3}{7}\tr_g(h)\varphi\in\Omega^3_1(L)$ and
$\eta^{(27)}=h^{(0)}_{ij}e^i\wedge\iota_{e^j}\varphi\in
\Omega^3_{27}(L)$. 

Given $\eta^{(S)}\in \Omega^3(L)$, the
corresponding symmetric tensor $h$ which satisfies the second equation
of \eqref{eta_param} is uniquely determined and given by the formula
\cite{FG, BryantEH, Bryant, Kflows}:
\ben
\label{Symtens1}
h_{ij}=-\frac{1}{2}\tr_g(h)g_{ij}-\frac{1}{4}\ast_L(\iota_{e^i}\varphi\wedge\iota_{e^j}\varphi\wedge\eta^{(S)}) =
-\frac{1}{2}\tr_g(h)g_{ij}-\frac{1}{4}\left[ \langle \iota_{e^i}\varphi , \iota_{e^j} \eta^{(S)} \rangle + (i\leftrightarrow j) \right]~~.
\een
Furthermore, $\omega^{(S)}$ and $\eta^{(S)}$ are Hodge dual to each other iff. 
$h$ and ${\hat h}$ are related through:
\be
{\hat h}_{ij}=h_{ij}-\frac{1}{4}\tr_g(h) g_{ij}\Longleftrightarrow h_{ij}=\hat{h}_{ij}-\frac{1}{3}\tr_g(\hat h)g_{ij}~~,
\ee
a relation which implies $\tr_g({\hat h})= -\frac{3}{4}\tr_g(h)
\Longleftrightarrow \tr_g(h)=-\frac{4}{3}~\tr_g(\hat{h})$. This
amounts to the requirement that $\omega^{(1)}$ and $\eta^{(1)}$ are Hodge dual
to each other and that the same holds for $\omega^{(27)}$ and
$\eta^{(27)}$. One also notices $\hat{h}^{(0)}=h^{(0)}$.

The torsion classes $\boldsymbol{\tau}_0\in\Omega^0_1(L)$,
$\boldsymbol{\tau}_1\in \Omega^1_7(L)$, $\boldsymbol{\tau}_2\in
\Omega^2_{14}(L)$ and $\boldsymbol{\tau}_3\in \Omega^3_{27}(L)$ of the
$G_2$ structure are uniquely specified through the following equations,
which follow the conventions of \cite{Bryant,Kflows}:
\beqan
\label{TorsionEqs}
\begin{split}
&\dd_\perp\varphi=\boldsymbol{\tau}_0\psi+3\boldsymbol{\tau}_1\wedge\varphi+\ast_\perp\boldsymbol{\tau}_3~~,\\
&\dd_\perp\psi=4\boldsymbol{\tau}_1\wedge\psi+\ast_\perp\boldsymbol{\tau}_2~~.\\
\end{split}
\eeqan
Since the torsion class $\boldsymbol{\tau_3}$ belongs to $\Omega^3_{27}(L)$, it can be
parameterized through a traceless symmetric tensor $t_{ij}$:                           :
\ben
\label{Tau3Param}
\boldsymbol{\tau}_3=t_{ij}e^i\wedge \iota_{e^j}\varphi~~,
\een
where $t_{ij}$ can be recovered from $\boldsymbol{\tau}_3$ through
relation \eqref{Symtens1}:
\ben
\label{t}
t_{ij}=-\frac{1}{4}\left[\langle \iota_{e_i}\varphi,\iota_{e_j}\tau_3\rangle + (i\leftrightarrow j)\right]~~.
\een
Under a conformal transformation with conformal factor $e^{2\alpha}$, we have: 
\ben
\label{G2conf}
\begin{split}
& g'_{ij}=e^{2\alpha}g_{ij}~~,~~\nu_L'=e^{7\alpha}\nu_L~~~~~~~~~~,
~~\varphi'=e^{3\alpha}\varphi~~~,~~\psi'=e^{4\alpha}\psi~,\\
&\boldsymbol{\tau}'_0=e^\alpha \boldsymbol{\tau}_0~~~,~~\boldsymbol{\tau}'_1=e^\alpha (\boldsymbol{\tau}_1+\dd\alpha)~~,~~\boldsymbol{\tau}'_2=e^\alpha \boldsymbol{\tau}_2~~,~~\boldsymbol{\tau}'_3=e^\alpha \boldsymbol{\tau}_3~~
\end{split}
\een
and $\ast'_L\omega=e^{(7-2\rk \omega)\alpha}\ast_L\omega$ for any pure rank form $\omega\in \Omega(L)$. 

For reader's convenience, we reproduce the following identities given
in \cite{Kflows}, where indices are raised and lowered using the
metric $g=g_\varphi$ and implicit summation over repeated indices is
understood:
\begin{itemize}
\itemsep0em 
\item{Contractions between $\varphi$ and $\varphi$}
\beqan
\varphi_{ijk}\varphi^{ijk} &=& 42~,\nn\\
\varphi_{ijk}\varphi_a{}^{jk} &=& 6g_{ia}~,\\
\varphi_{ijk}\varphi_{ab}{}^{k} &=& g_{ia}g_{jb}-g_{ib}g_{ja}-\psi_{ijab}~,\nn
\eeqan
\item{Contractions between $\varphi$ and $\psi$}
\beqan
\varphi_{ijk}\psi_{a}{}^{ijk} &=& 0~,\nn\\
\varphi_{ijk}\psi_{ab}{}^{jk} &=& -4\varphi_{iab}~,\\
\varphi_{ijk}\psi_{abc}{}^{k} &=& g_{ia}\varphi_{jbc}+g_{ib}\varphi_{ajc}+g_{ic}\varphi_{abj} -g_{aj}\varphi_{ibc}-g_{bj}\varphi_{aic}-g_{cj}\varphi_{abi}~,\nn
\eeqan
\item{Contractions between $\psi$ and $\psi$}
\beqan
\psi_{ijkl}\psi^{ijkl} &=& 168~,\nn\\
\psi_{ijkl}\psi_a{}^{jkl} &=& 24g_{ia}~,\nn\\
\psi_{ijkl}\psi_{ab}{}^{kl} &=& 4g_{ia}g_{jb}-4g_{ib}g_{ja}-2\psi_{ijab}~,\nn\\
\psi_{ijkl}\psi_{abc}{}^{l} &=& -\varphi_{ajk}\varphi_{ibc}-\varphi_{iak}\varphi_{jbc}-\varphi_{ija}\varphi_{kbc}  \\
&& + g_{ia}g_{jb}g_{kc} + g_{ib}g_{jc}g_{ka} + g_{ic}g_{ja}g_{kb} \nn\\
&& - g_{ia}g_{jc}g_{kb} - g_{ib}g_{ja}g_{kc} - g_{ic}g_{jb}g_{ka} \nn\\
&& - g_{ia}\psi_{jkbc} - g_{ja}\psi_{kibc} - g_{ka}\psi_{ijbc} \nn\\
&& + g_{ab}\psi_{ijkc} - g_{ac}\psi_{ijkb}~.\nn
\eeqan
\item For any 1-form $\alpha$ and any vector field $w$, the following identities hold \cite{Kthesis}:
\beqan
\label{Krel}
\varphi\wedge\ast_L(\alpha\wedge\varphi)=4\ast_L\alpha~~~~&,&~~~~\varphi\wedge (w~\lrcorner~\psi)=-4\ast_L w^\# ~,\nn\\
\psi\wedge\ast_L(\alpha\wedge\psi)=3\ast_L\alpha~~~~&,&~~~~\psi\wedge(w~\lrcorner \varphi)=3\ast_L w^\#~,\nn\\
\psi\wedge\ast_L(\alpha\wedge\varphi)=0~~~~&,&~~~~\psi\wedge (w~\lrcorner~\psi)=0~,\\
\varphi\wedge\ast_L(\alpha\wedge\psi)=-2~\alpha \wedge\psi~~~~&,&~~~~\varphi\wedge (w~\lrcorner~\varphi)=-2\ast_L (w~\lrcorner~\varphi) ~.\nn
\eeqan
\end{itemize}

\paragraph{Remark.} Formulas for contractions and for projectors on $G_2$ representations can also be found in \cite{GrigorianYau, Grigorian}. 

\

\noindent The contractions listed above imply that the canonically-normalized coassociative form $\psi$ satisfies the following identity in the \KA algebra of $L$:
\ben
\label{psisquared}
\psi^2=6\psi+7~~,
\een
which amounts to the statement that: 
\be
\label{PiDef}
\Pi\eqdef \frac{1}{8}(1+\psi)~~ 
\ee
is an idempotent in the \KA algebra. 

\paragraph{The right action of $\psi$ on 4-forms and 3-forms in the \KA algebra.} Let $\omega\in \Omega^4(L)$. Then:
\be
\cR_\psi(\omega)\eqdef \omega\psi=-\omega\btu_1\psi 
 -\omega\btu_2\psi+\omega\btu_3\psi+\omega\btu_4\psi~~.
\ee
Using the $G_2$-structure identities and the parameterization of 
$\omega^{(7)}$ and $\omega^{(S)}$ given in \eqref{omega_param}, one finds:
\ben
\label{Psi4}
\omega\psi=4\ast_L\alpha -4~\tr_g(\hat h)\psi -\omega^{(S)}+3\omega^{(7)} +4\iota_\alpha\varphi- 4~\tr_g(\hat h)~~.
\een
This implies $\ker({\cal R}_\psi|_{\Omega^4(L)})=0$. Similarly, 
for any $\eta\in \Omega^3(L)$, we have:
\be
\cR_\psi(\eta)\eqdef \eta\psi=\eta\wedge\psi+\eta\btu_1\psi-\eta\btu_2\psi-\eta\btu_3\psi~~.
\ee
This can be computed either directly using the identities of Appendix
\ref{app:G2} or by Hodge dualizing \eqref{Psi4} 
(using $\eta\psi=(\ast_L\omega)\psi=\omega\nu\psi=(\omega\psi)\nu$~). This
gives:
\ben
\label{Psi3}
\eta\psi=3 ~\tr_g(h)\nu-4\alpha\wedge\psi-\eta^{(S)}+3\eta^{(7)}+3~\tr_g(h)\varphi-4\alpha~,
\een
where $\eta$ is parameterized as in \eqref{eta_param} and implies
$\ker({\cal R}_\psi|_{\Omega^3(L)})=0$. Finally, we recall the following identities:
\be
(\ast_L)^2=\id_{\Omega(M)}~~,~~(\nu_L)^2=-1~~
\ee
and the fact that $\nu_L$ is central in the \KA algebra of $L$ (i.e. $\lambda\nu_L=\nu_L\lambda~,~\forall \lambda\in\Omega(L)$).

\

\section{Characterizing the extrinsic geometry of $\cF$}
\label{app:FE}

\subsection{Fundamental second order objects}
The vector field $n=\hat V^\sharp$ has unit norm and is orthogonal to
$\cF$. Any $X\in \Gamma(M,TM)$ decomposes as $X=g(n,X) n+X_\perp$,
where $X_\perp \perp n$. Since $g(n,n)=1$, one has $g(n,\nabla_X
n)=\frac{1}{2}[g(\nabla_X n, n)+g(n,\nabla_X
  n)]=\frac{1}{2}X(g(n,n))=0$, i.e. $\nabla_X n$ is orthogonal to
$n$. The second order data of $\cF$ and $\cF^\perp$ are encoded by the
following objects:

\

\noindent For the foliation $\cF$:
\begin{itemize}
\itemsep0em 
\item $\nabla^\perp:\Gamma(M,\cD)\times \Gamma(M,\cD)\rightarrow
  \Gamma(M,\cD)$, the Levi-Civita connection of the metric induced by
  $g$ on the leaves of $\cF$ (this is a partial connection on $M$, valued in $\cD$)
\item $B:\Gamma(M,\cD)\times \Gamma(M,\cD)\rightarrow \cinf$,
  $B(X_\perp,Y_\perp)\eqdef g(n,\nabla_{X_\perp} Y_\perp)$, the scalar
  second fundamental form of the foliation $\cF$ (the full second fundamental form is given by $B n$)
\item $A:\Gamma(M,\cD)\rightarrow \Gamma(M,\cD)$, $A(X_\perp)\eqdef
  -(\nabla_{X_\perp} n)^\perp $, the Weingarten operator at $n$
  of the leaves of $\cF$
\item $\delta:\Gamma(M,\cD)\rightarrow \cinf$, $\delta(X)=g(n,D_X^\cF
  n)$, where $D^\cF$ is the normal connection along the leaves of
  $\cF$.
\end{itemize}

\noindent For the foliation $\cF^\perp$:
\begin{itemize}
\itemsep0em 
\item $a\in \cinf$, the unique connection coefficient (with respect to
  the frame given by $n$) of the Levi-Civita connection of the metric
  induced by $g$ on the leaves of $\cF^\perp$
\item $H\eqdef \nabla_n n\in \Gamma(M,\cD)$, the value of the second
  fundamental form of $\cF^\perp$ on the pair $(n,n)$ of vector
  fields tangent to $\cF^\perp$
\item $W:\Gamma(M,\cD)\rightarrow \cinf$, $W(X_\perp) \eqdef
  -g(n,\nabla_n X_\perp )$, the value at $(n,n)$ of the covariant
  Weingarten tensor at $X$ of the leaves of $\cF^\perp$
\item $D_n:\Gamma(M,\cD)\rightarrow \Gamma(M,\cD)$, the normal
  covariant derivative with respect to $n$ along the leaves of
  $\cF^\perp$.
\end{itemize}
Gauss's theorem for $\cF$ amounts to the identity
$\nabla^\perp_{X_\perp}Y^\perp=(\nabla_{X_\perp} Y_\perp)^\perp$ plus
the fact that $B$ is a symmetric tensor.  Weingarten's theorem for
$\cF$ amounts to the identity
$g(A(X_\perp),Y_\perp)=B(X_\perp,Y_\perp)$ plus the statement that
$D^\cF$ preserves the metric induced on the normal bundle
$N\cF=\cD^\perp$, which is equivalent with the statement that the map
$\delta$ vanishes\footnote{The normal bundle to any leaf of $\cF$ is a
  line bundle trivialized by the section $n$, which is parallel with
  respect to $D^\cF$, so $D^\cF$ is the trivial connection for this
  trivialization.}. It follows that $\nabla^\perp$ and $A$ encode all
information contained in the second order data of $\cF$.
Gauss's theorem for $\cF^\perp$ amounts to the fact that $a$ 
vanishes\footnote{With respect to its flow parameter $t$, each
  leaf of $\cF^\perp$ is an integral curve of the unit norm vector field $n$. This vector
  field trivializes the tangent bundle to the leaf of $\cF^\perp$ and the Levi-Civita
  connection of the induced metric (which is of course flat) is
  the trivial connection of this trivialization. The Gauss identity
  encodes this fact since it requires that the connection coefficient
  $a$ equals $g(n,\nabla_nn)$, which vanishes. }; symmetry of the
second fundamental form is automatic in this case since the leaves are
one-dimensional.  Weingarten's theorem amounts to the identity
$W(X_\perp)=-g(H,X_\perp)$ plus the statement that $D_n$ is compatible
with the metric induced on $N(\cF^\perp)=\cD$. Hence $H$ and $D_n$ encode
all information contained in the second order data of
$\cF^\perp$. Summarizing, we can take $H$, $A$, $\nabla^\perp$ and $D_n$
to be the fundamental second order data of the foliation $\cF$ in the
presence of the metric $g$ and write its fundamental equations as in
\eqref{FE}.

\paragraph{Local expressions.} Let $e_a$ be a (generally non-holonomic) local frame of $M$ such that $e_1=n$ and $e_j\perp n$ and let $\Omega_{ab}^c$ be the connection coefficients 
in this frame:
\be
\nabla_{e_a}e_b=\Omega_{ab}^ce_c~~.
\ee
Expanding $H=H^je_j$, we have $g_{11}=1$ and $g_{1j=0}$ as well as:
\ben
\Omega_{11}^1=\Omega_{i1}^1=0~~,~~\Omega_{11}^j=H^j~~,~~\Omega_{i1}^j=-A_i^j~~,~~\Omega_{ij}^1=A_{ij}~~,~~\nabla^\perp_{e_i}e_j=\Omega_{ij}^ke_k~~.
\een
The last equation says that $\Omega_{ij}^k$ are the connection
coefficients of the leafwise partial connection $\nabla^\perp$ in the
(generally non-holonomic) frame $(e_i)_{i=2,\ldots,8}$ of $\cD$.  The
identity $\Omega_{abc}=-\Omega_{acb}$ satisfied by the quantity
$\Omega_{abc}\eqdef g_{bf}\Omega_{ac}^f$ amounts to the condition that
the tensor $A_{ij}\eqdef g(Ae_i,e_j)$ is symmetric. Notice the
relations:
\be
\Omega_{i1j}=-\Omega_{ij1}=A_{ij}~~,~~\Omega_{1j1}=H_j\eqdef g_{jk}H^k~~.
\ee

\subsection{The Naveira tensor of $\cP$}

Recall that every orthogonal (i.e. $g$-compatible) almost product
structure $\cP$ on $(M,g)$ corresponds to a $g$-orthogonal
decomposition $TM=\cD\oplus \cD^\perp$, where $\cD$, $\cD^\perp$ are
the eigen-subbundles of $TM$ corresponding to the eigenvalues $+1$ and
$-1$ of $\cP$ respectively (thus $\cP=\id_\cD\oplus (-\id_\cD)$). Such
almost product structures can be classified \cite{Naveira, GilMedrano,
  Miquel, Montesinos} using the Naveira tensor
$\cN_\cP=\nabla\Phi_\cP\in \Gamma(M,(T^\ast M)^{\otimes 3})$, which is
given by:
\be
\cN_\cP(X,Y,Z)=(\nabla_X\Phi_\cP)(Y,Z)=g((\nabla_X\cP)Y,Z)~~,~~\forall X,Y,Z\in \Gamma(M,TM)~~.
\ee
Here, $\Phi_\cP\in \Gamma(M,\Sym^2(T^\ast M))$ is the symmetric
covariant 2-tensor given by $\Phi_\cP(X,Y)\eqdef g(\cP X,Y)$, which
vanishes for $(X,Y)\in \Gamma(M,\cD\times \cD^\perp\oplus
\cD^\perp\times \cD)$ and whose restrictions to $\cD$ and $\cD^\perp$
equal plus and minus the restrictions of $g$, respectively. The
Naveira tensor is symmetric in its last two variables and satisfies
$\cN(X,\cP Y,\cP Z)=-\cN(X,Y,Z)$ and thus it vanishes when both $X,Y$
belong to $\cD$ or to $\cD^\perp$. When both $\cD$ and $\cD^\perp$ are
integrable, an easy computation gives:
\beqa
(\nabla_X\cP)Y&=&2h_\cD(X,Y)~~\mathrm{for}~~X,Y\in \Gamma(M,\cD)~~,\nn\\
(\nabla_X\cP)Y&=&2h_{\cD^\perp}(X,Y)~~\mathrm{for}~~X,Y\in \Gamma(M,\cD^\perp)~~,
\eeqa
where $h_{\cD}$ and $h_{\cD^\perp}$ are the second fundamental forms of $\cD$ and $\cD^\perp$. Thus: 
\beqan
\label{NavInt}
\cN_\cP(X,Y,Z)&=&2g(h_\cD(X,Y),Z)~~\mathrm{for}~~X,Y\in \Gamma(M,\cD)~,~Y\in \Gamma(M,\cD^\perp)~~,\nn\\
\cN_\cP(X,Y,Z)&=&2g(h_{\cD^\perp}(X,Y),Z)~~\mathrm{for}~~X,Y\in \Gamma(M,\cD^\perp)~~,~~Y\in \Gamma(M,\cD)~~.
\eeqan
For our distributions $\cD=T\cF$ and $\cD^\perp=T(\cF^\perp)$, we find:
\beqa
h_{\cD}(X_\perp,Y_\perp)&=&B(X_\perp,Y_\perp)n=g(AX_\perp,Y_\perp)n~~,\nn\\
h_{\cD^\perp}(n,n)~~&=&H~~,
\eeqa
so \eqref{NavInt} give the following relations, which completely specify the Naveira tensor of $\cP$: 
\beqa
\label{NavP}
\begin{split}
&\cN_\cP(X_\perp,Y_\perp,n)=B(X_\perp,Y_\perp)=g(AX_\perp,Y_\perp)~~,\\
&\cN_\cP(n,n,X_\perp)~~~=2g(H,X_\perp)~~.
\end{split}~~
\eeqa
Notice that $\cN_\cP$ contains the same information as $H$ and $A$ and hence determining the latter amounts to determining $\cN_\cP$.

\noindent When $\cD$ is integrable, with $\cD=T\cF$, one has the following formulas for the components of the covariant derivative, differential and codifferential 
of arbitrary forms $\omega\in \Omega(M)$.

\paragraph{The covariant derivative of forms longitudinal to $\cF$.}
Direct computation using \eqref{FE} gives:
\beqan
\label{NablaLongit}
(\nabla_n\omega)_\top=\iota_\hV~\nabla_n\omega=n\lrcorner\nabla_n\omega=-H\lrcorner\omega~~
&,&~~(\nabla_{_{X_\perp}}\omega)_\top=n\lrcorner(\nabla_{_{X_\perp}}\omega)=(AX_\perp)\lrcorner\omega~~,\nn\\
(\nabla_n\omega)_\perp=D_n\omega~~~~~~~~~~~~~~~~~~~~~~~~~~~~~~~~
&,&~~(\nabla_{X_\perp}\omega)_\perp=\nabla^\perp_{_{X_\perp}}\omega~~,~~\mathrm{for}~~\omega\in \Omega(\cD)~~.~~
\eeqan

\paragraph{The covariant derivative of arbitrary forms $\omega\in \Omega(M)$.}
Direct computation using \eqref{nablaV} and \eqref{NablaLongit} gives:
\ben
\label{DerOmega}
\begin{split}
&(\nabla_n\omega)_\top=D_n(\omega_\top)-H\lrcorner \omega_\perp~~~,~~~(\nabla_j\omega)_\top=\nabla_j^\perp(\omega_\top)+(Ae_j)\lrcorner \omega_\perp~~,\\
&(\nabla_n\omega)_\perp=D_n(\omega_\perp)+H_\sharp\wedge \omega_\top~~,~~~(\nabla_j\omega)_\perp=\nabla_j^\perp(\omega_\perp)-(Ae_j)_\sharp\wedge \omega_\top~~
\end{split}
\een
as well as: 
\!\!\!\ben
\label{ExtOmega}
\!\!\begin{split}
&(\dd\omega)_\top=D_n(\omega_\perp)+H_\sharp\wedge \omega_\top-\dd_\perp(\omega_\top)-A_{jk}e^j\wedge \iota_{e_k}(\omega_\perp)~~~,~~~(\updelta \omega)_\top=- \delta_\perp(\omega_\top)~~,\\
&(\dd\omega)_\perp=\dd_\perp(\omega_\perp)~~~,~~~(\updelta \omega)_\perp=-D_n(\omega_\top)+H\lrcorner \omega_\perp-\delta_\perp(\omega_\perp)-[A_{jk}e^j\wedge \iota_{e_k}-\tr A ]\omega_\top~~.\\
\end{split}~~
\een

\paragraph{Structure of the normal covariant derivative.} 

Consider the $\SO(7)$ group bundle whose fiber at $p\in M$ is
$\SO(\cD)\eqdef \SO(\cD_p,g_p|_\cD)$. The bundle $\End_a(\cD)$ of
$g$-antisymmetric endomorphisms of $\cD$ coincides with the
corresponding bundle of Lie algebras; its fiber
$\End_a(\cD_p)=\so(\cD_p, g_p)$ at $p$ is the space of $g_p$-antisymmetric
endomorphisms of $\cD_p$. The $G_2$ structure of $\cD$ defines a $G_2$
sub-bundle of the $\SO(7)$ group bundle, obtained by taking the
stabilizer of $\varphi_p$ inside $\SO(\cD_p, g_p)$ for every point $p\in
M$. Taking the tangent space to the origin in the fibers, we obtain a
$\g2$ sub-bundle $\cG\subset \End_a(\cD)$ of the bundle of Lie
algebras mentioned above. The Killing form of $\so(7)$ endows
$\End_a(\cD)$ with a symmetric and non-degenerate pairing which at
each $p\in M$ is given by $\langle A,B\rangle=\tr(AB)$, where
$A,B\in \End_a(\cD_p)$. We let $\cG^\perp$ denote the linear
sub-bundle of $\End_a(\cD)$ obtained by taking the complement of $\cG$
with respect to this pairing. We thus have a Whitney sum
decomposition:
\be
\End_a(\cD)=\cG\oplus \cG^\perp~~.
\ee
The normal connection $D_n$ decomposes as:
\ben
\label{NormalDecomp}
D_n={\hat D}_n+{\hat \Theta}~~,
\een
where ${\hat D}_n$ is a partial connection on $\cD$ which is adapted
to the $G_2$ structure of $\cD$ while ${\hat \Theta}\in
\Gamma(M,\cG^\perp)$ is a section of $\cG^\perp$.  The fact that
${\hat D}_n$ is adapted to the $G_2$ structure means that its parallel
transport along the leaves of $\cF^\perp$ takes $G_2$-frames of $\cD$
into $G_2$-frames, which means that this parallel
transport preserves the associative form $\varphi\in \Omega^3(\cD)$
and hence also the coassociative form $\psi\in \Omega^4(\cD)$:
\ben
\label{hDPhiPsi}
{\hat D}_n\varphi={\hat D}_n\psi=0~~.
\een
Consider the 2-form $\Theta\in \Omega^2(\cD)$ defined through:
\ben
\label{ThetaDef}
\Theta(X_\perp,Y_\perp)\eqdef g({\hat \Theta}(X_\perp), Y_\perp)~~,~~\forall X_\perp, Y_\perp \in \Gamma(M,\cD)~~.
\een
We have:
\be
{\hat \Theta}(X_\perp)=(X_\perp\lrcorner\Theta)^\sharp~~,~~\forall X_\perp\in \Gamma(M,\cD)~~,
\ee
which implies: 
\ben
\label{NormalOmega}
D_n\omega={\hat D}_n\omega+\Theta\btu_1\omega~~,~~\forall \omega\in \Omega(\cD)~~.
\een
The fact that ${\hat \Theta}$ is a section of $\cG^\perp$ amounts to
the condition that $\Theta$ belongs to the subspace
$\Omega^2_7(\cD)$. Thus \cite{Kthesis}:
\ben
\label{ThetaParam}
\Theta=\iota_\vartheta\varphi\in \Omega^2_7(\cD)~~,
\een
for some uniquely determined one-form $\vartheta\in \Omega^1(\cD)$. Notice that $\vartheta$ can be expressed in terms of $\Theta$ using the 
second relation in the second column of \eqref{Krel}, which gives:
\ben
\label{theta}
\vartheta=\frac{1}{3}\ast_\perp(\psi\wedge \Theta)~~.
\een
Using this parameterization of $\Theta$ and the identities of Appendix
\ref{app:G2}, relations \eqref{NormalOmega}, \eqref{NormalDecomp} and \eqref{hDPhiPsi} give:
\ben
\label{NPhiPsi}
D_n\varphi=\Theta \btu_1\varphi=3\iota_\vartheta\psi~~~,~~~D_n\psi=\Theta\btu_1\psi=-3\vartheta\wedge\varphi~~.
\een
\paragraph{Remark.} 
Let $\times$ denote the cross product defined by the $G_2$ structure
of $\cD$, i.e. $g(u\times v,w)=\varphi(u,v,w)$ for all vector fields
$u,v,w\in \Gamma(M,\cD)$. Then relation \eqref{ThetaParam} gives
$\Theta(X_\perp,Y_\perp)=\varphi(\vartheta^\sharp, X_\perp, Y_\perp)$ and
\eqref{ThetaDef} implies:
\be
{\hat \Theta}(X_\perp)=\vartheta^\sharp \times X_\perp~~.
\ee

\paragraph{Local coordinate expressions for the normal covariant derivative.} 
Let $(e^a)_{a=1\ldots 8}$ be a local orthonormal coframe of $M$ defined on an open subset
$U\subset M$ such that $e^1={\hat V}$ and such that $e^2,\ldots, e^8$
is a $G_2$-coframe of $\cD$,
i.e. $\varphi|_U=\frac{1}{6}\varphi_{ijk}e^i\wedge e^j\wedge
e^k$. Let $e_a$ be the dual orthonormal frame of $M$ (thus
$e_1=n$). Let $\cA_i^{\,\,j}\in \cC^\infty(U,\R)$ be the connection
coefficients of $D_n$ in such a frame:
\be
D_n(e_i)=\cA_i^{\,\,j} e_j~~
\ee
and set $\cA_{ij}\eqdef \cA_i^{\,\,k} g_{kj}=g(D_n(e_i), e_j)$ (notice that $\cA_{ij}= \cA_i^{\,\,j}$ since $g_{kj}=\delta_{kj}$).  Since
$D_n$ is $g$-compatible, we have $\cA_{ij}=-\cA_{ji}$, i.e. the
connection matrix ${\hat \cA}\eqdef (\cA_i^{\,\,j})_{i,j=2,\ldots 8}$
is valued in the Lie algebra $\so(7)$ of $\SO(7)$. Consider the
standard embedding $G_2\subset \SO(7)$, which is obtained by realizing
$G_2$ as the stabilizer in $\SO(7)$ of the 3-form
$\frac{1}{6}\epsilon_{ijk}\epsilon^i\wedge \epsilon^j\wedge
\epsilon^k\in \wedge^3(\R^7)^\ast$ , where $\epsilon^1,\ldots
\epsilon^7$ is the standard coframe of $\R^7$. This induces the
standard Lie algebra embedding $\g2\subset \so(7)$ and hence a
decomposition:
\be
\so(7)=\g2\oplus \g2^\perp~~,
\ee
where $\g2^\perp$ is the orthocomplement of $\g2$ in $\so(7)$ with
respect to the Killing form of $\so(7)$. We have $\dim\g2=\rk G_2=14$
and $\dim(\g2^\perp)=7$. Using this decomposition, we write:
\be
{\hat \cA}={\hat \cA}^{(14)}+{\hat \cA}^{(7)}~~\mathrm{where}~~
{\hat \cA}^{(14)}\in \cC^\infty(U,\g2)~~,~~{\hat \cA}^{(7)}\in \cC^\infty(U,\g2^\perp)~~.
\ee
Using a partition of unity to globalize, this gives the decomposition \eqref{NormalDecomp}, 
where ${\hat \Theta}\in \Gamma(M,\cG)$ is locally given by:
\be
{\hat \Theta}(e_i)= \Theta_i^{\,\,j}e_j~~\mathrm{with}~~\Theta_i^{\,\, j}=(\cA^{(7)})_i^{\,\,j}
\ee
while: 
\be
{\hat D}_n(X_\perp)=(\delta_{ij}\partial_n X^i+(\cA^{(14)})_{i}^{\,\,j} X^i)e_j~~,~~
\forall X_\perp=X_\perp^ie_i\in \Gamma(U,\cD)~~.
\ee
The 2-form $\Theta\in \Omega^2(\cD)$ has the local
expression\footnote{All local expressions are given in the so-called
  ``{\rm Det}'' convention for the wedge product \cite{Ricci}, which
  is the convention used, for example, in \cite{Spivak}. Thus
  $e^i\wedge e^j=e^i\otimes e^j-e^j\otimes e^i$ (without a prefactor
  of $\frac{1}{2}$).}:
\be
\Theta\eqdef \frac{1}{2}\Theta_{ij}e^i\wedge e^j~~,
\ee
where $\Theta_{ij}=\Theta_i^{\,\,k}g_{kj}$. 

\section{Other details}
\label{app:other}

\paragraph{Relation with the conventions of \cite{MartelliSparks}.}
Reference \cite{MartelliSparks} works with a Majorana spinor ${\hat
  \xi}=\sqrt{2}\xi$, which has $\cB$-norm equal to $\sqrt{2}$ at every
point of $M$ and considers the spinors:
\be
\epsilon^\pm\eqdef \frac{1}{\sqrt{2}}({\hat \xi}^+ \pm {\hat \xi}^-)=
\xi^+\pm \xi^-~~\mathrm{i.e.}~~\epsilon^+=\xi~~,~~\epsilon^-=\gamma^{(9)}\xi~~.
\ee
Loc. cit parameterizes our function $b$ through an angle $\zeta\in
[\frac{\pi}{2},\frac{\pi}{2}]$:
\be
b=\epsilon^+\epsilon^-=||\xi^+||^2-||\xi^-||^2=\sin\zeta~~.
\ee
The form $Y$ considered in \cite{MartelliSparks} agrees with the form
denoted through the same letter in this paper, while the vector ${\bar
  K}$ of loc. cit.  agrees with the vector denoted by $V$ here. The
vector $K$ of loc. cit. is what we denote by ${\hat V}$. Our constant
$\kappa$ is denoted by $m$ in \cite{MartelliSparks}.  Loc. cit. also
considers the 3-form:
\ben
\label{phiMS}
\phi\eqdef \frac{1}{\sqrt{1-b^2}}\overline{\phi}=-\frac{1}{||V||}\ast Z=
-\varphi\Longrightarrow \ast \phi=+\frac{1}{||V||} Z~~,
\een
where: 
\be
{\overline \phi}\eqdef \frac{1}{3!}\cB({\hat  \xi}^+, \gamma_{abc} {\hat  \xi}^-)\dd x^a \dd x^b \dd x^c =
-\frac{1}{3!}\cB(\xi, \gamma_{abc} \gamma^{(9)}\xi)\dd x^a \dd x^b \dd x^c=-\bcE_{\xi,\gamma(\nu)\xi}^{(3)}=-\bcE_{\xi,\xi}^{(5)}\nu=-\ast Z~~.
\ee
In terms of $\phi$, loc. cit. gives the following relation
(cf. \cite[eq. (3.5), page 10]{MartelliSparks}):
\ben
\label{YMSw}
Y=\pmb{-}\iota_{\hat V}(\ast \phi)+b\phi\wedge {\hat V}=\pmb{-}\iota_{\hat V}(\ast \phi)-b{\hat V}\wedge\phi
=\frac{1}{||V||^2}\left[ \pmb{-}\iota_V Z +b V \wedge (\ast Z) \right]~~,
\een
which differs from our relation \eqref{YVZ} (which was derived
directly in \cite{ga1}) in the sign of the term $\iota_V Z$. Relation
\eqref{YMSw} corresponds to replacing the second equation of the
second row of \eqref{SolMS} with:
\be
\iota_V Z=\pmb{-}Y-b\ast Y\Longleftrightarrow VZ=\pmb{-}Y-b\ast Y\Longleftrightarrow Y=\frac{1}{1-b^2}(\pmb{-}1+b\nu)VZ
\Longleftrightarrow Y=(\pmb{-}1+b\nu)\psi~~,
\ee
where in the first equivalence we used the fact that $V\wedge Z=0$ and
hence $VZ=V\wedge Z+\iota_V Z=\iota_V Z$, while in the last
equivalence we used the fact that $Z=V\psi$ (see
\eqref{psidef}). Equation \eqref{YMSw} would then lead to
$Z+Y=(\pmb{-}1+V+b\nu)\psi$ and hence to
$\check{E}=\frac{1}{16}[1+V+b\nu+(\pmb{-}1+V+b\nu)\psi]$,
which would {\em not} satisfy the condition $\check{E}^2=\check{E}$.

\paragraph{Remark.} 
One can check directly that the signs given in the second equation of the second row of \eqref{SolMS} (and hence in 
relation \eqref{YVZ}) are the only ones which can insure that the Fierz identities encoded by the condition 
$\check{E}^2=\check{E}$ indeed hold. For this, consider a modification
of that equation having the form:
\be
\iota_V Z=\epsilon_1 (Y-\epsilon_2 b\ast Y)\Longleftrightarrow Y=\epsilon_1 (1+\epsilon_2 b\nu)\psi~~\mathrm{with}~~\epsilon_1,\epsilon_2\in\{-1,1\}~~,
\ee
where $\epsilon_1=\epsilon_2=+1$ corresponds to our relation while
$\epsilon_1=\epsilon_2=-1$ corresponds to \eqref{YMSw}.  Then $Z+Y=2\epsilon_1 C
\psi$ where $C\eqdef \frac{1}{2}(1+\epsilon_1 V+\epsilon_2 b\nu)$ is easily
seen to satisfy $C^2=C$, $C\psi=\psi C$ and $\tau(C)=C$. Thus
$\check{E}=\frac{1}{16}(1+V+b\nu+Y+Z)=\frac{1}{8}(P+\epsilon_1 C \psi)$,
where (as mentioned below equation \eqref{Enr}) $P\eqdef \frac{1}{2}(1+V+b\nu)$ is an idempotent in the \KA
algebra.  Direct computation using relation \eqref{psisquared} (which
holds for any $G_2$ structure in seven dimensions, as a consequence of
the identities given in \cite{Kthesis}) then shows that
$\check{E}^2=\check{E}$ iff. $\epsilon_1=\epsilon_2=+1$. Notice that identity
\eqref{psisquared} was also derived directly in \cite{ga1} from the
condition $\check{E}^2=\check{E}$. In the same reference, we also
derived \eqref{YVZ} using \KA algebra methods.

\

\noindent The following relation is also given in \cite[eq. (3.16), page 11]{MartelliSparks}:
\be
e^{-6\Delta}\dd (e^{6\Delta}||V||\phi)=-\ast F+bF+4\kappa[\pmb{+}\iota_{\hat V}(\ast \phi)+b{\hat V}\wedge \phi]
\Longleftrightarrow e^{-6\Delta}\dd (e^{6\Delta}||V||\phi)=-(\ast F-b F)-4\kappa Y~~,
\ee
where in the equivalence we used \eqref{YMSw}. Modulo \eqref{phiMS},
the last form of this equation agrees with the fourth relation listed
in \eqref{Diff}.  Hence we find agreement with this relation as well,
up to the sign indicated in boldface in equation \eqref{YMSw}.

We also remark on the orientation of the leaves of the
foliation $\cF$. The following relation is given in \cite{MartelliSparks}
immediately below equation (3.14) of that reference:
\ben
\label{MSvol}
\vol_7=\frac{1}{7}\phi\wedge \iota_{\hat V}(\ast \phi)~~\Longleftrightarrow 
\varphi\wedge \iota_{\hat V}(\ast \varphi)=7\vol_7\Longleftrightarrow 
\vol_7=-\iota_{\hat V}\nu=-\nu_\top~~,
\een
where we used $\iota_{\hat V}\varphi=0$, the fact that
$\varphi\wedge \ast \varphi=||\varphi||^2\nu=7\nu$ and our definition
of the induced volume form $\nu_L\eqdef \nu_\top= \iota_{\hat V}\nu$
along the leaves of $\cF$ (a definition which fixes the
choice of orientation along those leaves). Recall from \cite{Kthesis}
that the $G_2$ structure defined by a coassociative 3-form $\varphi$
on a seven-manifold $L$ fixes the orientation of $L$ as well as a
compatible metric (up to normalization of the latter). Indeed, the
volume form $\nu_L$ of $L$ is determined by $\varphi$ through the
cubic relation \eqref{G2rel}. Changing the sign of $\varphi$
in that relation changes the sign of $\nu_L$, which is why the
3-form $\phi=-\varphi$ of \cite{MartelliSparks} leads loc. cit. to use
the opposite volume form $\vol_7=-\nu_\top$ along the leaves. Hence
\eqref{MSvol} agrees with our conventions if one keeps in mind that
loc. cit. works with the associative 3-form $\phi=-\varphi$, which is
opposite to the one used in this paper. Because $\vol_7=-\nu_\top$,
the Hodge operator $\ast_7$ used in loc. cit. along the leaves of
$\cF$ equals minus our longitudinal Hodge operator
$\ast_L=\ast_\perp$:
\be
\ast_7=-\ast_\perp~~.
\ee
Reference \cite{MartelliSparks} uses the decomposition $F=F_4+F_3\wedge {\hat V}=F_4-{\hat V}\wedge F_3$, where:
\ben
\label{F3F4}
F_4=F_\perp~~,~~F_3=-F_\top~~.
\een
Relations \eqref{TF1} and \eqref{TF7} were also given in
\cite{MartelliSparks}. Using the notations of loc. cit., we find that
\eqref{TF1} and \eqref{TF7} take the form:
\beqan
\label{FMS}
&&F^{(1)}_3=-\frac{4}{7}\tr_g({\hat \chi})\phi~~,~~F_4^{(1)}=-\frac{4}{7}\tr_g({\hat h})\iota_{\hat V}(\ast \phi)~~\nn\\
&&F^{(7)}_3=\iota_{\alpha_2}\iota_{\hat V}(\ast \phi)~~~,~~~F_4^{(7)}=-\alpha_1\wedge \phi~~,
\eeqan
where:
\beqan
\label{ConnectionWithMS}
&&\tr_g({\hat \chi})=-\frac{3}{14}(\partial_n\zeta-2\kappa)~~,~~\tr_g({\hat h})=-\frac{3}{14}[4\kappa b-e^{-3\Delta}\partial_n (e^{3\Delta} \cos \zeta)]\nn\\
&& \alpha_2=-\frac{1}{2}e^{-3\Delta}\dd_\perp(e^{3\Delta}\cos\zeta)~~,~~\alpha_1=\frac{1}{2}\dd_\perp \zeta~~.
\eeqan
Notice that \cite{MartelliSparks} denotes $\dd_\perp$ by
$\dd_7$. Substituting \eqref{ConnectionWithMS} into \eqref{FMS}, one
recovers relations (3.20) and (3.21) of loc. cit. except for the fact
that the second equation in \cite[(3.20)]{MartelliSparks} and the
first equation in \cite[(3.21)]{MartelliSparks} need an extra minus
sign in front of the right hand side. We suppose that these signs
arose in loc. cit. from the same sign issue which was discussed above
regarding equation \eqref{YMSw}.

\paragraph{Some consequences of the exterior differential relations.} 
Taking into account the sign issues mentioned above (which originate
from the single relation \eqref{YMSw}), we showed in \cite{ga1} that
the following relations (the first five of which were originally given
in \cite{MartelliSparks}) can be obtained as consequences of the
exterior differential and codifferential relations \eqref{EDS},
\eqref{CDS} and of the $\check{Q}$-constraints \eqref{QConstr}:
\beqan
\label{Diff}
\begin{split}
& \dd(e^{3\Delta} V)=0~~,\\
& e^{-3\Delta}\dd(e^{3\Delta}b) = f-4\kappa~V~~,\\
& V\wedge \dd(\frac{e^{6\Delta}}{1-b^2}\iota_V Z)=0~~,\\
& e^{-6\Delta}\dd(e^{6\Delta}(\ast Z))=\ast F-b F~+4\kappa~ Y~~,\\
&e^{-12\Delta}\dd(e^{12\Delta}||V||\nu_\top)=8\kappa b~\hV\wedge \nu_\top~~,\\
& \dd(e^{9\Delta} Z)=0~~.
\end{split}
\eeqan

\

\paragraph{Some useful identities and intermediate steps.}
We list some identities deduced using the package {\tt Ricci}
\cite{Ricci} from \eqref{omega_param}-\eqref{Krel},
which involve the components $F_\top$, $F_\perp$ defined in
\eqref{Fdecomp} and which were used extensively in this paper:
\beqan
\label{Reductions}
\iota_{F_\top}\varphi=\langle F_\top,\varphi \rangle=4\tr_g(\hat\chi)~~&,&
~~\iota_{F_\perp}\psi=\langle F_\perp,\psi \rangle=-4\tr_g(\hat h)~~\nn\\
\iota_\varphi F_\perp=4\alpha_1~~&,&~~\iota_{F_\top}\psi=-4\alpha_2~~\nn\\
~~F_\perp\btu_3\psi=4\iota_{\alpha_1}\varphi~~&,&
~~F_\top\btu_2\varphi=4\iota_{\alpha_2}\varphi~~\nn\\
F_\top\btu_1\psi=-4{\alpha_2}\wedge\psi~~&,&~~F_\perp\btu_1\varphi=4\alpha_1\wedge\psi~\nn\\
F_\perp\btu_2\psi=-3\alpha_1\wedge\varphi+4\tr_g(\hat h)\psi+F_\perp^{(S)}~~&,&
~~F_\top\btu_2\psi=3\iota_{\alpha_2}\psi+4\tr_g(\hat\chi)\varphi+F_\top^{(S)} \nn\\
F_\top\btu_1\varphi=-3\alpha_2\wedge\varphi+4\tr_g(\hat\chi)\psi+\ast_7 F_\top^{(S)}~~&,&
~~F_\perp\btu_2\varphi=3\iota_{\alpha_1}\psi+4\tr_g(\hat h)\varphi+\ast_7 F_\perp^{(S)} \nn\\
F_\top\btu_1\varphi=-3\alpha_1\wedge\varphi &+& \ast_7 F_\top^{(S)}+4\tr_g(\hat\chi)\varphi \nn\\
F_\top^{(S)}=F_\top^{(27)}-\frac{4}{7}\tr_g(\hat\chi)\varphi~~~&,&~~~F_\perp^{(S)}=F_\perp^{(27)}-\frac{4}{7}\tr_g(\hat h)\psi~\nn\\
\iota_{(e_j\lrcorner  F_\top)}\varphi=-2\iota_{e_j}\iota_{\alpha_2}\varphi 
  &-&(2\chi_{ij}+g_{ij}\tr_g(\chi))e^i ~~\nn\\
\iota_{(e_j\lrcorner F_\perp)}\varphi=-4 e_j\lrcorner\alpha_1~~,~~~\iota_{(e_j\lrcorner  F_\perp)}\psi&=&2\iota_{e_j}\iota_{\alpha_1}\varphi 
 -2(\hat h_{ij}+g_{ij}\tr_g(\hat h))e^i~~\\
(\iota_{e^j}F_\top)\btu_1\psi=-2\alpha_2\wedge\iota_{e^j}\psi &-& e^j\wedge F^{(S)}_\top+2e^j\wedge\iota_{\alpha_2}\psi-6\chi_{ij}e^i\wedge\varphi  \nn\\
(\iota_{e^j}F_\perp)\btu_1\varphi=-4(e_j\lrcorner\alpha_1)\psi+2\alpha_1\wedge\iota_{e^j}\psi &-& 2e^j\wedge\ast_\perp 
F^{(S)}_\perp -2\tr_g(\hat h)e^j\wedge\varphi-e^j\wedge\iota_{\alpha_1}\psi-6\hat h_{ij}e^i\wedge\varphi ~.\nn
\eeqan
Equations \eqref{nabla_psi} (which are equivalent with each other) amount to:
\beqan
\label{nablapsi}
(\nabla_n\psi)_\top &=& \frac{1}{2||V||}\iota_{(\iota_{F_\top}\psi)}\psi~,\nn\\
(\nabla_{n}\psi)_\perp &=& \frac{b}{2||V||} \left[-(\iota_{_{F_\top}}\varphi)\psi+F_\top\btu_1\varphi-
\ast_\perp~F_\top\right]+\frac{1}{2}f_\perp\wedge\varphi~,\nn\\
(\nabla_{j}\psi)_\top&=& \frac{1}{2||V||}\Big[ \iota_{\iota_{(e_j\lrcorner F_\perp)}\psi}\psi
 - b \iota_{\iota_{(e_j\lrcorner F_\top)}\varphi}\psi-||V||\iota_{\iota_{(e^j\wedge f_\perp)}\varphi}\psi 
+4\kappa ~b~{e_{j}}\lrcorner\psi\Big] ~,  \\
(\nabla_{j}\psi)_\perp &=& \frac{b}{2||V||}\Big[ -\left(\iota_{(e_{j}\lrcorner F_\perp)}\varphi \right)\psi
+(e_{j}\lrcorner F_\perp)\btu_1\varphi+ e^j\wedge\ast_\perp~(F_\perp) \Big]~\nn\\
&+& \frac{1}{2||V||}\Big[ -(e_{j}\lrcorner F_\top)\btu_1\psi   +  (4\kappa-||V||f_\top)e^j\wedge\varphi    \Big]~~.\nn
\eeqan
Using the relation:
\be
\nabla_m\varphi=\ast_\perp(\nabla_m\psi)_\perp-\hat V\wedge\ast_\perp[(\nabla_m\hat V)\wedge\psi]~~,
\ee
and the formulas given in Step 1 of the proof of Theorem 2, relations \eqref{nablapsi} give:
\beqan
\label{NablaPhi}
&& (\nabla_n\varphi)_\perp=\ast_\perp(\nabla_n\psi)_\perp=\frac{1}{||V||}\iota_{(\alpha_1-b\alpha_2)}\psi~,\nn\\
&& (\nabla_n\varphi)_\top=-\ast_\perp[(\nabla_n\hat V)\wedge\psi]=-\frac{2}{||V||}\iota_{\alpha_2}\varphi~,\nn\\
&& (\nabla_j\varphi)_\top=-\ast_\perp[(\nabla_j\hat V)\wedge\psi]=\frac{1}{||V||}\Big[  - h^{(0)}_{ij}+b\chi^{(0)}_{ij}
+\frac{1}{7}\left(14\kappa b-8\tr_g(\hat h)-6b~\tr_g(\hat\chi)\right)g_{ij}\Big]\iota_{e_i}\varphi ~,\nn\\
&&(\nabla_j\varphi)_\perp=\ast_\perp(\nabla_j\psi)_\perp=\frac{3}{2}(\dd\Delta)_\perp\wedge\iota_{e_j}\varphi
-\frac{3}{2}e^j\wedge\iota_{(\dd\Delta)_\perp}\varphi  \nn\\
&&~~~~~~~~~~~~+\frac{1}{||V||}\Big[ [b h^{(0)}_{ij}-\chi^{(0)}_{ij}+\frac{1}{7}\left(b~\tr_g(\hat h)-\tr_g(\hat\chi)+7\kappa\right)g_{ij}\Big]\iota_{e_i\psi} ~.
\eeqan
The exterior differential constraints take the form:
\beqa
&&\dd b=e_{j_\sharp}\wedge\partial_j b+n_\sharp\wedge\partial_n b~
=2||V||\big[\alpha_1+\left(\kappa-\tr_g(\hat\chi)\right)\hat V\big]=-4\kappa V+f-3b\dd\Delta~,\\
&&\dd V=e_{j_\sharp}\wedge\nabla_j V+n_\sharp\wedge\nabla_n V~
= 2 \hat V \wedge( \alpha_2 +b \alpha_1)=3 V\wedge(\dd\Delta)_\perp~,\\
&&\dd\psi=e_{j_\sharp}\wedge(\nabla_j \psi)_\perp+e_{j_\sharp}\wedge\hat V\wedge(\nabla_j \psi)_\top
+n_\sharp\wedge(\nabla_n \psi)_\perp~\\
&&~~~~=-6(\dd\Delta)_\perp\wedge\psi+\hat V\wedge\frac{1}{||V||}\Big[(\alpha_1-b\alpha_2)\wedge\varphi
-(F^{(27)}_\perp-b\ast_\perp F^{(27)}_\top) \nn\\
&&~~~~~~-\frac{4}{7}\big(14\kappa b- 8\tr_g(\hat h)-6b~\tr_g(\hat\chi)\big)\psi \Big]~,\nn\\
&&\dd\varphi=\hat V\wedge\frac{1}{||V||}\Big[\iota_{(b\alpha_2-\alpha_1)}\psi 
+(\ast_\perp F_\perp^{(27)}-b F^{(27)}_\top)-\frac{3}{7}\big(14\kappa b- 8\tr_g(\hat h)-6b~\tr_g(\hat\chi)\big)\varphi \Big]\nn\\
&&~~~~~-\frac{9}{2}(\dd\Delta)_\perp\wedge\varphi-\frac{1}{||V||}(b F^{(27)}_\perp-\ast_\perp F^{(27)}_\top)+
\frac{4}{7||V||}\big( b~\tr_g(\hat h)-\tr_g(\hat\chi)+7\kappa  \big)\psi ~,
\eeqa
while the codifferential constraints \eqref{CDS} become: 
\beqa
\updelta V &=&-n\lrcorner(\nabla_n V)_\parallel-e_j\lrcorner(\nabla_j V)_\perp-e_j\lrcorner(\nabla_j V)_\parallel =16\kappa b- 8\tr_g(\hat h)-8b~\tr_g(\hat\chi)=8\kappa b+12||V||(\dd\Delta)_\top~,\nn\\
\updelta \psi &=&-n\lrcorner(\nabla_n\psi )_\parallel-e_j\lrcorner(\nabla_j \psi )_\perp-e_j\lrcorner(\nabla_j \psi )_\parallel \nn\\
&=&\frac{2}{||V||}\iota_{\alpha_2}\psi
+\frac{9}{2}\iota_{(\dd\Delta)_\perp}\psi +\frac{1}{||V||}(F^{(27)}_\top-b\ast_\perp F^{(27)}_\perp)+\frac{4}{7||V||}\big( b~\tr_g(\hat h)-\tr_g(\hat\chi)+7\kappa  \big)\varphi~.
\eeqa

\section{The multivalued map defined by a closed nowhere-vanishing one-form}
\label{app:fibration}

Let $\pi:{\tilde M}\rightarrow M$ be the universal cover of $M$ and
$\momega$ be a closed nowhere-vanishing one-form on $M$ whose
cohomology class we denote by $\f$. Let $\cF$ be the foliation of $M$
defined by $\momega$. Define a smooth real-valued function $\tilde{\mathfrak{h}}$ on
${\tilde M}$ as follows. Fixing a base point $p_0$ of ${\tilde M}$,
set:
\ben
\label{fdef}
\tilde{\mathfrak{h}}(p)\eqdef \int_{\gamma_{p_0,p}}{\tilde \momega}~~,~~(p\in {\tilde M})
\een
where ${\tilde \momega}\eqdef \pi^\ast(\momega)$ and $\gamma_{p_0,p}$ is
any curve on ${\tilde M}$ starting at $p_0$ and ending at $p$. Then
${\tilde \momega}=\dd \tilde{\mathfrak{h}}$ and the level sets ${\tilde
  L}_t\eqdef \tilde{\mathfrak{h}}^{-1}(\{t\})$ are the leaves of the foliation
${\tilde \cF}$ of ${\tilde M}$ defined by the distribution $\ker
{\tilde \momega}$. Notice that ${\tilde \cF}$ coincides with the
pull-back of $\cF$ through $\pi$. Let ${\tilde \phi}$ be the flow of
the lift ${\tilde v}$ of $v$ to the universal covering space.
Composing $\gamma_{p_0,p}$ from the right with an integral curve of
${\tilde v}$, one easily sees from \eqref{fdef} that ${\tilde \phi}_s$
maps ${\tilde \cF}_t$ into ${\tilde \cF}_{t+s}$. It follows that $\tilde{\mathfrak{h}}$ is a fibration which presents the universal covering space
${\tilde M}$ as a trivial bundle over $\R$, i.e. as the direct product
${\tilde L}\times \R$.  Composing $\gamma_{p_0,p}$ from the right with
the lift of a closed curve in $M$, one easily sees that $\tilde{\mathfrak{h}}$
satisfies the following relation for all $p\in {\tilde M}$ and all
$\alpha \in \pi_1(M)$:
\ben
\label{tshift}
\tilde{\mathfrak{h}}(p\alpha) =\tilde{\mathfrak{h}}(p)+\per_\f(\alpha)~~,
\een
where on the left hand side we use the right action of $\pi_1(M)$ on
${\tilde M}$. It follows that $\tilde{\mathfrak{h}}$ descends to a map 
$\bar{\mathfrak{h}}:M\rightarrow \R/\Pi_\f$. When $\omega$ is projectively rational,
we have $\Pi_\f=\Z a_\f$ for $a_\f=\inf(\Pi_\f\cap \N^\ast)$ and the
quotient $\R/\Pi_\f=\R/\Z a_\f$ is diffeomorphic with the unit circle via
the map $\R/\Z a_\f\ni [t]\stackrel{\mu_\f}{\rightarrow} e^{\frac{2\pi
    i}{a_\f}t}\in S^1$. Thus $\bar{\mathfrak{h}}$ induces a smooth map:
\be
\mathfrak{h}\eqdef \mu_\f\circ \bar{\mathfrak{h}}:M\rightarrow S^1~~
\ee
which is a fibration since $\tilde{\mathfrak{h}}$ is. In this case, it is also
clear from the above that $M$ is diffeomorphic with the mapping torus
$\mathbb{T}_{\phi_{a_\f}}(M)$. Notice that $\phi$ is the parallel transport
of the Ehresmann connection on the bundle $\mathfrak{h}:M\rightarrow S^1$ whose
distribution of (one-dimensional) horizontal subspaces is generated by
$v$; in particular, $\phi_a$ is the holonomy of this Ehresmann
connection. It is also clear that $\tilde{\mathfrak{h}}$ descends to a
well-defined map $\hat{\mathfrak{h}}:{\hat M}_\f\rightarrow \R$, where ${\hat
  M}_\f\simeq L\times \R$ is the integration cover of $\per_\f$ and
that $\hat{\mathfrak{h}}$ induces the map $\bar{\mathfrak{h}}$ upon taking the
quotient of its domain and codomain through the action of $\Pi_\f$.


\end{document}